\documentclass[aps,preprint,prd,showpacs,showkeys,nofootinbib,superscriptaddress,tightenlines]{revtex4-1}
\usepackage{amsmath, amssymb, amsfonts}
\usepackage[margin=1in]{geometry}
\usepackage{graphicx}
\usepackage{color}
\usepackage{xcolor}
\usepackage{lineno}
\usepackage{bm}
\usepackage{ulem}

\newcommand{\bea}{\begin{eqnarray}}
\newcommand{\eea}{\end{eqnarray}}
\newcommand{\beq}{\begin{equation}}
\newcommand{\eeq}{\end{equation}}
\newcommand{\bqa}{\begin{eqnarray}}
\newcommand{\eqa}{\end{eqnarray}}

\graphicspath{{Figures/}{figures/}}

\begin{document}
\title{Charm-Meson $\bm{t}$-channel Singularities\\
    in an Expanding Hadron Gas}
\author{Eric Braaten}
\email{braaten.1@osu.edu}
\affiliation{Department of Physics,
    The Ohio State University, Columbus, OH\ 43210, USA}

\author{Roberto Bruschini}
\email{bruschini.1@osu.edu}
\affiliation{Department of Physics,
    The Ohio State University, Columbus, OH\ 43210, USA}

\author{Li-Ping He}
\email{heliping@hiskp.uni-bonn.de}
\affiliation{Helmholtz-Institut f\"ur Strahlen- und Kernphysik and Bethe Center for Theoretical Physics, Universit\"at Bonn, D-53115 Bonn, Germany}

\author{Kevin Ingles}
\email{ingles.27@buckeyemail.osu.edu}
\affiliation{Department of Physics,
    The Ohio State University, Columbus, OH\ 43210, USA}

\author{Jun Jiang}
\email{jiangjun87@sdu.edu.cn}
\affiliation{School of Physics, Shandong University, Jinan, Shandong 250100, China}
\date{\today}

\begin{abstract}
    We study the time evolution of the numbers of charm mesons
    after the kinetic freezeout of the expanding hadron gas
    produced by the hadronization of the quark-gluon plasma from a central heavy-ion collision.
    The $\pi D$  reaction rates have contributions from a $D^\ast$ resonance in the $s$ channel.
    The $\pi D^\ast$ reaction rates are enhanced
    by $t$-channel singularities from an intermediate $D$.
    The contributions to reaction rates from $D^\ast$ resonances and $D$-meson $t$-channel singularities
    are sensitive to thermal mass shifts and thermal widths.
    In the expanding hadron gas, the $t$-channel singularities are regularized by the thermal $D$ widths.
    After kinetic freezeout, the thermal $D$ widths are dominated by coherent pion forward scattering.
    The contributions to $\pi D^\ast$ reaction rates from $t$-channel singularities
    are inversely proportional to the pion number density, which decreases to 0 as the hadron gas expands.
The $t$-channel singularities produce small but significant changes in charm-meson ratios from those predicted
    using the known $D^\ast$-decay branching fractions.
\end{abstract}


\maketitle

\section{Introduction}
\label{sec:intro}

The charm mesons that are most easily observed in high-energy experiments are
the pseudoscalar mesons $D^+$ and $D^0$ and the vector mesons $D^{\ast +}$ and $D^{\ast 0}$. 
The $D$'s have very long lifetimes because they decay by the weak interactions.
The $D^*$'s are resonances whose widths are several orders of magnitude narrower than those of most hadron resonances.
This remarkable feature arises because the $D^\ast$-$D$ mass splittings are very close to the pion mass $m_\pi$, which limits the phase space available for those decays $D^\ast \to D \pi$
that are kinematically allowed. 
The soft pion in the rest frame of the $D^\ast$
 suppresses the rate for a hadronic decay $D^\ast \to D \pi$,
making it comparable to that for a radiative decay $D^\ast \to D \gamma$.

Another consequence of $D^\ast$-$D$ mass splittings being approximately equal to $m_\pi$ is
that there are charm-meson reactions with a $t$-channel singularity.
A $t$-channel singularity is a divergence in the rate for a reaction in which an unstable particle
decays and one of the particles from its decay  is scattered \cite{Grzadkowski:2021kgi}.
The singularity arises because the scattered particle can be on  shell.
The adjective ``$t$-channel'' refers to the fact that in the case of a $2 \to 2$ reaction,
the scattered particle is exchanged in the $t$ channel.
The existence of  $t$-channel singularities was first pointed out by Peierls in 1961
in the case of $\pi N^\ast$ scattering through the exchange of a nucleon \cite{Peierls:1961zz}.
An example of a reaction with a  $t$-channel singularity in the Standard Model of particle physics
is $\nu_e Z^0 \to \nu_e Z^0$,
which can  proceed through the exchange of $\bar \nu_e$,
which is one of the decay products in $Z^0 \to \nu_e\bar \nu_e$.
The tree-level cross section diverges when the center-of-mass energy is greater than $\sqrt{2}\, M_Z$,
because the $\bar \nu_e$ can be on shell.
Another reaction with a  $t$-channel singularity is $\mu^+ \mu^- \to  W^+ e^- \bar \nu_e$,
which can proceed through exchange of $\nu_\mu$,
which is among  the decay products in $\mu^- \to  \nu_\mu e^-  \bar{\nu}_e$.
Melnikov and Serbo solved the divergence problem by taking into account
the finite transverse sizes of the colliding $\mu^+$ and $\mu^-$ beams \cite{Melnikov:1996na}.

A general discussion of   $t$-channel singularities has been presented by
Grzadkowski, Iglicki, and Mr\'owczy\'nski \cite{Grzadkowski:2021kgi}.
They pointed out that if a reaction with a $t$-channel singularity occurs in a thermal medium,
the divergence is regularized by the thermal width  of the exchanged particle.
The most divergent term in the reaction rate is replaced by a term inversely proportional to the thermal width.
A general discussion of the thermal regularization of  $t$-channel singularities
has recently been presented by Iglicki \cite{Iglicki:2022jjf}.

The simplest charm-meson reactions with a $t$-channel singularity are $\pi D^\ast \to \pi D^\ast$.
There are $t$-channel singularities  in 6 of the 10  scattering channels:
the elastic scattering of  $\pi^0 D^{\ast +}$,  $\pi^+ D^{\ast +}$, and $\pi^0 D^{\ast 0}$
and the inelastic reactions
$\pi^0 D^{\ast +} \to \pi^+ D^{\ast 0}$, $\pi^+ D^{\ast 0} \to \pi^0 D^{\ast +}$, and $\pi^- D^{\ast +} \to \pi^0 D^{\ast 0}$.
These reactions can proceed through the decay $D^\ast \to D \pi$ followed by the inverse decay $\pi D \to D^\ast$.
The $t$-channel singularity arises  because the  intermediate $D$ can be on shell.
The  cross section diverges
when the center-of-mass energy squared, $s$, is in a narrow interval close to the threshold.
In the case of the elastic scattering reaction $\pi D^{\ast} \to \pi D^{\ast}$,
the $t$-channel singularity  region is
\beq
2 M_*^2 - M^2 + 2 m_\pi^2 < s < (M_*^2 - m_\pi^2)^2/M^2,
\label{t-sing:piD}
\eeq
where $M_*$ and $M$  are the masses of $D^\ast$ and  $D$.
The lower endpoint of the  interval is above the threshold  $(M_*+m_\pi)^2$ by
approximately $2M\delta$, where $\delta = M_*-M-m_\pi$.
The small energy difference $\delta$ is comparable to isospin splittings.
The difference between the upper and lower endpoints  is approximately $8(M_*/M) m_\pi \delta$,
which has a further suppression factor of $m_\pi/M$.
The interval in the center-of-mass energy $\sqrt{s}$ is largest 
for the reaction $\pi^0 D^{\ast 0} \to \pi^0 D^{\ast 0}$,
extending from 6.1~MeV to 8.1~MeV   
above the threshold $M_*+m_\pi=2141.8$~MeV.  
Since the $t$-channel singularity arises because the intermediate $D$ can be on-shell,
the divergence in the cross section could be regularized by taking into account
the tiny decay width $\Gamma$ of the $D$, which would replace
the divergent term by a term with a factor $1/\Gamma$.
However, the resulting  enormous cross section is unphysical. One reason is that the  widths of
the incoming and outgoing $D^{\ast}$ are  larger than $\Gamma$ by about 8 orders of magnitude,
so the $D^*$ widths are more relevant than the width of $D$.

An obvious question is whether the $t$-channel singularities in charm-meson reactions
have any observable  consequences.
One situation in which there may be observable consequences is  the production of charm mesons
in relativistic heavy-ion collisions.
A central heavy-ion collision is believed
to produce a hot dense region of quark-gluon plasma  in which quarks and gluons are deconfined.
The quark-gluon plasma expands and cools until it reaches the temperature  for the crossover
transition to a hadron resonance gas in which the quarks and gluons are confined into hadrons.
After hadronization, the  hadron resonance gas continues to expand and cool until it  reaches kinetic freezeout,
after which the momentum distributions of the hadrons are no longer affected by scattering.
After kinetic freezeout,  the hadron gas continues to expand
as the  hadrons free-stream away from the interaction region.
The $t$-channel singularities in charm-meson reactions could have significant effects
either during the expansion and cooling of the hadron resonance gas between hadronization
and kinetic freezeout or during the expansion of the hadron gas after  kinetic  freezeout.

In the hadron gas produced by a heavy-ion collision,
$t$-channel singularities are regularized by the thermal widths of the hadrons.
The divergent term in the rate for a reaction
with a $t$-channel singularity is replaced by a term inversely proportional to the thermal width
of the hadron that can be on shell.
Between hadronization and kinetic freezeout, the thermal widths are determined by the temperature.
After kinetic freezeout, the  thermal widths are determined by the temperature at kinetic freezeout
and by the density of the system, which decreases as the hadron gas expands.

In this paper, we restrict our study of the effects of $t$-channel singularities in charm-meson reactions
to the expanding hadron gas after kinetic freezeout.
The restriction to after kinetic freezeout offers many simplifications.
The only  hadrons in the hadron resonance gas that remain are the most stable ones
whose lifetimes $\tau$ are long enough  that $c \tau$ is larger than the size of the hadron gas, 
whose order of magnitude is 10~fm.
The most abundant hadrons by far are pions.
The temperature at kinetic freezeout is low enough that the interactions
of charm mesons and pions can be described by a chiral effective field theory.
The relevant charm mesons are $D^+$, $D^0$,  $D^{\ast +}$, and $D^{\ast 0}$.
The decays of $D^{\ast +}$ and $D^{\ast 0}$, whose lifetimes satisfy $c\tau > 2000$~fm, 
occur long after kinetic freezeout.
The dominant contribution to the thermal width of  a charm meson comes from 
the coherent forward scattering of pions
and is proportional to the pion number density $\mathfrak{n}_\pi$, 
which decreases to 0 as the hadron gas expands.
A $D$-meson $t$-channel singularity therefore gives a contribution to the reaction rate
inversely proportional to $\mathfrak{n}_\pi$.
The factor of $1/\mathfrak{n}_\pi$ can cancel a multiplicative factor of  $\mathfrak{n}_\pi$ in a term in a rate equation,
increasing the importance of that term at late times.
In Ref.~\cite{Braaten:2022qag}, we showed that $D$-meson $t$-channel singularities in the reactions
$\pi D^\ast \to \pi D^\ast$ produce significant modifications to the ratios of charm mesons produced by heavy-ion collisions.
In this paper, we present the details of the calculations that lead to this surprising result.

The rest of the paper is organized as follows.
In Section~\ref{sec:prelim}, we establish our notation for various properties of charm mesons.
In Section~\ref{sec:HeavyIons}, we describe the hadron resonance gas produced by a heavy-ion collision
and we present a simple model for its time evolution.
In Section~\ref{sec:PionGas}, we calculate the  mass shifts and thermal widths of charm meson in a pion gas.
In Section~\ref{sec:ReactionRates}, we calculate reaction rates of charm meson and pions in a pion gas.
In Section~\ref{sec:Evolution}, we solve the rate equations for the charm-meson number densities
in the expanding hadron gas produced by a heavy-ion collision after kinetic freezeout.
We show that $D$-meson $t$-channel singularities produce small but significant changes
in the ratios of charm-meson abundances.
We summarize our results in Section~\ref{sec:conclusion}.
In Appendix~\ref{sec:FeynmanRules}, we give the  Feynman rules for heavy-hadron $\chi$EFT
used in the calculations in Sections~\ref{sec:PionGas} and \ref{sec:ReactionRates}.
In Appendix~\ref{app:PionIntegral}, we calculate a thermal average  over the pion momentum distribution
that is sensitive to isospin splittings.


\section{Charm mesons}
\label{sec:prelim}

In this section, we introduce notation for the masses and decay widths of charm mesons.
We also describe simple relations between  
numbers of charm mesons that involve $D^*$ branching fractions.

\subsection{Masses and widths}
\label{sec:M,Gamma}

We denote the masses of the pseudoscalar charm mesons  $D^+$ and $D^0$ by $M_+$ and $M_0$
and the masses of the vector charm mesons $D^{\ast+}$ and $D^{\ast0}$  by $M_{\ast+}$ and $M_{\ast0}$.
We denote the $D^{\ast a}-D^b$ mass difference by $\Delta_{ab} = M_{*a} - M_b$.
The average of the four mass differences is $\Delta = 141.3$~MeV. 
We denote the masses of the pions $\pi^\pm$ and $\pi^0$ by $m_{\pi +}$ and  $m_{\pi 0}$.
We sometimes also denote the mass of the pion produced in the transition $D^{*a} \to  D^b \pi$ by $m_{\pi ab}$:
$m_{\pi ab} =m_{\pi 0}$  if $a=b$, $m_{\pi ab} =m_{\pi +}$ if $a \ne b$.
When isospin splittings can be neglected, we take the pion mass $m_\pi$ to be the
average over the three pion flavors: $m_\pi = 138.0$~MeV. 
Many reaction rates are sensitive to the difference between a $D^\ast$ mass and a $D \pi$ scattering threshold.
The differences for the transitions $D^\ast \to D \pi$ that conserve electric charge are
\begin{subequations}
    \bea
    \Delta_{00}-m_{\pi 0} &=& 7.04 \pm 0.03~\mathrm{MeV},
    \label{delta00}
    \\
    \Delta_{+0}-m_{\pi +} &=& 5.855 \pm 0.002~\mathrm{MeV},
    \label{delta0+}
    \\
    \Delta_{++}-m_{\pi 0} &=& 5.63\pm 0.02~\mathrm{MeV},
    \label{delta0+0}
    \\
    \Delta_{0+}-m_{\pi +} &=& - 2.38 \pm 0.03~\mathrm{MeV}.
    \label{delta+-}
    \eea
    \label{Deltaij}%
\end{subequations}

The negative value of $\Delta_{0+}-m_{\pi +}$ implies that the decay $D^{*0} \to  D^+ \pi^-$  is kinematically forbidden.

We denote the total decay widths of the vector charm mesons $D^{\ast+}$ and $D^{\ast0}$
by $\Gamma_{\ast+}$ and $\Gamma_{\ast0}$.
The decay width  of $D^{\ast+}$ is measured.
The decay width of $D^{\ast0}$ can be predicted using Lorentz invariance,
chiral symmetry, isospin symmetry, and measured $D^\ast$  branching fractions:
\begin{equation}
    \frac{\mathrm{Br}[D^{\ast0} \to D^0\pi^0]~ \Gamma_{\ast0}}{\mathrm{Br}[D^{\ast+} \to D^0\pi^+] \, \Gamma_{\ast+}}
    =  \frac{\lambda^{3/2}(M_{\ast0}^2, M_0^2,m_{\pi 0}^2)/M_{\ast0}^5}{2\, \lambda^{3/2}(M_{\ast+}^2, M_0^2,m_{\pi +}^2)/M_{\ast+}^5},
    \label{Gamma*ratio}
\end{equation}
where $\lambda(x,y,z) = x^2+y^2+z^2-2(xy+yz+zx)$.
The branching fractions for the decays $D^\ast  \to D \pi$ are
\begin{subequations}
    \begin{eqnarray}
        B_{+0} &\equiv&   \mathrm{Br}\big[D^{\ast +} \to D^0 \pi^+ \big] =  (67.7 \pm 0.5)\%,
        \label{Br*+0}
        \\
        B_{00} &\equiv&   \mathrm{Br}\big[D^{\ast 0} \to D^0 \pi^0 \big] =  (64.7 \pm 0.9)\%,
        \label{Br*00}
\\
    B_{++} &\equiv&   \mathrm{Br}\big[D^{\ast +} \to D^+ \pi^0 \big] =  (30.7 \pm 0.5)\%.
    \label{Br*++}
    \end{eqnarray}
    \label{Br*ab}%
\end{subequations}
The $D^\ast$ decay widths are
\begin{subequations}
    \begin{eqnarray}
        \Gamma_{\ast+} &\equiv&   \Gamma[D^{\ast+}] =  83.4 \pm 1.8~\mathrm{keV},
        \label{Gamma*1}
        \\
        \Gamma_{\ast0} &\equiv& \Gamma[D^{\ast0}] =  55.4 \pm 1.5~\mathrm{keV}.
        \label{Gamma*0}
    \end{eqnarray}
    \label{Gamma*}%
\end{subequations}
The $D^\ast$ radiative decay rates are
\begin{subequations}
    \bea
    \Gamma_{*+,\gamma} &\equiv& \Gamma\big[D^{\ast+} \to D^+ \gamma\big] = 1.3 \pm 0.3~\mathrm{keV} ,
    \label{GammaD*D+gamma}
    \\
    \Gamma_{*0,\gamma} &\equiv& \Gamma\big[D^{\ast0} \to D^0 \gamma\big] = 19.6  \pm 0.7~\mathrm{keV} .
    \label{GammaD*D0gamma}
    \eea
    \label{GammaD*Dgamma}%
\end{subequations}
The decay widths of the spin-0 charm mesons $D^+$ and $D^0$ are smaller than those for $D^\ast$
by about 8 orders of magnitude, because they only decay through weak interactions.

The interactions of low-energy pions with momenta at most comparable to $m_\pi$
can be described by chiral effective field theory ($\chi$EFT) \cite{Weinberg:1978kz}.
The self-interactions of pions in $\chi$EFT at leading order (LO)
are determined by the pion decay constant $f_\pi$.
It can be determined from the partial decay rate for $\pi^+$ into $\mu^+ \, \nu_\mu$:
\beq
\Gamma \big[ \pi^+ \to \mu^+ \nu_\mu \big]  =
\frac{1}{8 \pi} |V_{ud}|^2 G_F^2 f_\pi^2
\frac{m_\mu^2 (m_{\pi +}^2-m_\mu^2)^2}{m_{\pi +}^3}.
\label{Gammapi+}
\eeq
From the measured decay rate, we obtain  $f_\pi = 131.7$~MeV. 

The interactions of charm mesons with low-energy pions can be described by heavy-hadron $\chi$EFT
(HH$\chi$EFT) \cite{Burdman:1992gh,Wise:1992hn,Cheng:1992xi}.
The first-order corrections in HH$\chi$EFT include terms suppressed by $m_\pi/M$ and $\Delta/M$.
Isospin splittings can be treated as second-order corrections.
The partial decay rate for $D^* \to D \pi$  in HH$\chi$EFT
at LO is sensitive to isospin splittings through a multiplicative factor $(\Delta^2 - m_\pi^2)^{3/2}$.
Isospin splittings can be taken into account in the partial decay rate for $D^{\ast a} \to D^b \pi$
by replacing $\Delta$ by $\Delta_{ab}  = M_{\ast a} - M_b$ and $m_\pi$ by the mass $m_{\pi ab}$ of the emitted pion.
The resulting expression for the partial decay rate is
\beq
\Gamma \big[D^{\ast a} \to D^b \pi \big] =
\frac{g_\pi^2}{12\pi \, f_\pi^2} \, (2 - \delta_{ab})\,
\left( \Delta_{ab}^2 - m_{\pi ab}^2 \right)^{3/2} \theta(\Delta_{ab} - m_{\pi ab}).
\label{GammaD*Dpi}
\eeq
The dimensionless coupling constant $g_\pi$ can be determined from measurements of the decay
$D^{\ast+} \to D^0 \pi^+$: $g_\pi = 0.520 \pm 0.006$.

\subsection{Charm-meson numbers}
\label{sec:charmratios}

The numbers of  charm hadrons created in a high-energy  collision must be inferred from the numbers that are detected.
The decay of $D^{*0}$ always produces $D^0$.
The decay of $D^{*+}$ produces $D^0$ and $D^+$ with branching fractions $B_{+0}$ and $1-B_{+0}$.
We denote the numbers of $D^0$, $D^+$, $D^{\ast 0}$, and $D^{\ast +}$
observed in some kinematic region by
$N_{D^0}$, $N_{D^+}$, $N_{D^{\ast 0}}$, and $N_{D^{\ast +}}$.
The observed numbers of $D^0$  and $D^+$ can be predicted in terms of the
numbers $(N_{D^a})_0$ and $(N_{D^{\ast a}})_0$ before $D^\ast$ decays and  the branching fraction $B_{+0}$:
\begin{subequations}
    \bea
    N_{D^0} &=&
    \left( N_{D^0} \right)_0  + \left( N_{D^{\ast 0}} \right)_0  + B_{+0}  \left( N_{D^{\ast +}} \right)_0 ,
    \label{ND0}
    \\
    N_{D^+} &=&
    \left( N_{D^+} \right)_0 + 0  +  (1 - B_{+0}) \left( N_{D^{\ast +}} \right)_0 .
    \label{ND+}
    \eea
    \label{ND0,Dp}%
\end{subequations}
The last two terms in each equation come from $D^{\ast 0}$ and $D^{\ast +}$ decays, respectively.
The difference between the numbers of $D^0$ and $D^+$ can be expressed as
\beq
N_{D^0} - N_{D^+} = 2 B_{+0}  \left( N_{D^{\ast +}} \right)_0
+ \left( N_{D^0} -  N_{D^+} \right)_0  + \left( N_{D^{\ast 0}} - N_{D^{\ast +}} \right)_0 .
\label{ND+-ND-}
\eeq
The simple relations in Eqs.~\eqref{ND0,Dp} have been assumed in all previous analyses of charm-meson production.
We will show in this paper that  these relations can be modified by $t$-channel singularities.

In a high-energy hadron collision,
the numbers of $D^0$ and $D^+$ created in some kinematic region should be approximately equal by isospin symmetry:
$(N_{D^0})_0  \approx (N_{D^+}) _0$.
Similarly, the numbers of $D^{\ast0}$ and $D^{\ast+}$ created should be approximately equal:
$(N_{D^{\ast 0}})_0  \approx (N_{D^{\ast +}})_0$.
The deviations from isospin symmetry in the charm cross section should be negligible,
because isospin splittings are tiny compared to the energy  available for producing additional hadrons.
The  decays of bottom hadrons give isospin-violating contributions to charm-meson production,
but the bottom cross section is much smaller than the charm cross section at present-day colliders.

The charm mesons that are most  easily observed  at a hadron collider are $D^0$, $D^+$, and $D^{\ast+}$,
because they have significant decay modes with all charged particles.
If the only reactions of charm mesons after their production are $D^\ast$ decays,
the ratios of the observed numbers of $D^0$, $D^+$, and $D^{\ast+}$
are determined by the vector/pseudoscalar ratio before $D^*$ decays,
which we denote by $(N_{D^\ast}/N_D)_0$.  Assuming isospin symmetry,
that ratio can be expressed in terms of the observed numbers $N_{D^0}$, $N_{D^+}$, and $N_{D^{\ast +}}$:
\beq
\left(\frac{N_{D^\ast}}{N_D}\right)_{\!\! 0}  \approx \frac{2N_{D^{\ast+}}} {N_{D^0} + N_{D^+} - 2\,N_{D^{\ast+}}}.
\label{ND/ND*}
\eeq

Isospin symmetry also implies that
there is a combination of the three observed numbers that is completely determined by $B_{+0}$:
\beq
\frac{N_{D^0} - N_{D^+}}{N_{D^{\ast+}}} \approx 2\,B_{+0}= 1.35 \pm 0.01.
\label{DeltaND/ND*+}
\eeq
Deviations from this prediction must come
either from initial conditions that deviate from isospin symmetry or from charm-meson reactions
other than $D^\ast$ decays that also violate isospin symmetry.
Reactions with $t$-channel singularities are examples of such reactions.


\section{Heavy-ion collisions}
\label{sec:HeavyIons}

In this section, we present a simple model for the hadron resonance gas produced by a central relativistic heavy-ion collision.
We describe the Statistical Hadronization Model for the  abundances of hadrons produced by a heavy-ion collision.
Finally we describe the number densities of pions and charm mesons
both before and after the kinetic freezeout of the hadron gas.

\subsection{Expanding hadron gas}
\label{sec:Expansion}

The central collision of relativistic heavy ions is believed to produce
a quark-gluon plasma (QGP) consisting of deconfined quarks and gluons
which then evolves into a hadron resonance gas (HRG) consisting of hadrons.
A heavy-ion collision involves multiple stages:
the collisions of the  Lorentz-contracted nucleons in the nuclei,
the formation and thermalization of the QGP,
the expansion and cooling of the QGP,
the hadronization of the QGP into the HRG,
the expansion and cooling of the HRG as most of the resonances decay,
the kinetic freezeout of the HRG when its density becomes too low for
collisions to change momentum distributions,
and finally the expansion of the resulting hadron gas by the free-streaming of hadrons.
For each stage, complicated phenomenological models have been developed to provide quantitative descriptions
\cite{Bass:1998ca,Kolb:2003dz,Gelis:2010nm,Busza:2018rrf,Elfner:2022iae}.

A natural variable to describe the space-time evolution of the system created by the heavy-ion collision is
the proper time $\tau$ since the collision.
A simple phenomenological model that may describe the essential features of the system
between the equilibration of the QGP and the kinetic freezeout of the HRG is a homogeneous system
with volume $V(\tau)$ in thermal equilibrium at temperature $T(\tau)$.
We denote the proper time just after hadronization by $\tau_H$ and the proper time at kinetic freezeout by $\tau_\mathrm{kf}$.
The volume increases from $V_H$ at $\tau_H$ to $V_\mathrm{kf}$ at $\tau_\mathrm{kf}$,
while the temperature decreases from $T_H$ to $T_\mathrm{kf}$.
These proper times, volumes, and temperatures can be determined
by fitting the outputs of simplified hydrodynamic models for heavy-ion collisions.
Values of the volumes $V_H$ and $V_\mathrm{kf}$ and the temperatures $T_H$ and $T_\mathrm{kf}$
for various heavy-ion colliders are given in Refs.~\cite{ExHIC:2011say,ExHIC:2017smd}.
An explicit parametrization of the volume $V(\tau)$ can be  obtained by assuming
the boost-invariant longitudinal expansion proposed by Bjorken~\cite{Bjorken:1982qr}
and an accelerated transverse expansion caused by the pressure of the QGP  before hadronization
and by the pressure of the HRG after hadronization \cite{Ko:1998fs}.
The parametrization of $V(\tau)$ for the HRG  between hadronization and kinetic freezeout is \cite{Chen:2003tn}
\bea
V(\tau) = \pi \big[ R_H + v_H(\tau - \tau_H) + a_H(\tau - \tau_H)^2/2 \big]^2\, c\tau
\qquad
(\tau_H < \tau < \tau_\mathrm{kf}),
\label{V-tau<}
\eea
where $R_H$, $v_H$, and $a_H$ are the transverse radius, velocity, and acceleration at $\tau_H$.
If the transverse velocity  $v_H + a_H(\tau - \tau_H)$ reaches the speed of light before kinetic freezeout,
the subsequent transverse expansion proceeds at the constant velocity $c$.
The temperature $T(\tau)$ can be determined by assuming isentropic expansion.
The parametrization of $T(\tau)$ for the HRG between hadronization and  kinetic freezeout in  Ref.~\cite{Chen:2007zp} is
\bea
T(\tau) = T_H + (T_\mathrm{kf} - T_H) \left( \frac{\tau - \tau_H}{\tau_\mathrm{kf} - \tau_H} \right)^{4/5}
\qquad
(\tau_H < \tau < \tau_\mathrm{kf}).
\label{T-tau<}
\eea
The parameters in $V(\tau)$ and $T(\tau)$ for central Pb-Pb collisions at 5.02~TeV are given  in Ref.~\cite{Abreu:2020ony}.
Hadronization and kinetic freezeout occur at the proper times
$\tau_H=10.2~\mathrm{fm}/c$  and $\tau_\mathrm{kf} = 21.5$~fm/$c$. 
Between hadronization and kinetic  freezeout,
the temperature decreases from $T_H=156$~MeV to $T_\mathrm{kf}=115$~MeV.
The transverse radius increases from $R_H=13.0$~fm to 24.0~fm. 
The transverse speed increases from $v_H=0.78\,c$ to $c$ at $\tau = 12.7$~fm/$c$
and then remains constant at $v_\mathrm{kf} = c$.

After kinetic freezeout, the system continues to expand,
but the momentum distributions of the hadrons are those for a fixed  temperature:  $T(\tau)=T_\mathrm{kf}$.
A simple model for the volume $V(\tau)$ is continued longitudinal expansion at the speed of light
and transverse expansion at the same speed $v_\mathrm{kf}$ as at kinetic  freezeout:
\bea
V(\tau) = \pi \left[ R_\mathrm{kf} + v_\mathrm{kf}(\tau - \tau_\mathrm{kf}) \right]^2 c\tau
\qquad
(\tau > \tau_\mathrm{kf}).
\label{V-tau>}
\eea
We assume the system remains homogeneous throughout the expanding volume $V(\tau)$.
In the absence of further interactions,
the number density for each stable hadron would decrease in proportion to $1/V(\tau)$ as $\tau$ increases.

Charm  quarks and antiquarks are created
in the hard collisions of the nucleons that make up the heavy ions.
Charm quarks are assumed to quickly thermalize with the QGP
at the temperature $T(\tau)$. They are not in chemical equilibrium,
because the temperature of the QGP
is too low for gluon and light-quark collisions to create charm quark-antiquark pairs.
The low density of charm quarks suppresses the annihilation of charm quarks and antiquarks, so
the charm-quark and charm-antiquark numbers are essentially conserved.
Conservation of charm-quark number determines the charm-quark fugacity
$g_c(\tau)$ in terms of the temperature $T(\tau)$ and the volume $V(\tau)$.
After hadronization, charm  hadrons are in thermal equilibrium with the HRG
at the temperature $T(\tau)$.
Their number densities evolve according to rate equations consistent with the
conservation of charm-quark number.
The charm  hadrons are assumed to remain in thermal equilibrium until kinetic freezeout,
after which they free-stream to the detector.

\subsection{Statistical Hadronization Model}
\label{sec:SHM}

The Statistical Hadronization Model (SHM) is a model for the abundances of hadrons
produced by a heavy-ion collision  \cite{Braun-Munzinger:2003pwq}.
According to the SHM, the hadronization of the QGP into the HRG
occurs while they are in chemical and thermal equilibrium with each other
at a specific hadronization temperature $T_H$ 
that can be identified with the temperature of the crossover between the QGP and the HRG.
At hadronization, the number density of any spin state of a light hadron depends only on the hadron mass and
the temperature $T_H$.
(At sufficiently high rapidity or at lower heavy-ion collision energies,
a number density can also depend on the baryon chemical potential.)
The SHM takes into account the subsequent decays of hadron resonances,
which increase the abundances of the lighter and more stable hadrons.
The SHM does not take into account the scattering reactions that allow the HRG to remain in thermal equilibrium
after hadronization.

The SHM can also describe the abundances of charm hadrons produced by a heavy-ion collision \cite{Andronic:2003zv}.
According to the SHM, charm hadrons are created during hadronization
while the QGP and HRG are in thermal equilibrium at the temperature $T_H$.
At hadronization, the number density of any spin state of a charm hadron
is determined only by its mass, 
the hadronization temperature $T_H$, and multiplicative factors of the charm-quark fugacity $g_c$.
The number density of a charm hadron with a single charm quark or antiquark
is larger than the number density in chemical equilibrium by the   factor $g_c$.
The number density of a hadron whose heavy constituents consist of $n$ charm quarks and antiquarks
is  larger than the number density in chemical equilibrium by   $g_c^n$ \cite{Andronic:2021erx}.

The SHM gives simple predictions for charm-hadron ratios at hadronization.
Since the mass of a charm hadron is so large compared to $T(\tau)$,
its momentum distribution in the HRG can be approximated by a relativistic Boltzmann distribution.
The cham-hadron fugacity enters simply as a multiplicative factor.
At hadronization, the charm-hadron fugacity is the product of the  charm-quark fugacity $g_c$ and the number of spin states.
The factor of $g_c$ cancels in ratios of charm-hadron number  densities.
The ratio of the numbers of vector and pseudoscalar charm mesons at hadronization
is predicted to be
\beq
\frac{N_{D^\ast}}{N_D}  = 3\, \frac{M_\ast^2K_2(M_\ast/T_H)}{M^2K_2(M/T_H)},
\label{ND*/ND}
\eeq
where $M$ and $M_\ast$ are the masses of $D$ and $D^\ast$, 
which we take to be the isospin averages of the masses of the pseudoscalar and vector charm mesons, respectively.
At the hadronization temperature $T_H=156$~MeV,
the vector/pseudoscalar ratio is predicted to be $N_{D^\ast}/N_D = 1.339$. 
Ratios of the charm-hadron number  densities for isospin partners
are given by equations analogous to Eq.~\eqref{ND*/ND} but without the factor of 3.
The predicted ratio for pseudoscalar charm mesons at hadronization is $N_{D^0}/N_{D^+}= 1.028$. 
The predicted ratio for vector charm mesons at hadronization is  $N_{D^{\ast 0}}/N_{D^{\ast +}} =  1.020$. 
The SHM  predictions for charm-hadron ratios are modified from the simple predictions at hadronization
by the feeddown from the decays of higher charm-hadron resonances.

The SHM has been applied to Pb-Pb collisions for at nucleon-nucleon center-of-mass energy 
$\sqrt{s_{NN}}=5.02$~TeV in Ref.~\cite{Andronic:2021erx} for various centrality bins; we choose to
focus only on the most central collisions.
The charm-quark fugacity at hadronization has been determined to be $g_c = 29.6 \pm  5.2$. 
Predictions for the multiplicities $dN/dy$ for 4 charm mesons and 2 charm baryons
at  midrapidity ($|y|  < \tfrac12$) are given in Table 1 of Ref.~\cite{Andronic:2021erx}.
The expanding hadron gas is modeled by a ``core’’ in which the formation of charm hadrons is described by the SHM
and a ``corona’’ in which their formation is described by that in $pp$ collisions.
For collisions in the centrality range 0-10\%, the predicted multiplicities  $dN/dy$ from the core
for $D^0$, $D^+$, and $D^{\ast+}$ are 6.02, 2.67, and 2.36,  respectively,
with error bars consistent with those from a multiplicative factor of  $g_c = 29.6 \pm  5.2$.
The error bars on ratios of the multiplicities should be much smaller than 18\%, 
but they cannot be determined from the results presented in  Ref.~\cite{Andronic:2021erx}.
The predicted additional multiplicities $dN/dy$ from the corona
for $D^0$, $D^+$, and $D^{\ast+}$ are 0.396, 0.175, and 0.160,    respectively.
The effect of the corona is to increase all three multiplicities by about 7\%.  

An SHM prediction for the vector/pseudoscalar ratio before $D^*$ decays can be obtained
by inserting the predicted total multiplicities for $D^0$, $D^+$, and $D^{\ast+}$ into Eq.~\eqref{ND/ND*}:
$(N_{D^\ast}/N_D)_0 = 1.194$. 
This is significantly smaller than the ratio 1.339  at hadronization
predicted by Eq.~\eqref{ND*/ND}, but also includes feeddown effects from decays of higher resonances.
The SHM prediction for the ratio $(N_{D^0} - N_{D^+})/N_{D^{\ast+}}$ is 1.42.  
This is larger than the isospin-symmetry prediction 1.35  in Eq.~\eqref{DeltaND/ND*+} by about
 5\%.
 This, in turn, is larger then the thermal isospin-symmetry deviations at hadronization predicted by
 the SHM, which are less than 3\%.

\subsection{Pion momentum distributions}
\label{sec:Pions}

The  temperature $T$ of the HRG is comparable to the pion mass $m_\pi$.
By isospin symmetry, the pions $\pi^-$, $\pi^0$, and $\pi^+$ all have the same number density $\mathfrak{n}_\pi$.
The  number density for pions in chemical  and thermal equilibrium at temperature $T$ is
\begin{equation}
    \mathfrak{n}_\pi^\mathrm{(eq)} =
    \int  \frac{d^3q}{(2\pi)^3} \frac{1}{e^{\beta \omega_q}-1} ,
    \label{npieq}
\end{equation}
where $\omega_q = \sqrt{m_\pi^2 +q^2}$ and $\beta = 1/T$ is the inverse  temperature.
At the kinetic freezeout temperature $T_\mathrm{kf} = 115$~MeV,
the equilibrium number density  is $\mathfrak{n}_\pi^\mathrm{(kf)}=1/(3.95~\mathrm{fm})^3$. 

Between hadronization and kinetic freezeout, the pions are in chemical  and thermal equilibrium.
The temperature $T(\tau)$ of the HRG decreases as the proper time $\tau$ increases.
The momentum distribution $\mathfrak{f}_\pi$ of the pions is the Bose-Einstein distribution:
\begin{equation}
    \mathfrak{f}_\pi(\omega_q)=  \frac{1}{e^{\beta \omega_q}-1} \qquad (\tau_H < \tau < \tau_\mathrm{kf}),
    \label{fpi-k:<}
\end{equation}
where $\beta = 1/T(\tau)$.  The temperature $T(\tau)$ can be parametrized as in Eq.~\eqref{T-tau<}.

After kinetic freezeout,  the temperature remains constant: $T(\tau > \tau_\mathrm{kf}) = T_\mathrm{kf}$.
The pion number density decreases in inverse proportion to the volume $V(\tau)$ of the expanding hadron gas:
\bea
\mathfrak{n}_\pi(\tau) = \frac{V_\mathrm{kf}}{V(\tau)} \mathfrak{n}_\pi^\mathrm{(kf)}
\qquad
(\tau > \tau_\mathrm{kf}),
\label{npi-tau}
\eea
where $\mathfrak{n}_\pi^\mathrm{(kf)}$ is the equilibrium pion number density in Eq.~\eqref{npieq} at the temperature $T_\mathrm{kf}$
and $V_\mathrm{kf}$ is the volume of the hadron gas at kinetic freezeout.
The volume $V(\tau)$  can be parametrized as in Eq.~\eqref{V-tau>}.
The normalization of the momentum distribution $\mathfrak{f}_\pi$ of the pions 
is determined by the pion number density $\mathfrak{n}_\pi$:
\begin{equation}
    \mathfrak{f}_\pi(\omega_q)=
    \frac{\mathfrak{n}_\pi}{\mathfrak{n}_\pi^\mathrm{(kf)}}\,  \frac{1}{e^{\beta_\mathrm{kf}\,  \omega_q}-1}
    \qquad (\tau > \tau_\mathrm{kf}),
    \label{fpi-k:>}
\end{equation}
where $\beta_\mathrm{kf} = 1/T_\mathrm{kf}$.

We use angular brackets to denote the average over the momentum distribution of a pion.
The thermal average of a function $F(\bm{q})$ of the pion momentum is
\beq
\big\langle F(\bm{q}) \big\rangle =
\int \frac{d^3q}{(2\pi)^3}\,  \mathfrak{f}_\pi(\omega_q) \, F(\bm{q})
\bigg/  \int \frac{d^3q}{(2\pi)^3}\,  \mathfrak{f}_\pi(\omega_q).
\label{<F>}
\eeq
The thermal average depends on the temperature $T$.
After kinetic freezeout, the pion number density $\mathfrak{n}_\pi$
cancels in the thermal average in Eq.~\eqref{<F>}.
If the thermal average is sensitive to the flavor $i$ of the pion, the pion energy in Eq.~\eqref {<F>}
should be replaced by $\omega_{iq}  = \sqrt{m_{\pi i}^2+q^2}$.

The multiplicities of $\pi^+$ and $\pi^-$ 
produced by Pb-Pb collisions at the LHC with $\sqrt{s_{NN}}=5.02$~TeV
have been measured by the ALICE collaboration \cite{ALICE:2019hno}.
The pion multiplicity  averaged over $\pi^+$ and $\pi^-$ from collisions in the centrality range 0-10\% is
\beq
dN_\pi/dy = 769  \pm 34.
\label{dNpi/dy}
\eeq
The total pion multiplicity for $\pi^+$, $\pi^-$, and $\pi^0$ is 3 times larger.
A fit of the SHM to hadron abundances at midrapidity in Pb-Pb collisions at the LHC 
with $\sqrt{s_{NN}} = 2.76$~TeV has been presented in Ref.~\cite{Andronic:2017pug}.
The central values of the SHM fits 
for the multiplicities of $\pi^+$ and $\pi^-$ are lower than the data by about 10\% ,
which is comparable to the experimental error bars.

\subsection{Charm-meson momentum distributions}
\label{sec:CharmMesons}

We denote the number densities of the charm mesons $D^+$, $D^0$, $D^{\ast+}$, and  $D^{\ast0}$
in the hadron gas by $\mathfrak{n}_{D^+}$, $\mathfrak{n}_{D^0}$,
$\mathfrak{n}_{D^{\ast+}}$, and $\mathfrak{n}_{D^{\ast0}}$, respectively.
Since charm-meson masses are so much larger than the temperature $T$,
the momentum distributions of the charm mesons can be approximated by relativistic Boltzmann distributions.
If the charm mesons were in both chemical and thermal equilibrium,
their number densities would be determined by the temperature $T$:
\begin{subequations}
    \begin{eqnarray}
        \mathfrak{n}_{D^a}^{(\mathrm{eq})} &=&
        \int  \frac{d^3q}{(2\pi)^3} \exp\!\big(\!-\beta \sqrt{M_a^2+p^2}\, \big)
        = \frac{M_a^2 \, K_2(M_a/T)}{2\pi ^2/T} ,
        \label{nD:<}
        \\
        \mathfrak{n}_{D^{*a}}^{(\mathrm{eq})} &=&
        3 \int  \frac{d^3q}{(2\pi)^3} \exp\!\big(\!-\beta \sqrt{M_{\ast a}^2+p^2}\,\big)
        =\frac{3\, M_{\ast a}^2 \, K_2(M_{\ast a}/T)}{2\pi ^2/T}.
        \label{nD*:<}
    \end{eqnarray}
    \label{nD,nD*-pq}%
\end{subequations}
However the charm mesons in the expanding hadron gas are not in chemical equilibrium.
The number densities $\mathfrak{n}_{D^a}(\tau)$
and $\mathfrak{n}_{D^{*a}}(\tau)$ evolve with the proper time according to rate equations
consistent with the conservation of charm-quark number.
The momentum distributions of the charm mesons are
\begin{subequations}
    \begin{eqnarray}
        \mathfrak{f}_{D^a}(\bm{p}) &=&
        \frac{\mathfrak{n}_{D^a}}{\mathfrak{n}_{D^a}^{(\mathrm{eq})}} \, \exp\!\big(\!-\beta \sqrt{M_a^2+p^2}\, \big),
        \label{fD-pq}
        \\
        \mathfrak{f}_{D^{*a}}(\bm{p}) &=&
        3\, \frac{\mathfrak{n}_{D^{*a}}}{\mathfrak{n}_{D^{*a}}^{(\mathrm{eq})}} \,\exp\!\big(\!-\beta \sqrt{M_{\ast a}^2+p^2}\,\big).
        \label{fD*-pq}
    \end{eqnarray}
    \label{fD,fD*-pq}%
\end{subequations}
Before kinetic freezeout, the number densities $\mathfrak{n}_{D^a}(\tau)$ and $\mathfrak{n}_{D^{*a}}(\tau)$
evolve according to rate equations that take into account charm-meson reactions 
and the expanding volume $V(\tau)$.
After kinetic freezeout, the temperature remains constant at $T_\mathrm{kf}$, so $\beta =\beta_\mathrm{kf}$.
In the absence of further interactions,
$\mathfrak{n}_{D^a}(\tau)$ and $\mathfrak{n}_{D^{*a}}(\tau)$
would decrease in proportion to $1/V(\tau)$ as $\tau$ increases,
just like the pion number density in Eq.~\eqref{npi-tau}.
At very large proper times ($c\tau>2,000$~fm), the $D^\ast$'s decay into $D$'s.

The multiplicities of charm hadrons in central Pb-Pb collisions at $\sqrt{s_{NN}}=5.02$~TeV
have been predicted using SHM in Ref.~\cite{Andronic:2021erx}.
For collisions in the centrality range 0-10\%, the central values of the predicted multiplicities $dN/dy$  at midrapidity
for $D^0$, $D^+$, and $D^{\ast +}$ are 6.42, 2.84, and 2.52. 
No prediction was given for the multiplicity of $D^{\ast 0}$.
We can estimate the multiplicity of $D^{\ast 0}$ by assuming
that the ratio of the numbers of $D^{\ast 0}$ and  $D^{\ast +}$ is the same as at hadronization:
\beq
\frac{N_{\ast 0}}{N_{\ast +}} = \frac{M_{\ast 0}^2 \, K_2(M_{\ast 0}/T_H)}{M_{\ast +}^2 \, K_2(M_{\ast +}/T_H)}.
\label{ND*0/ND*+}
\eeq
For the hadronization temperature $T_H=156$~MeV, this ratio is 1.020. 
The estimated multiplicities for $D^{\ast 0}$ and  $D^{\ast +}$ are
\beq
(dN_{D^{\ast 0}}/dy)_0 = 2.57, \qquad (dN_{D^{\ast +}}/dy)_0 = 2.52.
\label{dND*/dy}
\eeq
The SHM predictions for the multiplicities for $D^0$ and $D^+$ 
take into account $D^\ast$ decays.
We obtain the predictions for the multiplicities before $D^*$ decays 
by using Eqs.~\eqref{ND0,Dp} with $B_{+0} = 67.7 \%$:
\beq
(dN_{D^0}/dy)_0 = 2.14, \qquad (dN_{D^+}/dy)_0  = 2.03.
\label{dND/dy}
\eeq
The ratio of a charm-meson multiplicity in Eqs.~\eqref{dND*/dy} or \eqref{dND/dy}
to the pion multiplicity in Eq.~\eqref{dNpi/dy}  can be identified with the ratio of the
charm-meson number density to the pion  number density at kinetic freezeout
\beq
\frac{(dN_{D^{(\ast)a}}/dy)_0}{dN_\pi/dy}
=
\frac{\mathfrak n_{D^{(\ast)a}}(\tau_\mathrm{kf})}{\mathfrak n_\pi(\tau_\mathrm{kf})}.
\eeq


\section{Mass shifts and thermal widths}
\label{sec:PionGas}

In this section, we determine the mass shifts and thermal widths of pions
and charm mesons in a  hadron gas at temperatures near that of kinetic freezeout.
The dominant effects from the hadronic medium come from coherent pion forward  scattering.

\subsection{Coherent pion forward scattering}
\label{app:pion-for-scatter}

When a particle propagates through a medium,
its properties are modified by the interactions with the medium.
The modifications can be described by the self-energy $\Pi(p)$, which
depends on the energy and momentum of the particle and also on the properties of the medium.
The real part of $\Pi(p)$ at $\bm{p}=0$ determines the shift in the rest mass of the particle.
The imaginary part of $\Pi(p)$ at $\bm{p}=0$ determines the thermal width of the particle at rest.

\begin{figure}[t]
    \includegraphics[width=0.7\textwidth]{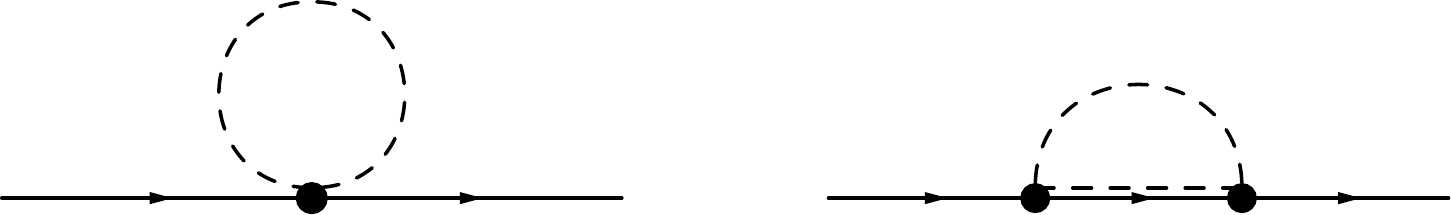}
    \caption{One-loop Feynman diagrams for the $D$ self-energy in  HH$\chi$EFT.
        The $D$, $D^\ast$, and $\pi$ are represented by solid, double (solid$+$dashed),
        and dashed lines, respectively.
    }
    \label{fig:Dselfenergy}
\end{figure}

If the particle is in thermal equilibrium with the medium, its self-energy can be calculated using thermal field theory.
To be more specific, we consider the self-energy $\Pi_D(p)$ of a pseudoscalar charm meson $D$.
The one-loop Feynman diagrams for $\Pi_D(p)$  in HH$\chi$EFT are shown in Fig.~\ref{fig:Dselfenergy}.
The first diagram can be expressed as the sum of a vacuum contribution and
a thermal contribution from pions. The second diagram can be expressed as the sum of a vacuum contribution,
a thermal contribution from pions, and a thermal contribution from vector charm mesons $D^\ast$.
At temperatures relevant to the hadron gas, thermal contributions from vector charm mesons
are severely suppressed by a Boltzmann factor $\exp(-M_\ast/T)$.
The thermal contributions from pions can be expressed as an integral over the pion momentum $\bm{q}$
weighted by the Bose-Einstein distribution $1/(e^{\beta \omega_q}-1)$,
where $\omega_q= \sqrt{m_\pi^2+q^2}$.

\begin{figure}[t]
    \includegraphics[width=.99\textwidth]{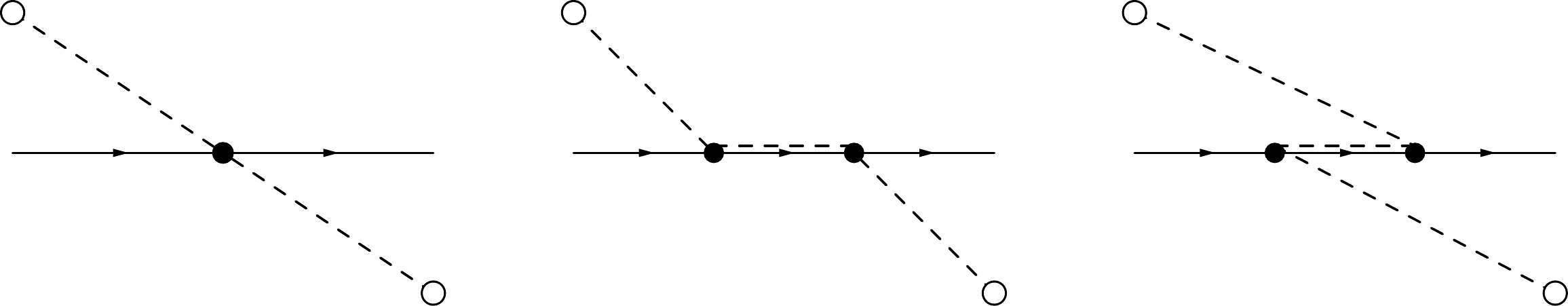}
    \caption{
        Feynman diagrams for the $D$ self-energy from coherent pion forward scattering in HH$\chi$EFT at LO.
        The  empty circles indicate an incoming and  outgoing pion with the same flavor and the same 3-momentum.
        These diagrams can be obtained by cutting the pion lines in the  diagrams in Fig.~\ref{fig:Dselfenergy}.
        The second diagram has a $D^\ast$ resonance contribution.
    }
    \label{fig:Dpiforwardscattering}
\end{figure}

The thermal contribution from pions to the $D$ self-energy can be calculated alternatively from the tree diagrams
for $\pi D$ scattering in Fig.~\ref{fig:Dpiforwardscattering}.  At this order,
the thermal contribution from pions comes from coherent pion forward scattering.
If a pion with flavor $k$ and momentum $\bm{q}$  is scattered back into the state with the same flavor $k$ and
momentum $\bm{q}$,
the initial many-body state is the charm meson plus the medium
(which includes the pion  with flavor $k$ and momentum $\bm{q}$),
and the final many-body state is also the charm meson plus the medium.
Since  the initial state  is the  same for all $\bm{q}$ and the final state is also the same,
the pion-forward-scattering amplitudes must be added coherently for all momenta $\bm{q}$ and all pion  flavors $k$.
The $D$ self-energy from coherent pion forward scattering
can be obtained from the negative of the $\mathcal T$-matrix element
by weighting it by $\mathfrak{f}_\pi(\omega_q)/(2 \omega_q)$,
where $\mathfrak{f}_\pi(\omega_q)$ is the pion momentum distribution and $1/(2 \omega_q)$ is a normalization factor,
integrating over the pion momentum $\bm{q}$ with measure $d^3q/(2\pi)^3$, and summing over the
three pion flavors.
If the pions are in chemical and thermal equilibrium at temperature $T$,
the pion momentum distribution is the Bose-Einstein distribution
in Eq.~\eqref{fpi-k:<}.
However, this prescription for the self-energy
from coherent pion forward scattering applies equally well to any medium
in which the pions have a momentum distribution $\mathfrak{f}_\pi(\omega_q)$.

The thermal contribution from pions to the $D$ self-energy
can be obtained directly from the $D$ self-energy diagrams in Fig.~\ref{fig:Dselfenergy}
by making a simple substitution for the pion propagator in the loop:
\beq
\frac{i}{q^2 - m_\pi^2 +i\epsilon}   \longrightarrow \mathfrak{f}_\pi (|q_0|)\, 2\pi \delta(q^2-m_\pi^2).
\label{pionpropsub}
\eeq
The delta  function can be expressed as
\beq
\delta(q^2-m_\pi^2)= \sum_\pm \theta(  \pm q_0)\, \frac{1}{2\omega_q}\, \delta (|q_0| - \omega_q).
\label{delta-pi}
\eeq
This substitution is referred to as the cutting of the pion line.
The cutting of the pion line in the first diagram in Fig.~\ref{fig:Dselfenergy} is 0,
because the vertex is 0 when the incoming and outgoing pions have the same flavor.
The cutting of the pion line in the second diagram in Fig.~\ref{fig:Dselfenergy}
gives the last two forward-scattering diagrams in Fig.~\ref{fig:Dpiforwardscattering}.
They come from the positive and negative regions of $q_0$, respectively.

\subsection{Pions}
\label{sec:pions}

The thermal mass shift and the thermal width for a pion in a pion gas can be calculated using $\chi$EFT.
The mass shift for a pion in thermal equilibrium was first calculated using $\chi$EFT   at LO
by Gasser and Leutwyler \cite{Gasser:1986vb}.
The pion thermal width
was calculated  in the low-density limit using $\chi$EFT   at NLO by Goity and Leutwyler \cite{Goity:1989gs}.
A complete calculation of the self-energy of a pion  in thermal equilibrium
in $\chi$EFT at NLO was presented by Schenk \cite{Schenk:1993ru}.
It was used to obtain the pion mass shift and the pion thermal width.
The pion mass shift at NLO has also been calculated by Toublan \cite{Toublan:1997rr}.

\begin{figure}[t]
    \includegraphics[width=0.35\textwidth]{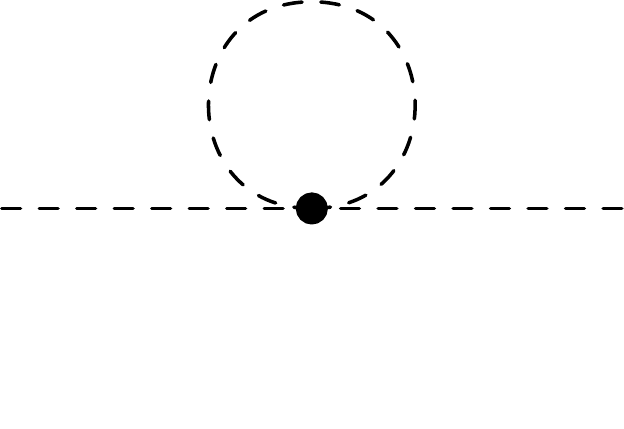}
    \hspace{2.5cm}
    \includegraphics[width=0.35\textwidth]{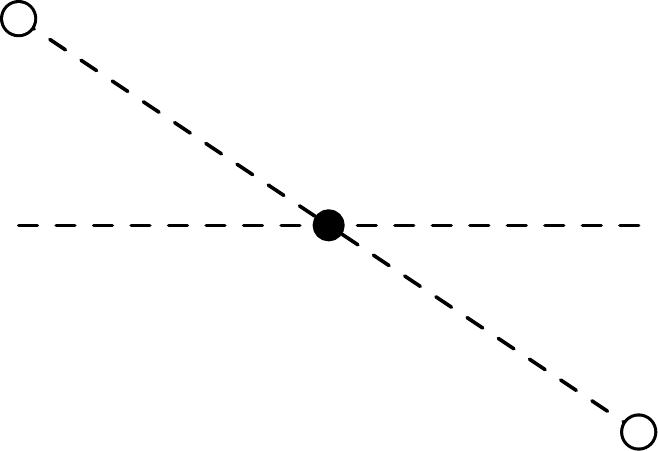}
    \caption{One-loop Feynman diagram for the pion self-energy in $\chi$EFT at LO  (left panel)
        and the corresponding Feynman diagram for the pion self-energy from coherent pion forward scattering (right panel).
    }
    \label{fig:pipiscattering}
\end{figure}

The pion self-energy in $\chi$EFT at LO is given by the one-loop Feynman diagram in the left panel of Fig.~\ref{fig:pipiscattering}.
The thermal contribution to the pion self-energy can also be obtained from
the Feynman diagram for coherent pion forward scattering  in the right panel of Fig.~\ref{fig:pipiscattering}.
The self-energy $\Pi_\pi(p_0,p)$ of a pion with 4-momentum $(p_0,\bm{p})$ 
can be obtained from the negative of the
amplitude $\mathcal{A}_{ik,jk}(p_0,\bm{p},\bm{q})$ for forward scattering of an on-shell pion
with flavor $k$ and 3-momentum $\bm{q}$ by weighting it by $\mathfrak{f}_\pi(\omega_q)/(2 \omega_q)$,
integrating over $\bm{q}$, and summing over the three pion  flavors $k$:
\begin{equation}
    \Pi_\pi(p_0,p)\,  \delta^{ij} = - \sum_k
    \int\!\!  \frac{d^3q}{(2\pi)^3 2 \omega_q}\,  \mathfrak{f}_\pi (\omega_q)\,
    \mathcal{A}_{ik,jk}(p_0,\bm{p},\bm{q}).
    \label{Pipi-forward}
\end{equation}
The amplitude $\mathcal{A}_{ik,jk}$ at LO does not depend on $\bm{q}$:
\begin{equation}
    \mathcal{A}_{ik,jk}(p_0,p) = - \frac{2 }{3 f_\pi^2}
    \big[ (2 p_0^2 - 2 p^2 + m_\pi^2)\, \delta^{ij} - 2(p_0^2 - p^2 + 2 m_\pi^2)\,\delta^{ik} \delta^{jk} \big] .
    \label{Apipi-forward}
\end{equation}
The pion self-energy at LO is
\begin{equation}
    \Pi_\pi(p_0,p) =  \frac{1 }{3 f_\pi^2} (4 p_0^2 - 4 p^2 - m_\pi^2) \, \mathfrak{n}_\pi
    \left\langle \frac{1}{\omega_q} \right\rangle_{\!\!\!\bm{q}}  ,
    \label{Pipi-LO}
\end{equation}
where the angular brackets represents the average over the Bose-Einstein distribution for the pion
defined in Eq.~\eqref{<F>}.

The pion mass shift $\delta m_\pi$ and the thermal width $\Gamma_\pi$ can be obtained
by evaluating the pion self-energy on the mass shell at  zero 3-momentum:
\begin{equation}
    \Pi_\pi(p_0=m_\pi,p=0) = 2 m_\pi \delta m_\pi - i\, m_\pi \Gamma_\pi.
    \label{Tpipi-forward}
\end{equation}
The pion mass shift  in $\chi$EFT at LO is
\begin{equation}
    \delta m_\pi=  \frac{m_\pi}{2 f_\pi^2}\, \mathfrak{n}_\pi
    \left\langle \frac{1}{\omega_q} \right\rangle_{\!\!\!\bm{q}}  .
    \label{deltampi}
\end{equation}
The pion thermal width $\Gamma_\pi$ is 0 in $\chi$EFT  at LO.
In a pion gas in chemical and thermal equilibrium at the temperature  $T_\mathrm{kf} = 115$~MeV,
the pion mass shift is 1.55~MeV.

\subsection{Pseudoscalar charm mesons}
\label{sec:D}

The contributions to the thermal mass shift and thermal width
of a pseudoscalar charm meson in a pion gas from  coherent pion forward scattering can be calculated using HH$\chi$EFT.

\subsubsection{$D$ self-energy}

\begin{figure}[t]
    \includegraphics[width=.99\textwidth]{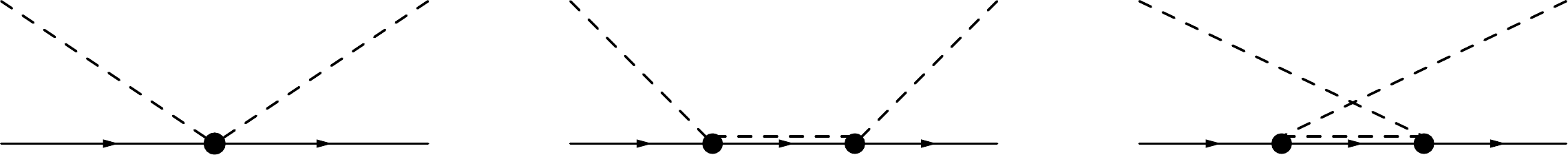}
    \caption{
        Feynman diagrams for   $\pi D \to \pi D$ in HH$\chi$EFT at LO.
        The second diagram has a $D^\ast$ resonance contribution.
    }
    \label{fig:piDscattering}
\end{figure}

In HH$\chi$EFT at  LO,  the reaction $\pi D \to \pi D$
proceeds through the three diagrams in  Fig.~\ref{fig:piDscattering}.
The 4-momentum of $D$ can be expressed as $P = M v + p$,
where $v$ is the velocity 4-vector and $p$ is the residual 4-momentum.
The amplitude for the transition $D^a(p) \pi^i(q) \to D^b(p^\prime) \pi^j(q^\prime)$ is
\bqa
\mathcal{A}_{ai,bj}(p,q,q^\prime) &=& \frac{1}{2 f_\pi^2} \, [\sigma^i, \sigma^j]_{ab} \, v \!\cdot \!(q + q^\prime)
\nonumber\\
&&\hspace{-1.5cm}
-\frac{g_\pi^2}{f_\pi^2} \left((\sigma^i \sigma^j)_{ab}  \,
\frac{- q \!\cdot\! q^\prime + (v \!\cdot\!q)\, ( v \!\cdot\!q^\prime)}{v \!\cdot\! (p+q)  -\Delta + i \Gamma_\ast/2}
+ (\sigma^j \sigma^i)_{ab}  \,
\frac{- q \!\cdot\! q^\prime + (v \!\cdot\!q)\, ( v \!\cdot\!q^\prime)}{v\!\cdot\! (p-q^\prime)  -\Delta + i \Gamma_\ast/2} \right).
\label{ADpi}
\eqa
We have inserted the $D^\ast$ width in the denominators to allow for the possibility
that the $D^\ast$ can be on shell.
In the case of  the forward scattering of $\pi^k(q)$ to $\pi^k(q)$,
the amplitude reduces to a function  of $v\!\cdot\! p$ and $v\!\cdot\! q$ and
it is diagonal in $a$ and $b$. The diagonal entry is
\begin{equation}
    \mathcal{A}_{ak,ak}(v \!\cdot\!p, v \!\cdot\!q)  =
    -\frac{2 g_\pi^2}{f_\pi^2} \,
    \frac{[(v \!\cdot\!q)^2 - m_\pi^2] (\Delta - v\!\cdot\! p)}
    {(v \!\cdot\! q)^2 -  (\Delta - v\!\cdot\! p)^2 + i \, (\Delta - v\!\cdot\! p)\Gamma_\ast } .
    \label{ADpi-forward}
\end{equation}
Since $\Gamma_\ast \ll \Delta$, we have omitted the terms proportional to $\Gamma_\ast$ in the numerator
and to $\Gamma_\ast^2$  in the denominator.

The $D$ self energy $\Pi_D(v \!\cdot\!p)$ in HH$\chi$EFT at LO is the sum of the two one-loop diagrams in Fig.~\ref{fig:Dselfenergy}.
The contribution from coherent pion forward scattering is the sum of the three tree diagrams in Fig.~\ref{fig:Dpiforwardscattering}.
The coherent sum of the first diagram over pion flavors is 0.
The $D$ self energy can be obtained from the  amplitude in Eq.~\eqref{ADpi-forward} by multiplying it by $-1/2$,
weighting it by $\mathfrak{f}_\pi (\omega_q)/(2 \omega_q)$,
integrating over the momentum $\bm{q}$, and summing over the three pion flavors $k$.
We choose the velocity 4-vector $v$ of the charm meson to  be the same as the 4-vector that defines the thermal frame
in which the pion momentum distribution is Eq.~\eqref{fpi-k:<}
before kinetic freezeout and Eq.~\eqref{fpi-k:>} after  kinetic freezeout.
The pion energy is $v \!\cdot\!q = \omega_q$.
The $D$ self-energy is
\begin{equation}
    \Pi_D(v \!\cdot\! p)  =
    \frac{3 g_\pi^2} {f_\pi^2} \,
    \int\!\!  \frac{d^3q}{(2\pi)^3  2 \omega_q}\mathfrak{f}_\pi (\omega_q)
    \frac{(\omega_q^2 - m_\pi^2) (\Delta - v \!\cdot\! p)}
    {\omega_q^2 -  (\Delta - v\!\cdot\! p)^2 + i \, (\Delta - v\!\cdot\! p)\Gamma_\ast } .
    \label{PiD-v.p}
\end{equation}

Since the charm-meson mass difference $\Delta= M_\ast-M$ is approximately equal to the pion mass $m_\pi$,
the self-energy is sensitive to isospin splittings when the $D$ is close to the mass shell  $v \cdot p = 0$.
The isospin splittings can be taken into account
by reintroducing a sum over the flavors $c$ of the intermediate $D^\ast$.
In the self energy  in Eq.~\eqref{PiD-v.p}, the factor $\sum_k(\sigma^k \sigma^k)_{aa} = 3 \delta_{aa}$ from the pion vertices
is replaced by $\sum_k\sum_c (\sigma^k)_{ac}(\sigma^k)_{ca} = \sum_c (2-\delta_{ac})$.
The mass difference  $\Delta$ is replaced by $\Delta_{ac}$
and the pion energy is replaced by $\omega_{caq} = \sqrt{m^2_{\pi ca}+ q^2}$.
The $D^a$ self energy  is
\begin{equation}
    \Pi_{D^a}(v \!\cdot\! p)  =
    \frac{g_\pi^2} {f_\pi^2} \, \sum_c (2-\delta_{ac}) \,
    \int\!\!  \frac{d^3q}{(2\pi)^3 2 \omega_{caq}}\mathfrak{f}_\pi (\omega_{caq})
    \frac{ q^2\,(\Delta_{ca} - v \!\cdot\! p)}
    {\omega_{caq}^2 -  (\Delta_{ca} - v\!\cdot\! p)^2 + i \, (\Delta_{ca} - v\!\cdot\! p)\Gamma_{\ast  c}} ,
    \label{PiDa-v.p}
\end{equation}
where $q^2$ is the square of the 3-momentum.

\subsubsection{Mass shift and thermal width}

The mass shift $\delta M_a$ and the thermal width $\delta \Gamma_a$ for the charm meson $D^a$ in HH$\chi$EFT at LO
are obtained by evaluating the $D^a$ self energy on the mass shell $v\!\cdot\! p=0$:
\begin{equation}
    \Pi_{D^a}(v\!\cdot\! p=0)  = \delta M_a - i\, \delta \Gamma_a/2 .
    \label{PiD-M,Gamma}
\end{equation}
The $D^a$ self energy with isospin splittings in Eq.~\eqref{PiDa-v.p} evaluated on the mass shell is
\begin{equation}
    \Pi_{D^a}(0)  =
    \frac{g_\pi^2} {f_\pi^2} \, \sum_c (2-\delta_{ac}) \, \Delta_{ca}
    \int\!\!  \frac{d^3q}{(2\pi)^3 2 \omega_{caq}}\mathfrak{f}_\pi (\omega_{caq})
    \frac{q^2}{q^2 - q_{ca}^2 + i \Delta_{ca} \Gamma_{\ast c}} ,
    \label{PiDa-0}
\end{equation}
where $q_{ca}^2=\Delta_{ca}^2 - m_{\pi ca}^2$.
Since $\Delta_{ca} \Gamma_{\ast c} \ll |q^2 - q_{ca}^2|$ except in a very narrow range of $q^2$,
the expressions for $\delta M_a$ and $\delta \Gamma_a$ can be simplified by taking the limit $\Gamma_{\ast c}  \to 0$.
The $D^a$ mass shift  in the limit $\Gamma_{\ast c}  \to 0$
can be expressed in terms of an average over the pion momentum distribution
of a function of $q$ that involves a principal-value distribution:
\begin{equation}
    \delta M_a =
    \frac{g_\pi^2} {2f_\pi^2} \, \mathfrak{n}_\pi\, \sum_c  (2-\delta_{ac}) \, \Delta_{ca}
    \left  \langle \frac{q^2}{\omega_{caq}}
    \mathcal{P} \frac{1}{q^2 -  q_{ca}^2}  \right\rangle_{\!\!\!\bm{q}} .
    \label{deltaMDa}
\end{equation}
The principal value is necessary only if $\Delta_{ca} > m_{\pi ca}$.
The $D^a$ thermal width in the limit $\Gamma_{\ast c}  \to 0$
can be evaluated analytically by using a delta function:
\begin{equation}
    \delta \Gamma_a =
    \frac{g_\pi^2} {4\pi \, f_\pi^2} \, \sum_c (2-\delta_{ac})  \, \mathfrak{f}_\pi(\Delta_{ca}) \,
    q_{ca}^3 \,  \theta(\Delta_{ca} - m_{\pi ca}).
    \label{deltaGammaDa}
\end{equation}
This thermal width comes from the second diagram in Fig.~\ref{fig:Dpiforwardscattering} with a $D^\ast$ in the $s$ channel.
The contribution from an intermediate $D^{\ast c}$ is nonzero only if $\Delta_{ca} > m_{\pi ca}$.
In a pion gas with temperature $T_\mathrm{kf} = 115$~MeV,
the mass shifts for $D^+$ and $D^0$ in Eq.~\eqref{deltaMDa} are
$\delta M_+ = 1.269$~MeV and $\delta M_0 = 1.418$~MeV. 
The thermal widths for $D^+$ and $D^0$ in Eq.~\eqref{deltaGammaDa} are
$\delta \Gamma_+ = 31.6$~keV  and $\delta \Gamma_0 = 110.3$~keV.

The mass shift and thermal width of $D^a$
can be expanded in powers of isospin splittings using the methods in Appendix~\ref{app:PionIntegral}.
The leading term in the expansion of the mass shift  is the same for $D^+$ and $D^0$:
\begin{equation}
    \delta M \approx
    \frac{3g_\pi^2}{2f_\pi^2}\, \mathfrak{n}_\pi \, m_\pi
    \left  \langle \frac{1}{\omega_q} \right\rangle_{\!\!\!\bm{q}} .
    \label{deltaMD-approx}
\end{equation}
The leading term in the expansion of the $D^a$ thermal width  is
\begin{equation}
    \delta \Gamma_a \approx 3 \,\mathfrak{f}_\pi(m_\pi)\,\sum_c\Gamma \big[D^{*c} \to D^a \pi \big],
    \label{deltaGammaDa-approx}
\end{equation}
where $\Gamma[D^{*c} \to D^a \pi]$ is the  partial decay rate of $D^{*c}$ in Eq.~\eqref{GammaD*Dpi}.
In a pion gas with temperature $T_\mathrm{kf} = 115$~MeV,
the $D$ mass shift in Eq.~\eqref{deltaMD-approx} is $\delta M = 1.257$~MeV. 
The thermal widths for $D^+$ and $D^0$ in Eq.~\eqref{deltaGammaDa-approx} are 
$\delta \Gamma_+ = 32.6$~keV  
and $\delta \Gamma_0 = 118.9$~keV.

\subsection{Vector charm mesons}
\label{sec:D*}

The contributions to the thermal mass shift and  thermal width
of a vector charm meson in a pion gas from  coherent pion forward scattering can be calculated using HH$\chi$EFT.

\subsubsection{$D^\ast$ self-energy}

\begin{figure}[t]
    \includegraphics[width=0.99\textwidth]{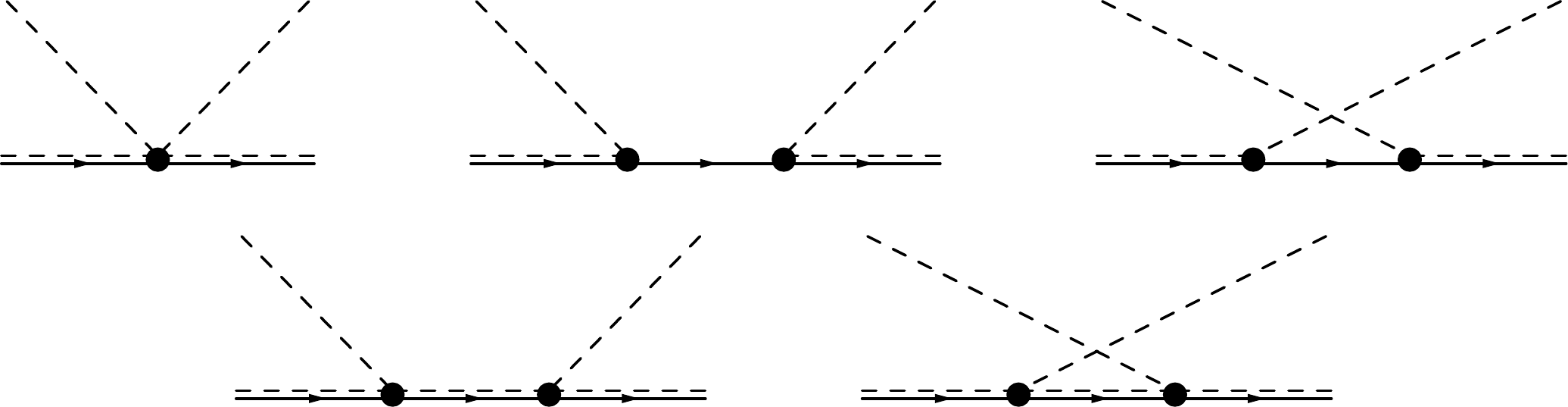}
    \caption{
        Feynman diagrams for   $\pi D^\ast \to \pi D^\ast$  in HH$\chi$EFT at LO.
        The  third diagram produces a $D$-meson $t$-channel singularity in the reaction rate.
    }
    \label{fig:piD*scattering}
\end{figure}

In HH$\chi$EFT at LO,  the reaction $\pi  D^\ast \to \pi D^\ast$ proceeds through the five diagrams in Fig.~\ref{fig:piD*scattering}.
The 4-momentum of $D^\ast$ can be expressed as $P = M v + p$,
where $v$ is the velocity 4-vector and $p$ is the residual 4-momentum.
The amplitude for the transition
$ \pi^i(q) D^{*a}(p)\to  \pi^j(q^\prime)D^{*b}(p^\prime)$ is
\bea
\mathcal A^{\mu\nu}_{ai,bj}
&=&
-\frac{1}{2 f_\pi^2}
g^{\mu\nu}
\left[\sigma^i,\sigma^j\right]_{ab}
v\cdot(q+q^\prime)
\nonumber\\
&&-\frac{g_\pi^2}{f_\pi^2}
\left[
    (\sigma^i \sigma^j)_{ab}
    \frac{q^\mu q^{\prime\nu}}{v\cdot(p+q)+i\Gamma/2}
    +(\sigma^j\sigma^i)_{ab}
    \frac{q^{\prime\mu}q^\nu}{v\cdot(p-q^\prime)+i\Gamma/2}
    \right]
\nonumber\\
&&+\frac{g_\pi^2}{f_\pi^2}
\epsilon^{\mu\rho\lambda}(v)\, \epsilon^{\nu\sigma}_{~~~\lambda}(v)
\left[
    (\sigma^i \sigma^j)_{ab}
    \frac{q_\rho q^{\prime}_{\sigma}}{v\cdot(p+q)-\Delta}
    + (\sigma^j \sigma^i)_{ab}
    \frac{q^{\prime}_\rho q_\sigma}{v\cdot(p-q^\prime)-\Delta}
    \right],
\label{ApiD*}
\eea
where  $\epsilon^{\mu\rho\lambda}(v) = \epsilon^{\mu\rho\lambda\alpha} v_\alpha$.
We have inserted the $D$  width in the denominators of the $D$ propagators to allow for the possibility
that the $D$ can be on shell. In the case of  the forward scattering of $\pi^k(q)$ to $\pi^k(q)$,
the amplitude is diagonal in $a$ and $b$. The diagonal entry  is
\begin{eqnarray}
    \mathcal{A}^{\mu \nu}_{ak,ak} &=&
    \frac{2g_\pi^2}{f_\pi^2} \left[ q^\mu q^\nu
        \frac{v\cdot p}{(v \!\cdot\! q)^2 -  (v\!\cdot\! p)^2 - i \, v\!\cdot\! p\,\Gamma} \right.
        \nonumber\\
        && \hspace{1cm} \left. -
        \epsilon^{\mu\lambda}(v,q)\,  \epsilon^\nu_{~~\lambda}(v,q) \,
        \frac{v\cdot p - \Delta}{(v \!\cdot\! q)^2 -  (\Delta - v\!\cdot\! p)^2} \right] ,
    \label{AD*pi-forward}
\end{eqnarray}
where $\epsilon^{\mu\lambda}(v,q) = \epsilon^{\mu\lambda\alpha\beta} v_\alpha  q_\beta$.
Since $\Gamma \ll \Delta$, we have omitted terms proportional to $\Gamma$ in the numerator
and to $\Gamma^2$  in the denominator.

\begin{figure}[t]
    \includegraphics[width=0.99\textwidth]{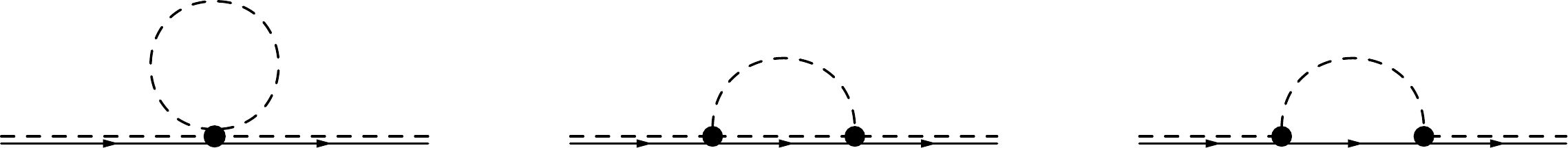}
    \caption{One-loop Feynman diagrams for the $D^\ast$ self-energy in HH$\chi$EFT.
    }
    \label{fig:D*selfenergy}
\end{figure}

\begin{figure}[t]
    \includegraphics[width=0.70\textwidth]{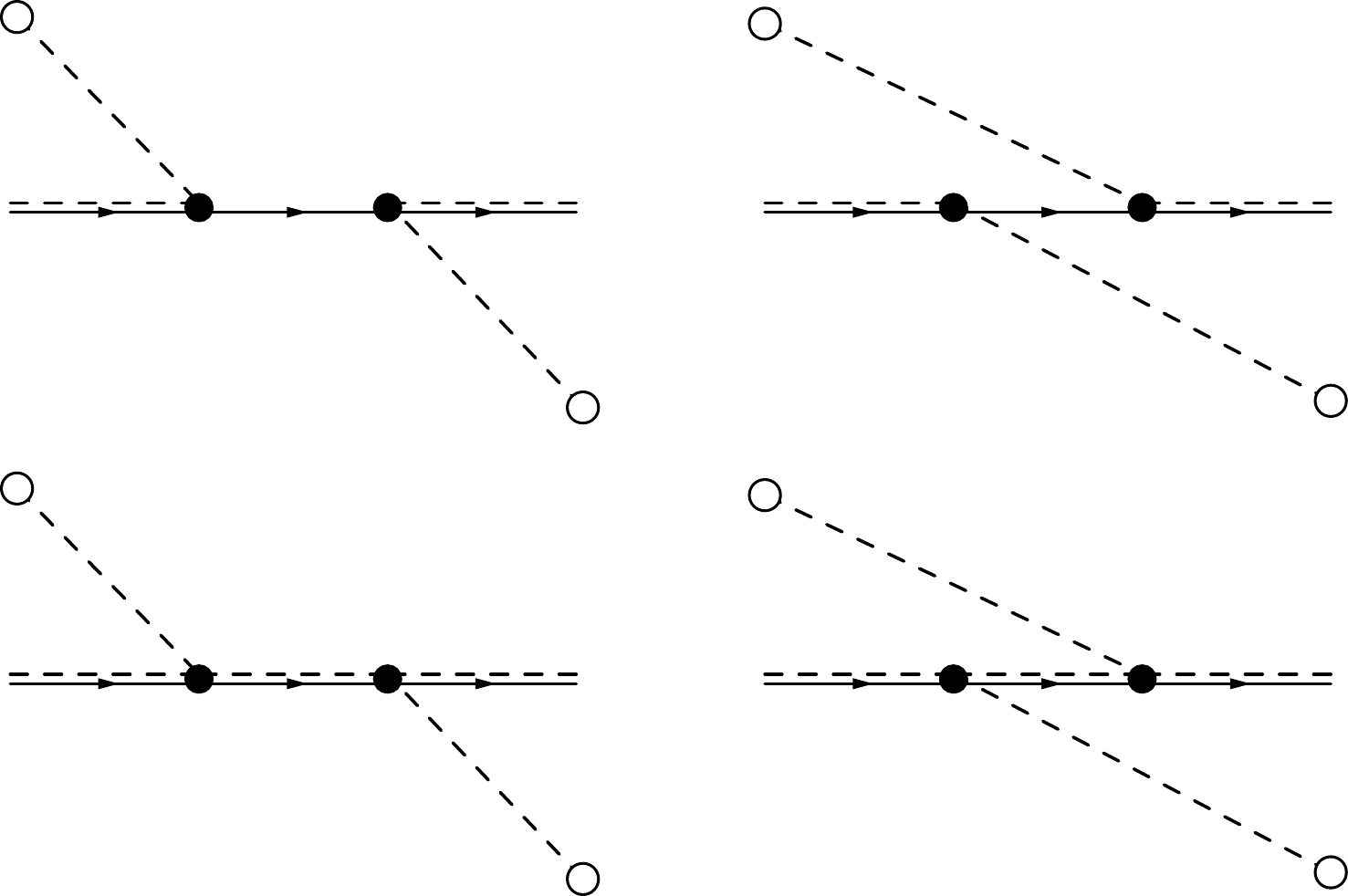}
    \caption{Feynman diagrams for the $D^*$ self energy from coherent pion forward scattering in HH$\chi$EFT at LO.
        These diagrams can be obtained by cutting the pion lines in the last two diagrams in Fig.~\ref{fig:D*selfenergy}.
        The second diagram in the first row produces a $D$-meson $t$-channel singularity.
    }
    \label{fig:D*piforwardscattering}
\end{figure}

The self-energy tensor $\Pi^{\mu\nu}$ of a vector meson $D^\ast$   in HH$\chi$EFT at  LO
is the sum of the three one-loop Feynman diagrams in Fig.~\ref{fig:D*selfenergy}.
The contribution from coherent pion forward scattering  can be
obtained by cutting the pion lines using the prescription in Eq.~\eqref{pionpropsub}.
The cut of the first diagram in Fig.~\ref{fig:D*selfenergy} is zero, 
because the coherent sum over pion flavors is 0.
The cuts of the last two diagrams in Fig.~\ref{fig:D*selfenergy} give the four tree diagrams in Fig.~\ref{fig:D*piforwardscattering}.
By rotational symmetry, the contribution to $\Pi^{\mu\nu}$ from the coherent forward scattering of a pion
with 4-momentum  $q$ is a linear combination of
$g^{\mu \nu}$, $q^\mu q^\nu$,  $v^\mu q^\nu+q^\mu v^\nu$, and $v^\mu v^\nu$.
However the tensor structure of the $D^\ast$ propagator in Eq.~\eqref{D*prop}
ensures that only the $- g^{\mu \nu} + v^\mu v^\nu$ component contributes to the $D^\ast$ self energy
$\Pi_{D^\ast}(v \!\cdot\!p)$.
That component can be obtained from the tensor $\mathcal{A}^{\mu \nu}_{ak,ak}$ in Eq.~\eqref{AD*pi-forward}
by contracting it with $(- g^{\mu \nu} + v^\mu v^\nu)/3$.
The $D^\ast$ self energy can be obtained from that component by multiplying it by $-1/2$,
weighting it by $\mathfrak{f}_\pi (\omega_q)/(2 \omega_q)$, integrating over $\bm{q}$,
and summing over the three pion  flavors $k$:
\bea
\Pi_{D^\ast}(v \!\cdot\!p) &=&
- \frac{g_\pi^2}{f_\pi^2}
\int\!\!  \frac{d^3q}{(2\pi)^3 2 \omega_q} \mathfrak{f}_\pi (\omega_q)  \, (\omega_q^2 - m_\pi^2)
\left(  \ \frac{v \!\cdot\! p}{\omega_q^2 - (v \!\cdot\! p)^2 - i \,  v \!\cdot\! p\, \Gamma}  + \frac{2(v \!\cdot\! p-\Delta)}{\omega_q^2 - (v \!\cdot\! p - \Delta)^2}  \right).
\nonumber\\
\label{PiD*-0}
\eea

Since the charm-meson mass difference $\Delta= M_\ast-M$ is approximately equal to the pion mass $m_\pi$,
the self energy is sensitive to isospin splittings when the $D^\ast$ is close to the mass shell $v \!\cdot\! p =\Delta$.
The isospin splittings can be taken into account
by reintroducing a sum over the flavors $c$ of the intermediate $D$ or $D^\ast$.
In the self energy  in Eq.~\eqref{PiD*-0}, the factor $\sum_k(\sigma^k \sigma^k)_{aa} = 3 \delta_{aa}$ from the pion vertices
is replaced by $\sum_k\sum_c (\sigma^k)_{ac}(\sigma^k)_{ca} = \sum_c (2-\delta_{ac})$.
The isospin splittings in the denominators of the propagators can be taken into account
in the first term in Eq.~\eqref{PiD*-0} by replacing $v \!\cdot\! p$ by $ v \!\cdot\! p - M_c+M$ and  $\Gamma$ by $\Gamma_c$.
They can be taken into account
in the second term by replacing $v \!\cdot\! p - \Delta$  by $ v \!\cdot\! p - M_{*c}+M$.
The mass-shell condition $ v \!\cdot\! p = \Delta$ is modified to $v \!\cdot\! p = M_{*a}-M$.

\subsubsection{Mass shift and thermal width}

The mass shift $\delta M_{\ast a}$ and the thermal width $\delta  \Gamma_{\ast a}$ for the charm meson $D^{*a}$
in HH$\chi$EFT at LO are obtained by evaluating the self-energy on the mass shell:
\begin{equation}
    \Pi_{D^{*a}}(v \!\cdot\!p = \Delta)  = \delta M_{*a} - i\, \delta \Gamma_{*a}/2 .
    \label{PiD*-M,Gamma}
\end{equation}
If isospin splittings are taken into account in Eq.~\eqref{PiD*-0}, the $D^{\ast a}$ self-energy on the mass shell is
\bea
\Pi_{D^{\ast a}}(\Delta)  &=&
-\frac{ g_\pi^2} {3f_\pi^2}  \sum_c (2-\delta_{ac}) \!
\int\!\!  \frac{d^3q\, }{(2\pi)^3 2 \omega_{acq}}\mathfrak{f}_\pi (\omega_{acq}) \!\!
\left( \frac{q^2 \Delta_{ac}}{q^2 - q_{ac}^2- i \Delta_{ac} \Gamma_c} +  \frac{2q^2(M_{*a}-M_{*c})}{\omega_{acq}^2} \right),
\nonumber\\
\label{PiD*a-0}
\eea
where $\omega_{acq} = \sqrt{ m_{\pi ac}^2 + q^2}$ and $q_{ac}^2=\Delta_{ac}^2 - m_{\pi ac}^2$.
In the second term inside the  parentheses, we have omitted the term $-(M_{*a}-M_{*c})^2$ in the denominator,
because $M_{*+}-M_{*0} \ll m_\pi$.

Since $\Delta_{ac} \Gamma_c \ll \lvert q^2 - q_{ac}^2\rvert$ except in a very narrow range of $q^2$,
the expressions for $\delta M_{\ast a}$ and $\delta \Gamma_{\ast a}$ can be simplified  by taking the limit $\Gamma_c  \to 0$.
The resulting $D^{*a}$ mass shift is
\bea
\delta M_{*a} &=&
- \frac{g_\pi^2} {6f_\pi^2} \, \mathfrak{n}_\pi
\sum_c   (2-\delta_{ac}) \left[\Delta_{ac}
\left\langle \frac{q^2}{\omega_{acq}} \mathcal{P} \frac{1}{q^2 - q_{ac}^2}  \right\rangle_{\!\!\!\bm{q}}
+ 2 (M_{*a}-M_{*c})\,
\left\langle \frac{q^2}{\omega_{acq}^3}\right\rangle_{\!\!\!\bm{q}} \right].
\label{deltaMD*a}
\eea
Note that the sum $\delta M_{*+} + \delta M_{*0}$ is equal to the sum $\delta M_+ + \delta M_0$ from Eq.~\eqref{deltaMDa}
multiplied by $-1/3$. Thus the spin-weighted average of the $D$ and $D^\ast$ mass shifts is 0.
The resulting $D^{*a}$ thermal width can be evaluated analytically:
\begin{equation}
    \delta \Gamma_{*a} =
    \frac{g_\pi^2} {12\pi \, f_\pi^2} \, \sum_c (2-\delta_{ac})  \, \mathfrak{f}_\pi(\Delta_{ac})\,
    q_{ac}^3\,  \theta(\Delta_{ac} - m_{\pi ac}).
    \label{deltaGammaD*a}
\end{equation}
This thermal width comes from the coherent pion forward scattering diagram with a $D$ in the $t$ channel
in Fig.~\ref{fig:D*piforwardscattering}.
Note that the sum $\delta \Gamma_{*+} + \delta \Gamma_{*0}$  is equal to the 1/3 of the sum
$\delta \Gamma_+ + \delta \Gamma_0$ from Eq.~\eqref{deltaGammaDa-approx}.
In a pion gas with temperature $T_\mathrm{kf} = 115$~MeV,
the  mass shifts for $D^{\ast+}$ and $D^{\ast0}$ are 
$\delta M_{\ast+} = -0.478$~MeV and $\delta M_{\ast0} = -0.417$~MeV. 
The thermal widths  for $D^{\ast+}$ and $D^{\ast0}$ are 
$\delta \Gamma_{\ast+} = 32.7$~keV and $\delta \Gamma_{\ast0} = 14.6$~keV.

The mass shift and thermal width of  $D^{\ast a}$
can be expanded in powers of isospin splittings using the methods in Appendix~\ref{app:PionIntegral}.
The leading term in the expansion of the mass shift  is the same for $D^{*+}$ and $D^{*0}$
and it differs from the mass shift $\delta M$ for $D^+$ and $D^0$ in Eq.~\eqref{deltaMD-approx}
by the multiplicative factor $-1/3$:
\begin{equation}
    \delta M_* \approx -\delta M/3,
    \label{deltaMD*-approx}
\end{equation}
The leading term in the expansion of the  $D^{*a}$ thermal width is
\begin{equation}
    \delta \Gamma_{*a} \approx \mathfrak{f}_\pi(m_\pi)\,\sum_c\Gamma \big[D^{*a} \to D^c \pi \big],
    \label{deltaGammaD*a-approx}
\end{equation}
where $\Gamma[D^{*a} \to D^c \pi]$ is the $D^\ast$ partial decay rate in Eq.~\eqref{GammaD*Dpi}.
In a pion gas with temperature $T_\mathrm{kf} = 115$~MeV,
the $D^\ast$ mass shift in Eq.~\eqref{deltaMD*-approx} is
$\delta M_* = -0.419$~MeV  
and the $D^{*+}$ and $D^{*0}$ thermal widths in Eq.~\eqref{deltaGammaD*a-approx} are 
$\delta \Gamma_{*+} = 35.2$~keV  
and $\delta \Gamma_{*0} = 15.3$~keV.

\subsection{Expanding hadron gas}
\label{sec:deltaM,Gamma}

Thermal mass shifts and thermal widths have significant effects on some reaction rates
for pions and charm mesons in the expanding hadron gas created by a heavy-ion collision.
The pion mass shift $\delta m_\pi$ in $\chi$EFT at LO is given in Eq.~\eqref{deltampi}.
The charm-meson mass shifts $\delta M_a$ for $D^a$ and $\delta M_{*a}$ for $D^{\ast a}$ in HH$\chi$EFT at LO
are given in Eqs.~\eqref{deltaMDa} and \eqref{deltaMD*a}.  We will use the
simpler approximations for the charm-meson mass shifts in Eqs.~\eqref{deltaMD-approx} and \eqref{deltaMD*-approx}.
The mass shifts in the hadron gas before kinetic freezeout are determined by the temperature $T$.
The mass shifts after kinetic freezeout are determined by the pion number density $\mathfrak{n}_\pi$.
Some reaction rates are sensitive to mass differences  through a factor of $M_\ast - M - m_\pi$ raised to a power.
The four relevant  mass differences in the vacuum are given in Eqs.~\eqref{Deltaij}.
The thermal mass shifts for $D^\ast$, $D$, and $\pi$ are given in Eqs.~\eqref{deltaMD*-approx}, \eqref{deltaMD-approx}, and \eqref{deltampi}.
The mass differences in the hadron gas after kinetic freezeout decrease linearly with  $\mathfrak{n}_\pi$
with the same slope:
\begin{subequations}
    \bea
    \Delta_{00}-m_{\pi 0} &\approx& +7.04~\mathrm{MeV}
    - \big(3.23~\mathrm{MeV}\big) \, \mathfrak{n}_\pi/\mathfrak{n}_\pi^\mathrm{(kf)},
    \label{delta00-npi}
    \\
    \Delta_{+0}-m_{\pi +} &\approx& +5.86~\mathrm{MeV}
    - \big(3.23~\mathrm{MeV}\big) \, \mathfrak{n}_\pi/\mathfrak{n}_\pi^\mathrm{(kf)},
    \label{delta0+-npi}
    \\
    \Delta_{++}-m_{\pi 0} &\approx& +5.63~\mathrm{MeV}
    - \big(3.23~\mathrm{MeV}\big) \, \mathfrak{n}_\pi/\mathfrak{n}_\pi^\mathrm{(kf)},
    \label{delta0+0-npi}
    \\
    \Delta_{0+}-m_{\pi +} &\approx& - 2.38~\mathrm{MeV} -
    \big(3.23~\mathrm{MeV}\big) \, \mathfrak{n}_\pi/\mathfrak{n}_\pi^\mathrm{(kf)},
    \label{delta+--npi}
    \eea
    \label{deltaij-npi}%
\end{subequations}
where $\mathfrak{n}_\pi^\mathrm{(kf)}$ is the pion number density at kinetic freezeout.
The signs of the mass differences in Eqs.~\eqref{deltaij-npi} imply that the
decays $D^{\ast 0} \to D^0 \pi^0$, $D^{\ast +} \to D^+ \pi^0$, and $D^{\ast +} \to D^+ \pi^0$
are always kinematically allowed in the expanding hadron gas after kinetic freezeout,
while the decay $D^{\ast 0} \to D^+ \pi^-$ is always  forbidden.

The partial widths of the charm mesons  from the decays $D^\ast \to D \pi$ are given in Eq.~\eqref{GammaD*Dpi}:
\begin{subequations}
    \bea
    \Gamma_{D^{\ast +} \to D^+ \pi} &=&
    \frac{g_\pi^2}{12\pi \, f_\pi^2}
    \left( \Delta_{++}^2 - m_{\pi 0}^2 \right)^{3/2}.
    \label{GammaD*+D+pi}
    \\
    \Gamma_{D^{\ast +} \to D^0 \pi} &=&
    \frac{g_\pi^2}{6\pi \, f_\pi^2} \left( \Delta_{+0}^2 - m_{\pi +}^2 \right)^{3/2} ,
    \label{GammaD*pD0pi}
    \\
    \Gamma_{D^{\ast 0} \to D^0 \pi} &=&
    \frac{g_\pi^2}{12\pi \, f_\pi^2}
    \left( \Delta_{00}^2 - m_{\pi 0}^2 \right)^{3/2},
    \label{GammaD*0D0pi}
    \\
    \Gamma_{D^{\ast 0} \to D^+ \pi} &=& 0,
    \label{GammaD*0D+pi}
    \eea
    \label{GammaD*aDbpi}%
\end{subequations}
where $\Delta_{ab} = M_{\ast a} - M_b$ is the $D^{\ast a}$-$D^b$ mass difference.
In the vacuum, the masses $M_{\ast a}$, $M_b$, and $m_{\pi ab}$ are constants.
In the hadron gas, the mass shifts from coherent pion forward scattering can be  taken into account
by replacing $\Delta_{ab}$ in Eqs.~\eqref{GammaD*aDbpi} by $\Delta_{ab} + \delta M_* - \delta M$,
where $\delta M$  and $\delta M_*$
are the charm-meson mass shifts in Eqs.~\eqref{deltaMD-approx} and \eqref{deltaMD*-approx},
and replacing $m_{\pi i}$ by $m_{\pi i} + \delta m_\pi$,
where $\delta m_\pi$  is  the pion  mass shift in Eq.~\eqref{deltampi}.
In the expanding hadron gas after kinetic freezeout,
the terms $\Delta_{ab}^2 - m_{\pi ab}^2$ in Eqs.~\eqref{GammaD*aDbpi} are quadratic functions of  $\mathfrak{n}_\pi$.

The thermal width $\Gamma_a$ of $D^a$ from coherent pion forward scattering is given in Eq.~\eqref{deltaGammaDa-approx}.
The thermal widths for $D^+$ and $D^0$ are
\begin{subequations}
    \bea
    \Gamma_+ &=&
    3 \,  \mathfrak{f}_\pi(m_\pi)\,
    \Gamma_{D^{\ast +} \to D^+ \pi} ,
    \label{GammaD+}
    \\
    \Gamma_0 &=&
    3 \,  \mathfrak{f}_\pi(m_\pi)\,
    \big( \Gamma_{D^{\ast 0} \to D^0 \pi} + \Gamma_{D^{\ast +} \to D^0 \pi} \big).
    \label{GammaD0}
    \eea
    \label{GammaD}%
\end{subequations}
In the hadron gas before kinetic freezeout, the factor $ \mathfrak{f}_\pi(m_\pi)$ depends on the temperature $T$.
In the hadron gas after kinetic freezeout at the temperature $T_\mathrm{kf} = 115$~MeV,
$\mathfrak{f}_\pi(m_\pi)= 0.431\,  \mathfrak{n}_\pi/\mathfrak{n}_\pi^\mathrm{(kf)}$, 
where $\mathfrak{n}_\pi^\mathrm{(kf)}$ is the pion number density at kinetic freezeout.
The thermal widths $\Gamma_a$ in Eqs.~\eqref{GammaD} also depend on $T$ or $\mathfrak{n}_\pi$
through the factors of $(\Delta_{ab}^2 - m_{\pi ab}^2)^{3/2}$ in $\Gamma_{D^{\ast a} \to D^b \pi}$.

The thermal correction $\delta \Gamma_{*a}$ to the width  for $D^{\ast a}$ is given in Eq.~\eqref{deltaGammaD*a-approx}.
The total widths  for $D^{*+}$ and $D^{*0}$ are
\begin{subequations}
    \bea
    \Gamma_{*+}  &=& \big[1+  \mathfrak{f}_\pi(m_\pi)  \big]\,
    \big( \Gamma_{D^{\ast +} \to D^+ \pi} + \Gamma_{D^{\ast +} \to D^0 \pi} \big) +  \Gamma_{*+,\gamma} ,
    \label{GammaD*+}
    \\
    \Gamma_{*0} &=&
    \big[1+  \mathfrak{f}_\pi(m_\pi) \big]\, \Gamma_{D^{\ast 0} \to D^0 \pi} +  \Gamma_{*0,\gamma},
    \label{GammaD*0}
    \eea
    \label{GammaD*}%
\end{subequations}
where $\Gamma_{*+,\gamma}$ and $\Gamma_{*0,\gamma}$ are the radiative decay rates in Eqs.~\eqref{GammaD*Dgamma}.
The terms with the factor $ \mathfrak{f}_\pi(m_\pi)$ come from coherent pion forward scattering.
In the hadron gas before kinetic freezeout, $ \mathfrak{f}_\pi(m_\pi)$ depends on $T$.
In the hadron gas after kinetic freezeout, $\mathfrak{f}_\pi(m_\pi)= 0.431\,  \mathfrak{n}_\pi/\mathfrak{n}_\pi^\mathrm{(kf)}$.
The thermal widths $\Gamma_{*a}$ in Eqs.~\eqref{GammaD*}  also depend on $T$ or $\mathfrak{n}_\pi$
through the factors of $(\Delta_{ab}^2 - m_{\pi ab}^2)^{3/2}$ in $\Gamma_{D^{\ast a} \to D^b \pi}$.

\begin{figure}[t]
    \includegraphics[width=0.7\textwidth]{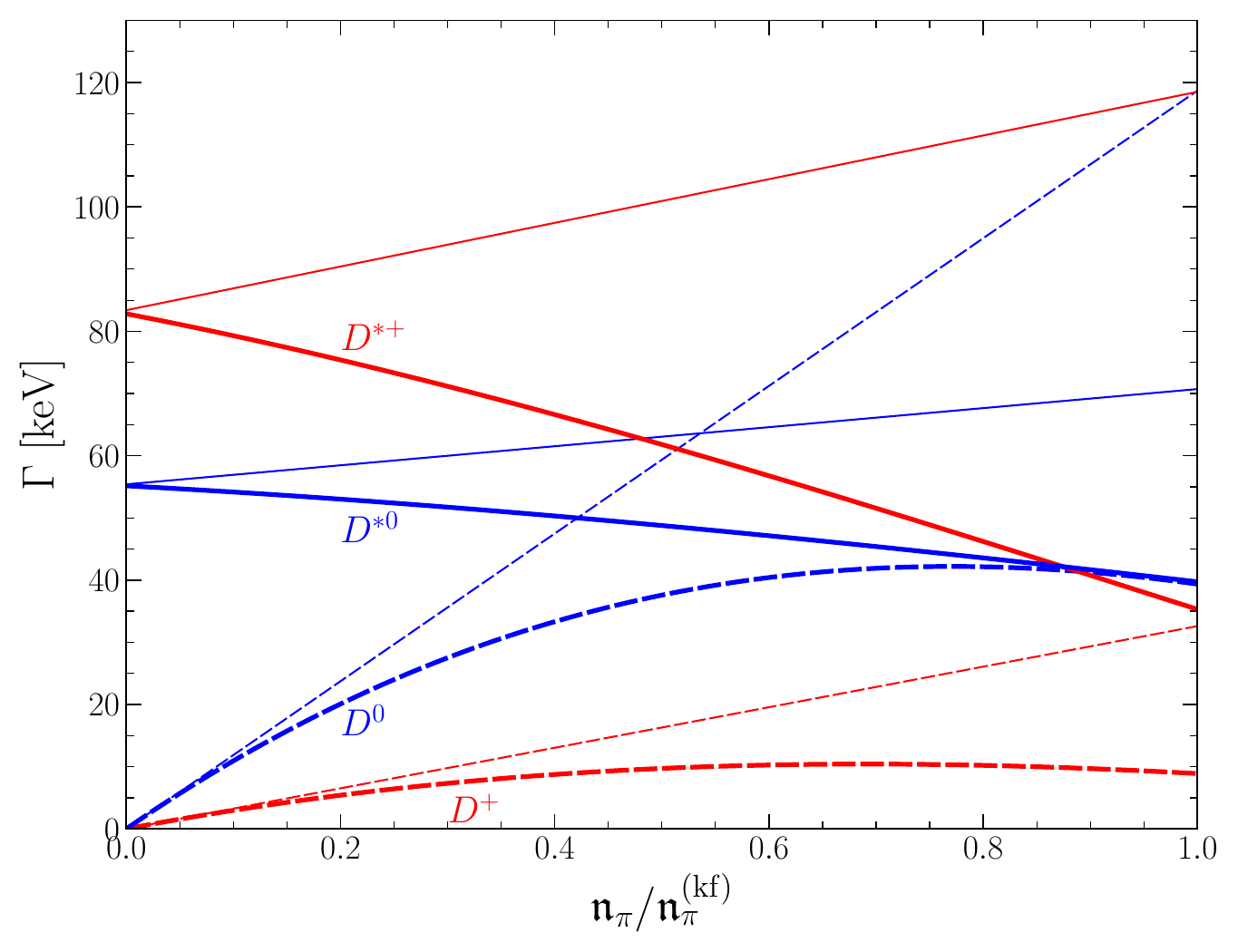}
    \caption{
        Thermal widths for the charm mesons in the hadron gas after kinetic freezeout
        as functions of the pion number density $\mathfrak{n}_\pi$:
        $D^+$ (dashed red), $D^0$ (dashed blue), $D^{\ast+}$ (solid red), and $D^{\ast0}$ (solid blue).
        The thicker  curves include the effects of mass shifts from coherent pion forward scattering.
        The thinner straight lines ignore the thermal mass shifts.
    }
    \label{fig:Gamma-npi}
\end{figure}

The thermal widths for the charm mesons after kinetic freezeout at the temperature $T_\mathrm{kf} = 115$~MeV
are shown as functions of the pion number density $\mathfrak{n}_\pi$ in Fig.~\ref{fig:Gamma-npi}.
The thermal widths of $D^+$ and $D^0$ are given in Eqs.~\eqref{GammaD}.
The thermal widths of $D^{\ast +}$ and $D^{\ast 0}$ are given in Eqs.~\eqref{GammaD*}.
The thicker curves in Fig.~\ref{fig:Gamma-npi}
take into account the thermal mass shifts of pions and charm mesons
in the partial decay rates for $D^{\ast a} \to D^b \pi$ in Eqs.~\eqref{GammaD*aDbpi}.
The thinner straight  lines in Fig.~\ref{fig:Gamma-npi} are obtained by setting the masses of pions and charm mesons
in those partial decay rates  equal to their vacuum values.
The effects of the thermal mass shifts are large.
At kinetic freezeout, the thermal widths of $D^+$ and $D^0$ are  9.1~keV 
and  40.2\,keV .
As $\mathfrak{n}_\pi$ decreases from $\mathfrak{n}_\pi^\mathrm{(kf)}$ to 0,
those decay rates increase to the maximum values
10.6\,keV and  42.9\,keV near $0.74\, \mathfrak{n}_\pi^\mathrm{(kf)}$ 
and then decrease to 0.
At kinetic freezeout, the thermal widths of $D^{*+}$ and $D^{*0}$  are  35.6\,keV and 39.9\,keV. 
As $\mathfrak{n}_\pi$ decreases from $\mathfrak{n}_\pi^\mathrm{(kf)}$ to 0,
those decay rates increase to the vacuum values in Eqs.~\eqref{Gamma*} to within errors.
The decrease in the  thermal widths of $D^{*+}$ and $D^{*0}$ with increasing  $\mathfrak{n}_\pi$ may be counterintuitive,
but it  is a consequence of the decreasing phase space available for the decay 
because of the decreasing mass differences in Eqs.~\eqref{deltaij-npi}.


\section{Reaction Rates}
\label{sec:ReactionRates}

In this section, we calculate reaction rates for charm mesons in a pion gas.
The results are applied to the hadron gas from a heavy-ion collision after kinetic freezeout.

\subsection{$\bm{D^\ast \leftrightarrow D\pi}$}
\label{sec:D*decay}

The decays of $D^\ast$ into $D\pi$  are 1-body reactions that give contributions
to the rate equations for the number densities of $D^\ast$ in a  pion gas
that are not suppressed by any powers of the pion number density.
The partial decay rate  in the vacuum for $D^{\ast a} \to D^b \pi$
in HH$\chi$EFT at LO  is given in Eq.~\eqref{GammaD*Dpi}.
This rate is nonzero only if $\Delta_{ab}  > m_{\pi ab}$, and it is sensitive to the masses through the factor
of $(\Delta_{ab}^2 -m_{\pi ab}^2)^{3/2}$.
This expression can also be used for the partial decay rate in the pion gas
by taking into account the mass shifts from coherent pion forward scattering.
The charm-meson mass difference $\Delta_{ab}$ is shifted by $\delta M_* - \delta M$,
where $\delta M_*$ and $\delta M$ are given by Eqs.~\eqref{deltaMD*-approx} and \eqref{deltaMD-approx}.
The pion mass $m_{\pi ab}$ is shifted by $\delta m_\pi$, which is given in Eq.~\eqref{deltampi}.

The radiative decays of $D^\ast$ into $D \gamma$ are also 1-body reactions.
The partial decay rates for $D^\ast \to D\gamma$ in the vacuum are not sensitive to masses,
because the $D^\ast$-$D$ mass differences are much larger than the mass shifts.
The radiative  decay rates in the pion gas can therefore be approximated by their values in the vacuum
in Eqs.~\eqref{GammaD*Dgamma}.

\begin{figure}[hbt]
    \includegraphics*[width=0.3\linewidth]{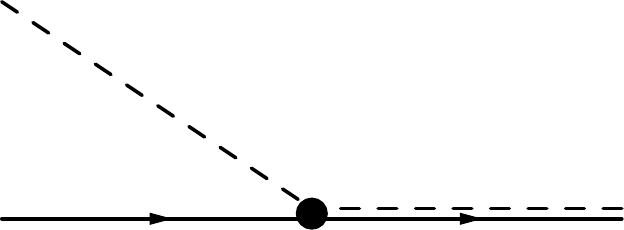}
    \caption{Feynman diagram for $\pi D \to D^\ast$  in HH$\chi$EFT at LO.
        The dashed line is a $\pi$,  the  solid line  is a $D$, and the double (solid+dashed) line is a $D^\ast$.}
    \label{fig:piDtoD*}
\end{figure}

A vector charm meson $D^\ast$ can be produced in a pion gas by the inverse decay  $\pi D \to D^\ast$.
The reaction rate in the vacuum for $D^a \pi  \to D^{\ast b}$ averaged over the three pion flavors  is
\beq
v\sigma \big[ \pi D^a  \to D^{\ast b}  \big] =
\frac{\pi g_\pi^2}{6 f_\pi^2} \,  (2 - \delta_{ab}) \, \frac{q_{ba}^2}{\Delta_{ba}} \,  \delta(\omega_{baq} - \Delta_{ba}),
\label{vsigmaDpitoD*}
\eeq
$\omega_{baq} = \sqrt{m_{\pi ba}^2 + q^2}$, and $q_{ba}^2 = \Delta_{ba}^2 - m_{\pi ba}^2$.
The reaction rate in the pion gas is
obtained by averaging over the momentum distributions of the incoming $D$ and $\pi$.
The average over the $D$ momentum distribution has no effect,
 because the reaction rate in Eq.~\eqref{vsigmaDpitoD*} does not depend on the charm-meson momentum.
 The average over the pion momentum distribution can be evaluated using the delta function in Eq.~\eqref{vsigmaDpitoD*}:
\bea
\left\langle v\sigma \big[ \pi D^a \to D^{\ast b}  \big] \right\rangle=
\big( \mathfrak{f}_\pi(\Delta_{ba})/\mathfrak{n}_\pi \big)\,
\Gamma \big[D^{\ast b} \to D^a \pi \big],
\label{<vsigmaDpitoD*>}
\eea
where $\Gamma [D^{\ast  b} \to D^a \pi]$ is the decay rate in Eq.~\eqref{GammaD*Dpi}.
Since $\Delta_{ba}$ is large compared to isospin splittings,
it can be approximated by the average $\Delta$ over the four $D^{\ast b} \to D^a \pi$ transitions
or alternatively by the pion mass $m_\pi$.
The reaction rates for $\pi D^a  \to D^{\ast b}$ in the hadron gas near or after kinetic freezeout are
\begin{subequations}
    \bea
    \left\langle v\sigma_{\pi D^+ \to D^{\ast +}} \right\rangle
    &=& \big[\mathfrak{f}_\pi(m_\pi)/\mathfrak{n}_\pi\big] \, \Gamma_{D^{\ast +} \to D^+ \pi},
    \label{vsigmapiD+toD+gamma}
    \\
    \left\langle v\sigma_{\pi D^0  \to D^{\ast +}} \right\rangle
    &=&  \big[\mathfrak{f}_\pi(m_\pi)/\mathfrak{n}_\pi\big] \, \Gamma_{D^{\ast +} \to D^0 \pi} ,
    \label{vsigmapiD0toD+gamma}
    \\
    \left\langle v\sigma_{\pi D^0 \to D^{\ast 0}} \right\rangle
    &=&  \big[\mathfrak{f}_\pi(m_\pi)/\mathfrak{n}_\pi\big] \, \Gamma_{D^{\ast 0} \to D^0 \pi} ,
    \label{vsigmapiD0toD0gamma}
    \\
    \left\langle v\sigma_{\pi D^+ \to D^{\ast 0}} \right\rangle
    &=& 0.
    \label{vsigmapiD+toD0gamma}
    \eea
    \label{vsigmapiDtoD*}%
\end{subequations}
Before kinetic freezeout, the factor $\mathfrak{f}_\pi(m_\pi)/\mathfrak{n}_\pi$ is determined by the temperature $T$.
After kinetic freezeout at the temperature $T_\mathrm{kf}=115$~MeV,
that factor has the constant value $0.431/\mathfrak{n}_\pi^\mathrm{(kf)}$ independent of $\mathfrak{n}_\pi$.

\subsection{$\bm{\pi D \to \pi D}$}
\label{sec:piDtopiD}

The reaction $\pi D^a \to \pi D^b$ can change the  flavor of a pseudoscalar charm meson.
The Feynman diagrams for this reaction in HH$\chi$EFT at LO are shown in Fig.~\ref{fig:piDscattering}.
The reaction rate has a $D^*$ resonance contribution from the second diagram in Fig.~\ref{fig:piDscattering}
that is sensitive to isospin splittings and to the $D^\ast$ width.
A  simple expression for the nonresonant contribution to the reaction rate can be obtained by
setting the $D^\ast$-$D$ mass splitting $\Delta$ equal to the pion mass $m_\pi$
and then taking the limit as the $D^\ast$ width $\Gamma_*$ approaches 0.
A simple expression for the resonant contribution to the reaction rate can be obtained by isolating the term with the factor $1/\Gamma_*$.
We approximate the reaction rate by the sum of the nonresonant reaction rate and the resonant reaction rate.

The $\mathcal T$-matrix element for $\pi ^iD^a \to \pi^j D^b$ in the zero-width limit
is  obtained from the amplitude in Eq.~\eqref{ADpi} by setting $\Gamma_*=0$ and
by putting the external legs on shell
by setting $v \cdot p=0$, $v \cdot q= \omega_q$, and $v \cdot q^\prime= \omega_{q^\prime}$:
\bqa
\mathcal{T}_{ai,bj} = \frac{1}{2 f_\pi^2} \, [\sigma^i, \sigma^j]_{ab} \,  (\omega_q +\omega_{q^\prime})
-\frac{g_\pi^2}{f_\pi^2} \left[ (\sigma^i \sigma^j)_{ab}  \,
    \frac{\bm{q} \!\cdot\! \bm{q}^\prime}{\omega_q - \Delta}
    - (\sigma^j \sigma^i)_{ab}  \,
    \frac{\bm{q} \!\cdot\! \bm{q}^\prime}{\omega_{q^\prime}+\Delta}  \right].
\label{TDpi}
\eqa
We can ignore the recoil of $D$ and set $|\bm{q}^\prime| = |\bm{q}|$.
The non-resonant reaction rate can be obtained by taking the  limit $\Delta \to m_\pi$.
The non-resonant reaction rate for $\pi D^a \to \pi D^b$ averaged over incoming pion flavors is
\bqa
v\sigma[\pi D^a \to \pi D^b]_\mathrm{nonres} = \frac{1}{12 \pi f_\pi^4}
\big[ 2(2 - \delta_{ab})\,(1 + g_\pi^4/3)\, \omega_q ^2 + \delta_{ab}\, g_\pi^4 \,m_\pi^2 \big] \frac{q}{\omega_q},
\label{vsigmaDpi}
\eqa
where $q$ is the 3-momentum of the incoming pion.
The reaction rate in the pion gas can be obtained by averaging over the momentum distributions
of the incoming $D$ and $\pi$:
\bqa
\big\langle v\sigma[\pi D^a \to \pi D^b]_\mathrm{nonres} \big\rangle =
\frac{1}{12 \pi f_\pi^4} \left[ 2(2 - \delta_{ab})\,(1 + g_\pi^4/3)\,  \big\langle \omega_q q \big\rangle_{\!\bm{q}}
    + \delta_{ab}\, g_\pi^4 \,m_\pi^2  \left\langle \frac{q}{\omega_q} \right\rangle_{\!\!\!\bm{q}} \right],
\label{vsigmaD+pi-}
\eqa
where the angular brackets represents the average over the Bose-Einstein distribution for the pion
defined in Eq.~\eqref{<F>}.

The second diagram in Fig.~\ref{fig:piDscattering}
with $D^{*c}$ in the $s$ channel  gives a resonance contribution to the reaction rate
proportional to $1/\Gamma_{*c}$ if $\Delta_{ac}  > m_{\pi ac}$.
In the square of the matrix element for the scattering of a $D$ with momentum $Mv+p$ and a $\pi$ with momentum $q$,
the resonance contribution can be isolated by making a simple substitution for the product of the
$D^\ast$ propagator and its complex conjugate:
\beq
\frac{1}{v \!\cdot\! (p+q) - \Delta + i \Gamma_*/2}
\left( \frac{1}{v \!\cdot \!(p+q) -\Delta + i \Gamma_*/2} \right)^{\!\!*}
\longrightarrow
\frac{2 \pi}{\Gamma_*} \, \delta \big( v \!\cdot \!(p+q) - \Delta \big).
\label{1/GammaD*}
\eeq
The resonant reaction rate for $\pi D^a \to \pi D^b$ averaged over the flavors of the incoming pion  is
\bea
v\sigma \big[ \pi D^a \to \pi D^b \big]_\mathrm{res} &=& \frac{g_\pi^4}{72 f_\pi^4}
\sum_c  (2 - \delta_{ac}) (2 - \delta_{bc })\frac{q_{ca}^2 \, q_{cb}^3}{\Delta_{ca}\Gamma_{*c}}\,
\theta(\Delta_{cb} -  m_{\pi cb})\, \delta(\omega_{caq}  - \Delta_{ca}) ,
\eea
where $q$ is the 3-momentum of the incoming pion.
Using the expressions for $\Gamma [D^\ast  \to D \pi]$ in Eq.~\eqref{GammaD*Dpi}
and $v\sigma[\pi D \to D^\ast]$ in Eq.~\eqref{vsigmaDpitoD*},
the singular term in the reaction rate can  be expressed as
\beq
v\sigma \big[ \pi D^a \to \pi D^b \big]_\mathrm{res} =
\sum_c \frac{1}{\Gamma_{\ast c}}\, v \sigma[\pi  D^a \to D^{\ast c}] \,\Gamma[D^{\ast c} \to D^b \pi].
\label{vsigmaDpiDpi}
\eeq
The  reaction rate in the pion gas
can be evaluated by using the thermal average of $v \sigma [\pi  D \to D^*]$ in  Eq.~\eqref{<vsigmaDpitoD*>}:
\bea
\left\langle v\sigma \big[ \pi D^a \to \pi D^b \big]_\mathrm{res} \right\rangle=
\frac{\mathfrak{f}_\pi(\Delta)}{\mathfrak{n}_\pi} \,
\sum_c \frac{\Gamma \big[D^{\ast c} \to D^a \pi \big] \,\Gamma \big[D^{\ast c} \to D^b \pi \big]}{\Gamma_{\ast c}}.
\label{<vsigmaDpitoDpi>}
\eea

The reaction rate for  $\pi D^a  \to \pi D^b$ in the pion gas
can be approximated by the sum of the non-resonant reaction rate  in Eq.~\eqref{vsigmaD+pi-}
and the resonant reaction rate in Eq.~\eqref{<vsigmaDpitoDpi>}.
The reaction rates in the hadron gas near or after kinetic freezeout are
\begin{subequations}
    \bea
    \left\langle v\sigma_{\pi D^0 \to \pi D^0} \right\rangle
    &=&(0.496+0.188 \,g_\pi^4)\, \frac{m_\pi^2}{ f_\pi^4}
    +  \frac{\mathfrak{f}_\pi(m_\pi)}{\mathfrak{n}_\pi} \,
    \left(\frac{\Gamma_{D^{\ast 0} \to D^0 \pi}^2}{\Gamma_{*0}}
    + \frac{\Gamma_{D^{\ast +} \to D^0 \pi}^2}{\Gamma_{*+}}\right) ,
    \label{vsigmaD0toD0}
    \\
    \left\langle v\sigma_{\pi D^0 \to \pi D^+} \right\rangle
    &=& (0.991 + 0.330 \,g_\pi^4)\,\frac{m_\pi^2}{ f_\pi^4}
    +  \frac{\mathfrak{f}_\pi(m_\pi)}{\mathfrak{n}_\pi} \,
    \frac{\Gamma_{D^{\ast +} \to D^0 \pi}\, \Gamma_{D^{\ast +} \to D^+ \pi} }{\Gamma_{*+}},
    \label{vsigmaD0toD+}
    \\
    \left\langle v\sigma_{\pi D^+ \to  \pi D^0} \right\rangle
    &=& (0.991 + 0.330 \,g_\pi^4)\,\frac{m_\pi^2}{ f_\pi^4}
    +  \frac{\mathfrak{f}_\pi(m_\pi)}{\mathfrak{n}_\pi} \,
    \frac{\Gamma_{D^{\ast +} \to D^0 \pi}\, \Gamma_{D^{\ast +} \to D^+ \pi} }{\Gamma_{*+}},
    \label{vsigmaD+toD0}
    \\
    \left\langle v\sigma_{\pi D^+ \to  \pi D^+} \right\rangle
    &=&  (0.496+0.188 \,g_\pi^4)\, \frac{m_\pi^2}{ f_\pi^4}
    +  \frac{\mathfrak{f}_\pi(m_\pi)}{\mathfrak{n}_\pi} \, \frac{\Gamma_{D^{\ast +} \to D^+ \pi}^2 }{\Gamma_{*+}}.
    \label{vsigmaD+toD+}
    \eea
    \label{vsigmaDtoD}%
\end{subequations}
The dimensionless numbers in the first terms depend only on $m_\pi/T$, which we have evaluated at $T_\mathrm{kf}=115$~MeV.
Before kinetic freezeout, the factor $\mathfrak{f}_\pi(m_\pi)/\mathfrak{n}_\pi$ is determined by  $T$.
After kinetic freezeout at $T_\mathrm{kf}=115$~MeV,
that factor has the constant value $0.431/\mathfrak{n}_\pi^\mathrm{(kf)}$ independent of $\mathfrak{n}_\pi$.
There can also be dependence on $T$ or $\mathfrak{n}_\pi$ through
the mass shifts in $\Gamma_{D^{\ast a} \to D^b \pi}$ and through the factors of $1/\Gamma_{*c}$.

The $D^\ast$ resonance terms in the reaction rates for  $\pi D^a  \to \pi D^b$  in Eqs.~\eqref{vsigmaDtoD}
can be obtained from the reaction rates for $\pi D^a  \to D^{\ast c}$ from Eqs.~\eqref{vsigmapiDtoD*}
by multiplying by the branching fraction $\Gamma_{D^{\ast c} \to D^b \pi}/\Gamma_{*c}$
and summing over the two flavors of $D^{\ast c}$.
Thus if the reaction rates for $\pi D  \to D^\ast$ in Eqs.~\eqref{vsigmapiDtoD*}
and  $\pi D  \to \pi D$  in Eqs.~\eqref{vsigmaDtoD} are both included in a rate equation,
the contributions of $\pi D  \to D^\ast$ in which $D^\ast$ subsequently decays to $D \pi$ are double counted.
The only contributions of $\pi D  \to D^\ast$ that are not double counted are
those in which $D^\ast$ subsequently decays to $D \gamma$.
The double counting can be avoided by replacing the reaction rates for $\pi D  \to D^\ast$ in Eqs.~\eqref{vsigmapiDtoD*}
by the contributions from the subsequent radiative decay of $D^\ast$:
\begin{subequations}
    \bea
    \left\langle v\sigma_{\pi D^+ \to D^+ \gamma} \right\rangle
    &=& \big[\mathfrak{f}_\pi(m_\pi)/\mathfrak{n}_\pi\big] \, \Gamma_{D^{\ast +} \to D^+ \pi} \,
    \big(  \Gamma_{\ast +, \gamma} / \Gamma_{\ast +} \big),
    \label{vsigmapiD+toD*+}
    \\
    \left\langle v\sigma_{\pi D^0  \to D^+\gamma} \right\rangle
    &=&  \big[\mathfrak{f}_\pi(m_\pi)/\mathfrak{n}_\pi\big] \, \Gamma_{D^{\ast +} \to D^0 \pi} \,
    \big(  \Gamma_{\ast +, \gamma}/  \Gamma_{\ast +} \big)\, ,
    \label{vsigmapiD0toD*+}
    \\
    \left\langle v\sigma_{\pi D^0 \to D^0\gamma} \right\rangle
    &=&  \big[\mathfrak{f}_\pi(m_\pi)/\mathfrak{n}_\pi\big] \,\Gamma_{D^{\ast 0} \to D^0 \pi} \,
    \big(  \Gamma_{\ast 0, \gamma} / \Gamma_{\ast 0} \big),
    \label{vsigmapiD0toD*0}
    \\
    \left\langle v\sigma_{\pi D^+ \to D^0\gamma} \right\rangle
    &=& 0,
    \label{vsigmapiD+toD*0}
    \eea
    \label{vsigmapiDtoDgamma}%
\end{subequations}
where $\Gamma_{\ast +, \gamma}$ and $\Gamma_{\ast 0, \gamma}$ are the radiative decay rates in the vacuum
in Eqs.~\eqref{GammaD*Dgamma}.

\subsection{$\bm{\pi D^\ast \leftrightarrow \pi D}$}
\label{sec:piD*topiD*}

\begin{figure}[t]
    \includegraphics[width=0.7\textwidth]{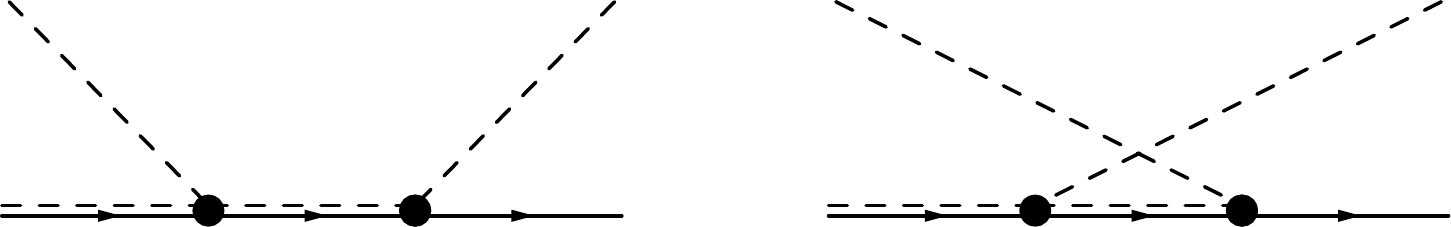}
    \caption{
        Feynman diagrams for $\pi D^\ast \to \pi D$ in HH$\chi$EFT at LO.
        The diagrams for $\pi D \to \pi D^\ast$ are the mirror images of these diagrams.
    }
    \label{fig:piD*topiD}
\end{figure}

The reactions $\pi D^\ast \leftrightarrow \pi D$ can change vector charm mesons 
into pseudoscalar charm mesons and vice versa.
The reaction $\pi  D^\ast \to \pi D $ is exothermic, releasing a  mass energy comparable to $m_\pi$.
Since this is large compared to isospin splittings,  isospin splittings  can be neglected.
Relatively simple expressions for the reaction rates can be obtained by taking the limit $\Delta \to m_\pi$.
The square of the matrix element  for $\pi(q)  D^{\ast a} \to  \pi(q^\prime) D^b$
averaged over $D^\ast$ spins and averaged/summed over pion flavors is
\beq
\overline{|\mathcal{M}|^2} = \frac{g_\pi^4}{9 f_\pi^4}
\frac{(\bm{q} \times \bm{q}^\prime)^2}{\omega_q^2 \omega_{q^\prime}^2}
\big[  2(2- \delta_{ab})   (\omega_q - \omega_{q^\prime})^2 
+ 3\delta_{ab}  (\omega_q + \omega_{q^\prime})^2 \big] .
\label{M^2:piD*topiD}
\eeq
The reaction rate can be reduced to
\beq
v\sigma \big[ \pi D^{\ast a}  \to \pi D^b \big] =
\frac{g_\pi^4}{216 \pi f_\pi^4}
\frac{q^2 \big[(\omega_q + \Delta)^2 -m_\pi^2 \big]^{3/2}}{\omega_q^3 (\omega_q + \Delta)^2}
\big[ 3\delta_{ab}  (2\omega_q + \Delta)^2 + 2(2- \delta_{ab})   \Delta^2 \big] .
\eeq
The reaction rate in the pion gas in the limit $\Delta \to m_\pi$ is
\bea
\left\langle v\sigma \big[ \pi D^{\ast a}  \to \pi D^b \big] \right\rangle &=&
\frac{g_\pi^4}{216 \pi f_\pi^4}
\left\langle \frac{q^2}{(\omega_q + m_\pi)^2}
\left( \frac{\omega_q+2m_\pi}{\omega_q} \right)^{3/2}\right.
\nonumber\\
&& \hspace{1cm} \left. \phantom{\left( \frac{1}{1} \right)^{3/2}}
\times
\big[ 3\delta_{ab}  (2\omega_q +m_\pi)^2 + 2(2- \delta_{ab})  m_\pi^2 \big]
\right\rangle_{\!\!\!\bm q}.
\eea

The square of the matrix element  for $ \pi  D^a \to \pi  D^{\ast b}$ summed over $D^\ast$ spins and
averaged/summed over the pion flavors can be obtained from $\overline{|\mathcal{M}|^2}$
for $\pi D^{\ast a} \to \pi D^b$  in Eq.~\eqref{M^2:piD*topiD} by multiplying by 3.
The reaction rate in the pion gas in the limit $\Delta \to m_\pi$ is
\bea
\left\langle v\sigma \big[ \pi D^a \to \pi D^{\ast b} \big] \right\rangle &=&
\frac{g_\pi^4}{72 \pi f_\pi^4}
\left\langle \frac{q^2}{(\omega_q - m_\pi)^2}
\left( \frac{\omega_q-2m_\pi}{\omega_q} \right)^{3/2} \right.
\nonumber\\
&& \hspace{0cm} \left.  \phantom{\left( \frac{1}{1} \right)^{3/2}}
\times \big[ 3\delta_{ab}  (2\omega_q - m_\pi)^2 + 2(2- \delta_{ab})   m_\pi^2 \big]
\theta(\omega_q - 2 m_\pi) \right\rangle_{\!\!\!\bm q}.
\eea
The theta function restricts the pion energy to be above the threshold $2m_\pi$ for producing $\pi D^\ast$,
which requires $q > \sqrt{3}\, m_\pi$.

The reaction rates for  $\pi D^a  \to \pi D^{*b}$  in the hadron  gas after kinetic freezeout are
\begin{subequations}
    \bea
    \left\langle v\sigma_{\pi D^0 \to \pi D^{*0}} \right\rangle = \left\langle v\sigma_{\pi D^+ \to \pi D^{*+}} \right\rangle
    &=&  0.215\,g_\pi^4\, m_\pi^2/f_\pi^4 ,
    \label{vsigmaD0toD*0}
    \\
    \left\langle v\sigma_{\pi D^0 \to \pi D^{*+}} \right\rangle
    =\left\langle v\sigma_{\pi D^+ \to  \pi D^{*0}} \right\rangle
    &=&  0.005\, g_\pi^4\, m_\pi^2/f_\pi^4  .
    \label{vsigmaD0toD*+}
    \eea
    \label{vsigmaDtoD*}%
\end{subequations}
The reaction rates for  $\pi D^{*a}  \to \pi D^b$ in the hadron  gas after kinetic freezeout are
\begin{subequations}
    \bea
    \left\langle v\sigma_{\pi D^{*0} \to \pi D^0} \right\rangle = \left\langle v\sigma_{\pi D^{*+} \to \pi D^+} \right\rangle
    &=&  0.243\, g_\pi^4\, m_\pi^2/f_\pi^4    ,
    \label{vsigmaD*0toD0}
    \\
    \left\langle v\sigma_{\pi D^{*0} \to \pi D^+} \right\rangle =\left\langle v\sigma_{\pi D^{*+} \to  \pi D^0} \right\rangle
    &=& 0.006\, g_\pi^4\, m_\pi^2/f_\pi^4.
    \label{vsigmaD*0toD+}
    \eea
    \label{vsigmaD*toD}%
\end{subequations}
The dimensionless numerical factors depend only on $m_\pi/T$, which has been evaluated at $T_\mathrm{kf}=115$~MeV.

\subsection{$\bm{\pi D^\ast \to \pi D^\ast}$}

The  reaction $\pi D^\ast \to \pi D^\ast$  can change the flavor of a vector charm meson.
The five Feynman diagrams for this reaction in HH$\chi$EFT at LO are shown in Fig.~\ref{fig:piD*scattering}.
The third diagram, which proceeds through an intermediate $D$,
produces a $t$-channel singularity in the reaction rate that is proportional to $1/\Gamma$ in the limit $\Gamma \to 0$.
The singularity comes from the decay $D^\ast \to \pi D$ followed by the inverse decay $\pi D \to D^\ast$.
A relatively simple expression for the nonsingular contribution to the reaction rate
is obtained by setting $\Delta =m_\pi$ and then taking the limit $\Gamma \to 0$.
A simple expression for the resonant contribution to the reaction rate can be obtained by isolating the term with the factor  $1/\Gamma$.
We approximate the reaction rate by the sum of the nonsingular reaction rate and the singular reaction rate.

The $\mathcal T$-matrix element for $\pi^i D^{*a} \to \pi^j D^{* b}$ in the zero-width limit
is  obtained from the amplitude in Eq.~\eqref{ApiD*} by
contracting it with the $D^\ast$ polarization vectors, setting $\Gamma=0$, and
then putting the external legs on shell by setting $v \cdot p = \Delta$,
$v \cdot q= \omega_q$, and $v \cdot q^\prime= \omega_{q^\prime}$:
\bea
\varepsilon_\mu \mathcal {T}^{\mu \nu}_{ai,bj} \varepsilon^{\prime\ast}_\nu &=&
-\frac{1}{2 f_\pi^2} 
\left[\sigma^i,\sigma^j\right]_{ab} (\omega_q +\omega_{q^\prime}) (\bm{\varepsilon} \cdot \bm{\varepsilon}^{\prime\ast})
\nonumber\\
&&-\frac{g_\pi^2}{f_\pi^2}
\left[
    (\sigma^i \sigma^j)_{ab}
    \frac{ (\bm{\varepsilon} \cdot \bm{q}) \,  (\bm{q}^\prime \cdot  \bm{\varepsilon}^{\prime\ast})}{\omega_q+\Delta}
    - (\sigma^j\sigma^i)_{ab}
    \frac{ (\bm{\varepsilon} \cdot \bm{q}^\prime) \,  (\bm{q} \cdot  \bm{\varepsilon}^{\prime\ast})}{\omega_{q^\prime}-\Delta}
    \right]
\nonumber\\
&&+\frac{g_\pi^2}{f_\pi^2}
\left[
    (\sigma^i \sigma^j)_{ab}
    \frac{ (\bm{\varepsilon} \times \bm{q})\cdot ( \bm{q}^\prime \times  \bm{\varepsilon}^{\prime\ast})}{\omega_q}
    - (\sigma^j \sigma^i)_{ab}
    \frac{ (\bm{\varepsilon} \times \bm{q}^\prime)\cdot ( \bm{q} \times  \bm{\varepsilon}^{\prime\ast})}{\omega_{q^\prime}}
    \right].
\eea

We can ignore the recoil of $D^*$ and set $|\bm{q}^\prime| = |\bm{q}|$.
The non-singular contribution to the reaction rate can be obtained by taking the  limit $\Delta \to m_\pi$.
The reaction rate averaged over incoming pion flavors and incoming $D^*$ spins is
\bqa
v\sigma[\pi D^{*a} \to \pi D^{*b}]_\mathrm{nonsing} & = &
\frac{1}{36 \pi f_\pi^4}
\Big\{ (2 - \delta_{ab}) \big[ 6\omega_q^2 + 2 g_\pi^4 (  m_\pi^2+q^4/\omega_q^2) \big]
\nonumber \\
&& \hspace{1.5cm} 
+\,  \delta_{ab} \, g_\pi^4 (3 \omega_q^2 + q^4/\omega_q^2 )\Big\}
\frac{q}{\omega_q}.
\label{vsigmapiD*}
\eqa
The  reaction rate in the pion gas is obtained by averaging over the momentum distributions of the incoming $D$ and $\pi$:
\bqa
\big\langle v\sigma[\pi D^{*a} \to \pi D^{*b}]_\mathrm{nonsing} \big\rangle &= &
\frac{1}{36 \pi f_\pi^4}
\bigg(
\big[ 6(2 - \delta_{ab})  +  2(2+\delta_{ab}) g_\pi^4 \big] \langle \omega_q q \rangle_{\bm{q}}
\nonumber \\
&& \hspace{2cm}
-4 g_\pi^4 m_\pi^2 \left\langle\frac{q}{\omega_q}\right\rangle_{\!\!\!\bm q}
+(4-\delta_{ab})g_\pi^4 m_\pi^4
\left\langle\frac{q}{\omega_q^3}\right\rangle_{\!\!\!\bm q} \bigg).
\label{vsigmaD*pi:Delta=mpi}
\eqa

The third diagram in Fig.~\ref{fig:piD*scattering} produces a $t$-channel singularity,
because the intermediate $D$ can be on shell.
In the square of the matrix element for an incoming $D^\ast$ with momentum $(M + \Delta) v+p$
and an outgoing pion with momentum $q^\prime$,
the $t$-channel singularity can be isolated by making a simple substitution
for the product of the $D$ propagator and its complex conjugate:
\beq
\frac{1}{v \!\cdot\! (\Delta v+p-q^\prime) + i \Gamma/2}
\left( \frac{1}{v \!\cdot\!  (\Delta v+p-q^\prime)  + i \Gamma/2} \right)^{\!\!\ast}
\longrightarrow
\frac{2 \pi}{\Gamma} \, \delta \big( \Delta + v \!\cdot \! p - v \!\cdot\!q^\prime \big).
\label{1/GammaD}
\eeq
The $t$-channel singularity contribution to the reaction rate for $\pi(q) D^{*a}(p) \to \pi(q^\prime) D^{*b}(p^\prime)$
averaged over incoming  $D^\ast$ spins and  over incoming pion flavors  is
\bea
v\sigma\big[ \pi D^{*a} \to \pi D^{*b} \big]_\mathrm{sing} &=& \frac{g_\pi^4}{72\,  f_\pi^4}
\sum_c(2 - \delta_{ac}) (2 - \delta_{bc })  \frac{q_{bc}^2 \, q_{ac}^3}{\Delta_{bc} \Gamma_c} \,
\delta(\omega_{bcq} - \Delta_{bc}) \theta( \Delta_{ac} - m_{\pi ac}).~~~
\eea
Using the expressions for $\Gamma [D^\ast  \to D \pi]$ in Eq.~\eqref{GammaD*Dpi}
and $v\sigma[D \pi \to D^\ast]$ in Eq.~\eqref{vsigmaDpitoD*},
the singular term in the reaction rate can  be expressed as
\beq
v\sigma \big[ \pi D^{\ast a}  \to \pi D^{\ast b} \big]_\mathrm{sing} =
\sum_c \frac{1}{\Gamma_c} \, \Gamma \big[ D^{\ast a} \to D^c \pi \big] \, v\sigma \big[D^c \pi \to D^{\ast b} \big] .
\eeq
The reaction rate in the pion gas can be evaluated
using the thermal average of  $v\sigma[D \pi \to D^\ast]$ in Eq.~\eqref{<vsigmaDpitoD*>}:
\bea
\left\langle v\sigma \big[  \pi D^{\ast a} \to \pi D^{\ast b} \big]_\mathrm{sing} \right\rangle=
\frac{\mathfrak{f}_\pi(\Delta)}{\mathfrak{n}_\pi}
\sum_c \frac{\Gamma \big[D^{\ast a} \to D^c \pi \big] \,\Gamma \big[D^{\ast b} \to D^c \pi \big]}{\Gamma_c}.
\label{<vsigmaD*pitoDpi>}
\eea
This differs from the resonant term in the reaction rate for $\pi D \to \pi D$ in Eq.~\eqref{<vsigmaDpitoDpi>}
in that the sum is over $D$  flavors instead of $D^\ast$ flavors and that
the product of $D^\ast$ partial decay rates is divided by a $D$ decay rate $\Gamma_c$ instead of
a $D^*$ decay rate $\Gamma_{*c}$.

The reaction rates for  $\pi D^{*a}  \to \pi D^{*b}$ in the hadron gas near or after kinetic freezeout
can be approximated by the sum of the nonsingular reaction rates in Eq.~\eqref{vsigmaD*pi:Delta=mpi}
and the $t$-channel singularity reaction rate in Eq.~\eqref{<vsigmaD*pitoDpi>}:
\begin{subequations}
    \bea
    \left\langle v\sigma_{\pi D^{*0} \to \pi D^{*0}} \right\rangle
    &=& (0.496 +0.469\, g_\pi^4)\, \frac{ m_\pi^2}{ f_\pi^4}
    +  \frac{\mathfrak{f}_\pi(m_\pi)}{\mathfrak{n}_\pi} \, \frac{\Gamma_{D^{\ast 0} \to D^0 \pi}^2}{\Gamma_0} ,
    \label{vsigmaD*0toD*0}
    \\
    \left\langle v\sigma_{\pi D^{*0} \to \pi D^{*+}} \right\rangle
    &=& (0.991 + 0.306\, g_\pi^4) \, \frac{m_\pi^2}{ f_\pi^4}
    +  \frac{\mathfrak{f}_\pi(m_\pi)}{\mathfrak{n}_\pi} \, \frac{\Gamma_{D^{\ast 0} \to D^0 \pi}\, \Gamma_{D^{\ast +} \to D^0 \pi}}{\Gamma_0} ,
    \label{vsigmaD*0toD*+}
    \\
    \left\langle v\sigma_{\pi D^{*+} \to  \pi D^{*0}} \right\rangle
    &=& (0.991 + 0.306\, g_\pi^4) \, \frac{m_\pi^2}{ f_\pi^4}
    +  \frac{\mathfrak{f}_\pi(m_\pi)}{\mathfrak{n}_\pi} \, \frac{\Gamma_{D^{\ast 0} \to D^0 \pi}\, \Gamma_{D^{\ast +} \to D^0 \pi}}{\Gamma_0} ,
    \label{vsigmaD*+toD*0}
    \\
    \left\langle v\sigma_{\pi D^{*+} \to  \pi D^{*+}} \right\rangle
    &=&(0.496 +0.469\, g_\pi^4)\, \frac{m_\pi^2}{ f_\pi^4}
    +  \frac{\mathfrak{f}_\pi(m_\pi)}{\mathfrak{n}_\pi} \,
    \left(\frac{\Gamma_{D^{\ast +} \to D^0 \pi}^2}{\Gamma_0} + \frac{\Gamma_{D^{\ast +} \to D^+ \pi}^2}{\Gamma_+}\right) .
    \label{vsigmaD*+toD*+}
    \eea
    \label{vsigmaD*toD*}%
\end{subequations}
The dimensionless numbers in the first terms depend only on $m_\pi/T$, which we have evaluated at $T_\mathrm{kf}=115$~MeV.
Before kinetic freezeout, the factor $\mathfrak{f}_\pi(m_\pi)/\mathfrak{n}_\pi$ is determined by  $T$.
After kinetic freezeout at $T_\mathrm{kf}=115$~MeV,
that factor has the constant value $0.431/\mathfrak{n}_\pi^\mathrm{(kf)}$ independent of $\mathfrak{n}_\pi$.
There can also be dependence on $T$ or $\mathfrak{n}_\pi$ through
the mass shifts in $\Gamma_{D^{\ast a} \to D^b \pi}$ and through the factors of $1/\Gamma_c$.

After kinetic freezeout, the most dramatic dependences on $\mathfrak{n}_\pi$ can be made explicit 
by inserting the expressions for the thermal widths of $D^+$ and $D^0$ in Eqs.~\eqref{GammaD}.
The resulting expression for the reaction rate for $\pi D^{*0} \to \pi D^{*+}$ 
(or $\pi D^{*+} \to \pi D^{*0}$) is 
\beq
\left\langle v\sigma_{\pi D^{*0} \to \pi D^{*+}} \right\rangle
= (0.991 + 0.306\, g_\pi^4) \, \frac{m_\pi^2}{ f_\pi^4}
    +  \frac{1}{3 \mathfrak{n}_\pi} \, \frac{\Gamma_{D^{\ast 0} \to D^0 \pi}\, \Gamma_{D^{\ast +} \to D^0 \pi}}{\Gamma_{D^{\ast 0} \to D^0 \pi} + \Gamma_{D^{\ast +} \to D^0 \pi}} .
\label{vsigmaD*0toD*+-npi}
\eeq
At kinetic freezeout, the $t$-channel singularity term is smaller than the nonsingular term 
by the factor 0.0003. 
However the multiplicative factor of $1/\mathfrak{n}_\pi$ 
makes the $t$-channel singularity term increase dramatically as the hadron gas expands.
It becomes equal to the nonsingular term 
when $\mathfrak{n}_\pi$ decreases by the factor 0.0009. 
Since the volume $V(\tau)$ of the hadron gas increases roughly as $\tau^3$,
this corresponds to an increase in its linear dimensions by about a factor of 10.

\begin{figure}[t]
    \includegraphics[width=0.8\textwidth]{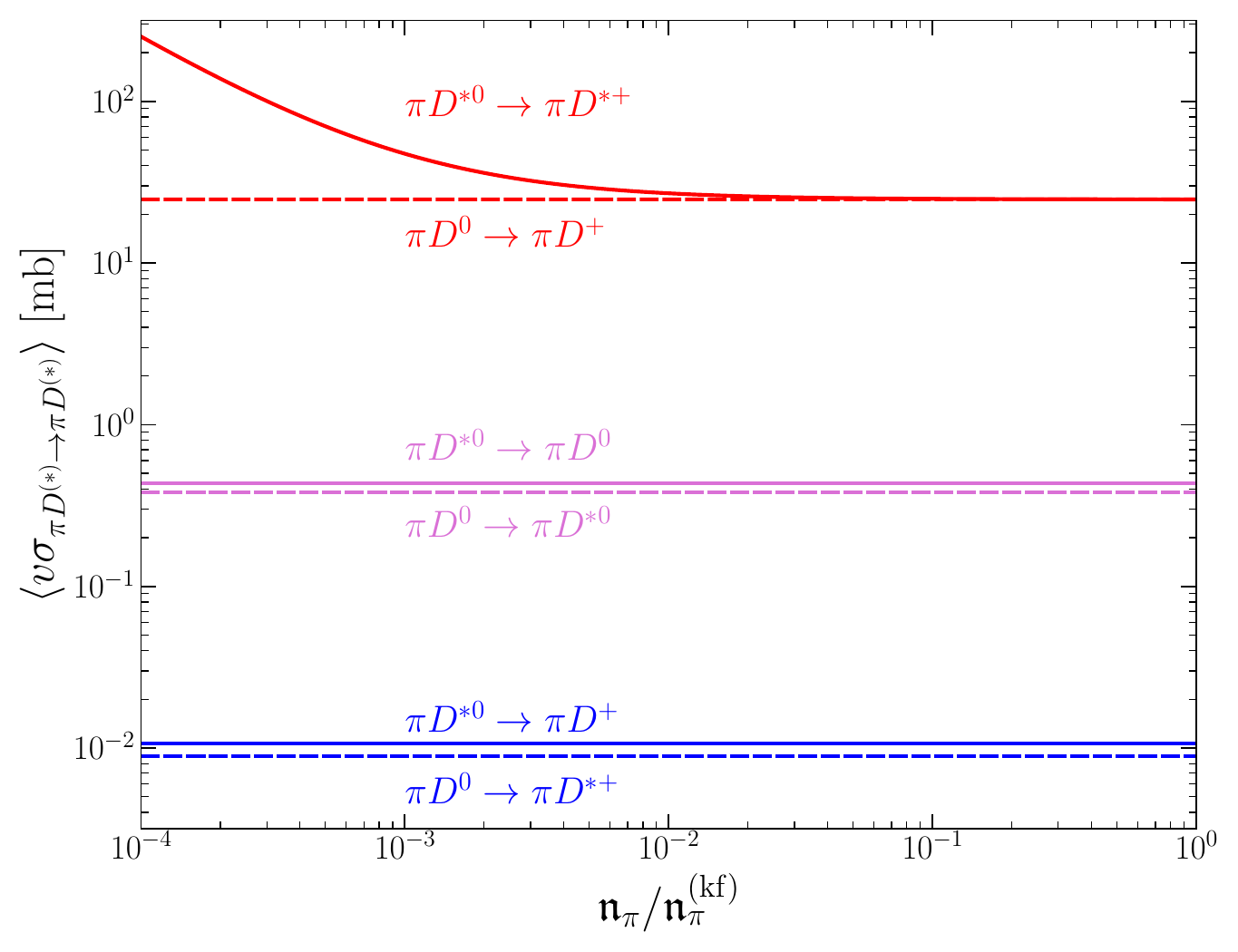}
    \caption{
    Reaction rates $\langle v\sigma_{\pi D^{(*)} \to  \pi D^{(*)}} \rangle$
    for the scattering of an incoming neutral charm meson $D^0$ or $D^{\ast 0}$
    and a  pion in the hadron gas after kinetic freezeout as functions of the pion number density $\mathfrak{n}_\pi$:
    $\pi D^0 \to  \pi D^+$, $\pi D^0 \to  \pi D^{\ast 0}$, $\pi D^0 \to  \pi D^{\ast +}$
    (dashed curves:  higher red, intermediate purple, and lower blue),
    $\pi D^{\ast 0} \to  \pi D^{\ast +}$, $\pi D^{\ast 0} \to  \pi D^0$, $\pi D^{\ast 0} \to  \pi D^+$
    (solid curves: higher red, intermediate purple, and lower blue).
    The increase in the reaction rate for $\pi D^{\ast 0} \to  \pi D^{\ast +}$ as   $\mathfrak{n}_\pi \to 0$
    comes from a $D$-meson $t$-channel singularity.
    }
    \label{fig:piD(*)topiD(*)}
\end{figure}

Many of the $\pi D^{(*)} \to \pi D^{(*)}$ scattering reactions change the flavor or spin of the charm meson.
The reaction rates in the hadron gas after kinetic freezeout at $T_\mathrm{kf}=115$~MeV
for incoming neutral charm mesons $D^0$ or $D^{\ast 0}$
are shown as functions of the pion number density $\mathfrak{n}_\pi$ in Fig.~\ref{fig:piD(*)topiD(*)}.
For each of these reactions, there is another one with an incoming charged charm meson $D^+$ or $D^{\ast +}$
that has the same reaction rate.
The reaction rate for $\pi D^0 \to  \pi D^+$ is given in Eq.~\eqref{vsigmaD0toD+}.
The reaction rates for $\pi D^0 \to  \pi D^{\ast 0}$ and  $\pi D^0 \to  \pi D^{\ast +}$
are given in  Eqs.~\eqref{vsigmaD0toD*0} and \eqref{vsigmaD0toD*+}.
The reaction rates for  $\pi D^{\ast 0} \to  \pi D^0$ and $\pi D^{\ast 0} \to  \pi D^+$
are given in  Eqs.~\eqref{vsigmaD*0toD0} and \eqref{vsigmaD*0toD+}.
The largest reaction rates in Fig.~\ref{fig:piD(*)topiD(*)} are for reactions that change the charm-meson flavor only.
The reaction rate for $\pi D^0 \to  \pi D^+$ has a $D^{*+}$ resonance contribution
that makes it increase as $\mathfrak{n}_\pi$ decreases.
However the $D^{*+}$ resonance contribution is about 3 orders of magnitude smaller than the nonresonant term,
so the decrease is not visible in  Fig.~\ref{fig:piD(*)topiD(*)}.
The reaction rate for $\pi D^{\ast 0} \to  \pi D^{\ast +}$ has a $D^0$ $t$-channel singularity contribution
that makes it diverge as  $\mathfrak{n}_\pi \to 0$.
The other reaction rates in  Fig.~\ref{fig:piD(*)topiD(*)} are constant functions of $\mathfrak{n}_\pi$.
The  rates in  Fig.~\ref{fig:piD(*)topiD(*)} for reactions  that change the charm-meson spin
are suppressed by more than 1.5 orders of magnitude.
The  rates for reactions  that change both the flavor and spin of the charm meson
are suppressed by more than 3 orders of magnitude.


\section{Evolution of charm-meson ratios}
\label{sec:Evolution}

In this section, we calculate the evolution of the charm-meson number densities
in the expanding hadron gas from a heavy-ion collision after kinetic freezeout.

\subsection{Rate equations}
\label{sec:evolutioneqs}

The evolution of the number density $\mathfrak{n}_{D^{(\ast)}}(\tau)$ of a charm meson in the expanding hadron gas
with the proper time $\tau$ can be described by a first order differential equation.
The number density decreases because of the increasing volume $V(\tau)$, but it can also be changed by reactions.
The time derivative of $\mathfrak{n}_{D^{(\ast)}}$
has positive contributions from reactions with $D^{(\ast)}$ in the final state
and negative contributions from reactions  with $D^{(\ast)}$ in the initial state.
Near kinetic freezeout, the most important reactions involve pions,
because pions are by far the most abundant hadrons in the  hadron gas.

After kinetic freezeout, most interactions have a negligible effect
on  the number density $\mathfrak{n}_{D^{(\ast)}}(\tau)$ of a charm meson.
The charm-meson number density decreases in proportion to $1/V(\tau)$,
like the pion number density $ \mathfrak{n}_\pi(\tau)$ in Eq.~\eqref{npi-tau}.
The effect of the increasing volume can be removed from the differential equation
by considering the rate equation for the ratio of number densities  $\mathfrak{n}_{D^{(\ast)}}/ \mathfrak{n}_\pi$.
The remaining terms in the rate equation come from reactions that change the spin or flavor of the charm meson.
The reaction rate is multiplied by a factor of the number density for every particle in the initial state.
The  reaction rate is determined by the temperature, which is fixed at the kinetic freezeout temperature
$T_\mathrm{kf}$, the charm-meson number density $\mathfrak n_{D^{(\ast)}}$, and the pion number density
$\mathfrak n_\pi(\tau)$, which decreases as $1/V(\tau)$.
Some reaction rates in the expanding hadron gas are sensitive to the
thermal mass shifts  and  the thermal widths of the hadrons.
The greatest sensitivities to the thermal mass shifts for charm mesons and pions
are in reactions whose rates are proportional to a power of
$M_{D^\ast} - M_D - m_\pi$, such as $D^* \to D \pi$ decay rates.
The greatest sensitivity to the thermal width  $\Gamma_a$ of a  pseudoscalar charm meson $D^a$
comes from $D$-meson $t$-channel singularities, which can produce contributions to reaction rates proportional
to $1/\Gamma_a$ in the limit $\Gamma_a \to 0$.

The most relevant reactions  for charm mesons in the expanding hadron gas include
the decays $D^\ast \to D \pi$ and $D^\ast \to D \gamma$,
the scattering reactions $\pi D \to \pi D$, $\pi D \to D\gamma$, and $\pi D^\ast  \to  \pi D^\ast$
that change the charm-meson flavor,
and the scattering reactions $\pi D \to \pi D^\ast$ and $\pi D^\ast \to \pi D$ that change the charm-meson spin.
The  rate equations for the number densities of the pseudoscalar charm mesons $D^a$
and the vector charm mesons $D^{\ast a}$ are
\begin{subequations}
    \begin{eqnarray}
        \mathfrak{n}_\pi \frac{d \ }{d \tau} \left( \frac{\mathfrak{n}_{D^a}}{\mathfrak{n}_\pi} \right)  &=&
        [1 + \mathfrak{f}_\pi(m_\pi)] \sum_b \Gamma_{D^{\ast  b} \to D^a \pi} \,\mathfrak{n}_{D^{\ast  b}}
        + \Gamma_{\ast  a,\gamma} \,\mathfrak{n}_{D^{\ast  a}}
        \nonumber\\
        && \hspace{-1cm}
        + 3\sum_{b \neq a}
        \big[ \left\langle  v\sigma_{\pi D^b \to \pi D^a} \right\rangle \, \big( \mathfrak{n}_{D^b} -  \mathfrak{n}_{D^a} \big)
            + \left\langle  v\sigma_{\pi D^b \to D^a \gamma} \right\rangle \,  \mathfrak{n}_{D^b}
            - \left\langle  v\sigma_{\pi D^a \to D^b \gamma} \right\rangle \,  \mathfrak{n}_{D^a} \big]\,
        \mathfrak{n}_{\pi}
        \nonumber\\
        && \hspace{-1cm}
        + 3\sum_b   \big[ \left \langle  v\sigma_{\pi D^{\ast b} \to \pi D^a} \right\rangle \,\mathfrak{n}_{D^{\ast  b}}
        - \left \langle  v\sigma_{\pi D^a \to \pi  D^{\ast b}} \right\rangle \mathfrak{n}_{D^a} \big] \, \mathfrak{n}_{\pi}
        +\ldots,
        \label{dnDa/dt}
        \\
        \mathfrak{n}_\pi \frac{d \ }{d \tau} \left( \frac{\mathfrak{n}_{D^{\ast a}}}{\mathfrak{n}_\pi} \right) &=&
        -  \Big( [1 + \mathfrak{f}_\pi(m_\pi)] \sum_b \Gamma_{D^{\ast  a} \to D^b \pi}
        +  \Gamma_{\ast  a,\gamma} \Big) \mathfrak{n}_{D^{\ast  a}}
        \nonumber\\
        &&  \hspace{-1cm}
        + 3\sum_{b \neq a} \left\langle  v\sigma_{\pi D^{*b} \to \pi D^{\ast a}} \right\rangle \, \big( \mathfrak{n}_{D^{*b}}
        -   \mathfrak{n}_{D^{\ast a}} \big) \mathfrak{n}_\pi
        \nonumber\\
        && \hspace{-1cm}
        +3 \sum_b   \big[ \left \langle  v\sigma_{\pi D^b \to \pi D^{\ast a}} \right\rangle \,\mathfrak{n}_{D^b}
        - \left \langle  v\sigma_{\pi D^{\ast a} \to \pi  D^b} \right\rangle \mathfrak{n}_{D^{*a}}  \big] \mathfrak{n}_\pi
        +\ldots.
        \label{dnD*a/dt}
    \end{eqnarray}
    \label{dnD(*)/dt}%
\end{subequations}
The partial decay rates $ \Gamma_{D^{\ast  a} \to D^b \pi}$ and $ \Gamma_{\ast  a,\gamma} $
are given in Eqs.~\eqref{GammaD*aDbpi} and \eqref{GammaD*Dgamma}.
The reaction rates $\langle  v\sigma_{\pi D^a \to \pi D^b} \rangle$ and $\langle  v\sigma_{\pi D^a \to D^b\gamma} \rangle$,
which have $D^*$ resonance contributions, are given in Eqs.~\eqref{vsigmaDtoD} and \eqref{vsigmapiDtoDgamma}.
The reaction rates $\langle  v\sigma_{\pi D^a \to \pi D^{\ast b}} \rangle$ and  $\langle  v\sigma_{\pi D^{\ast a} \to \pi  D^b} \rangle$
are given in Eqs.~\eqref{vsigmaDtoD*} and \eqref{vsigmaD*toD}.
The reaction rates $\langle  v\sigma_{\pi D^{*a} \to \pi D^{\ast b}} \rangle$, which have $D$-meson $t$-channel singularities,
are given in Eqs.~\eqref{vsigmaD*toD*}.
The  rate equations in Eqs.~\eqref{dnD(*)/dt} are consistent with the conservation of charm-quark number,
which implies that the sum of the ratios of the number densities for all four charm mesons remains constant:
\beq
\mathfrak{n}_\pi \frac{d \ }{d \tau}
\left(  \frac{\mathfrak{n}_{D^0} +  \mathfrak{n}_{D^+} +  \mathfrak{n}_{D^{*0}} + \mathfrak{n}_{D^{*+}}}{\mathfrak{n}_\pi} \right) = 0.
\label{dndt}
\eeq

Given initial conditions on  the ratios $\mathfrak n_{D^{(\ast)}}(\tau)/\mathfrak n_\pi (\tau)$
of charm-meson and pion number densities,
the rate equations in Eqs.~\eqref{dnD(*)/dt} can be integrated to determine the ratios at larger $\tau$.
As our initial conditions on the ratio at kinetic freezeout,
we take the ratio of the multiplicity of the charm meson before $D^*$ decays and the pion multiplicity:
\begin{subequations}
    \bea
    \frac{\mathfrak{n}_{D^a}(\tau_\mathrm{kf})}{\mathfrak{n}_{\pi}(\tau_\mathrm{kf})} &=&
    \frac{(dN_{D^a}/dy)_0}{dN_\pi/dy} ,
    \label{nD-K}
    \\
    \frac{\mathfrak{n}_{D^{\ast a}}(\tau_\mathrm{kf})}{\mathfrak{n}_{\pi}(\tau_\mathrm{kf})} &=&
    \frac{(dN_{D^{\ast a}}/dy)_0}{dN_\pi/dy} .
    \label{nD*-K}
    \eea
    \label{nD*,D-K}%
\end{subequations}
In the case of Pb-Pb collisions in the centrality range 0-10\% at $\sqrt{s_{NN}}=5.02$\,TeV,
the multiplicity $dN_\pi/dy$ for a single pion flavor measured by  the ALICE collaboration
is given in Eq.~\eqref{dNpi/dy} \cite{ALICE:2019hno}.
The multiplicities $(dN_{D^{\ast a}}/dy)_0$  and $(dN_{D^a}/dy)_0$ for charm mesons before $D^*$ decays
inferred from SHM predictions are given in Eqs.~\eqref{dND*/dy}  and \eqref{dND/dy}.
The resulting initial values of the ratios of charm-meson and pion number densities at kinetic freezeout
for $D^0$, $D^+$, $D^{\ast 0}$, and $D^{\ast +}$ are
0.00278, 0.00264, 0.00334, and 0.00328,  respectively.

\begin{figure}[t]
    \includegraphics[width=0.8\textwidth]{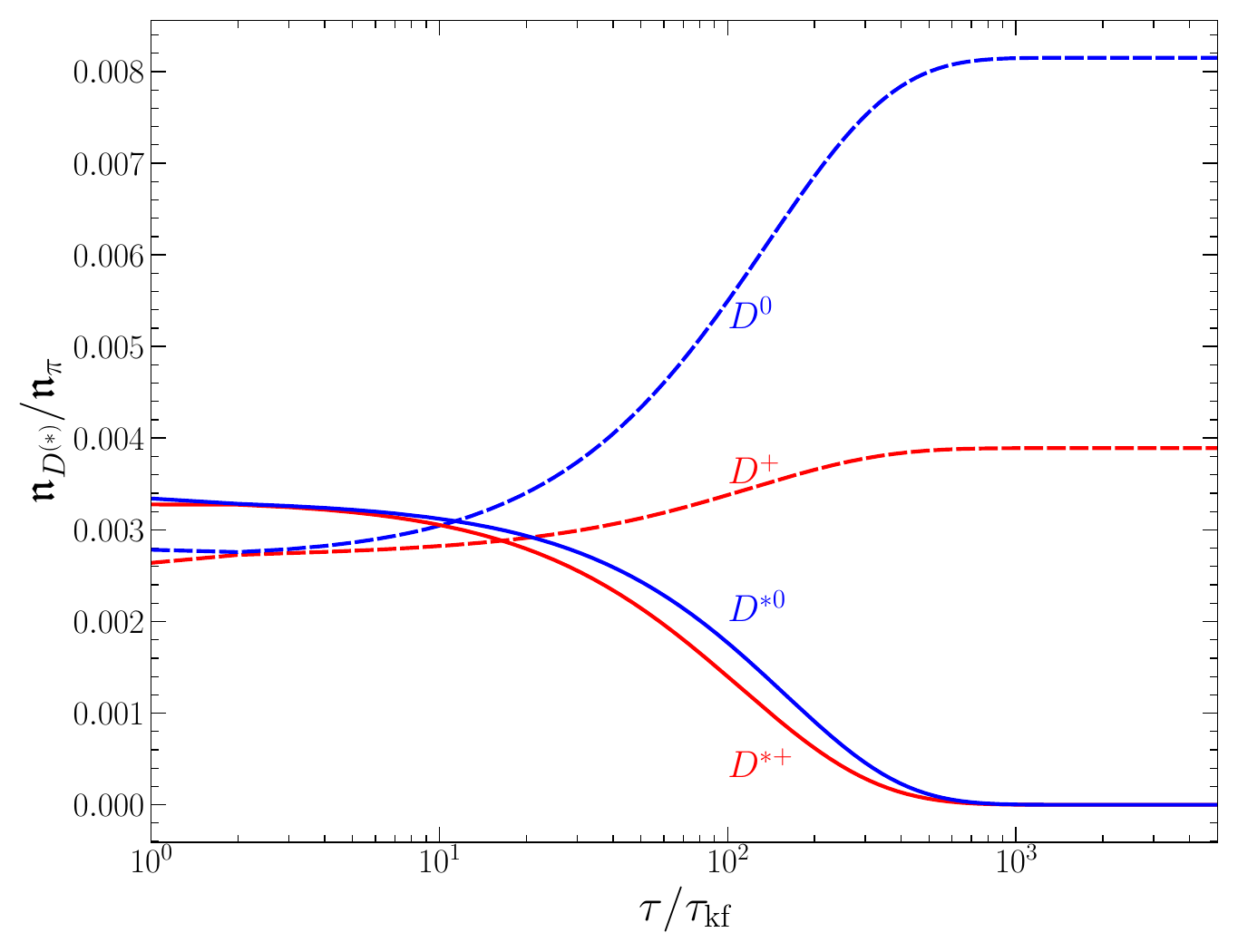}
    \caption{
        Proper-time evolution of the  ratios of number densities of charm meson and pions
        from solving the  rate equations in Eqs.~\eqref{dnD(*)/dt}:
        $\mathfrak n_{D^0}/\mathfrak n_\pi$, $ \mathfrak n_{D^+}/\mathfrak n_\pi$ (dashed curves: higher blue, lower red),
        $\mathfrak n_{D^{\ast0}}/\mathfrak n_\pi$, $\mathfrak n_{D^{\ast+}}/\mathfrak n_\pi$ (solid curves: higher blue, lower red).
    }
    \label{fig:density-diff-evol-1}
\end{figure}

The solutions to the rate equations in Eqs.~\eqref{dnD(*)/dt}
with the initial conditions in Eqs.~\eqref{nD*,D-K} are shown in Fig.~\ref{fig:density-diff-evol-1}.
The ratios $\mathfrak{n}_{D^{*0}}/\mathfrak{n}_\pi$ and $\mathfrak{n}_{D^{*+}}/\mathfrak{n}_\pi$
decrease exponentially to 0 on time scales comparable to the $D^\ast$ lifetimes.
The ratio $N_{D^0}/N_{D^+}$ of the numbers of $D^0$ and $D^+$ is predicted to
increase from 1.053 at kinetic freezeout to about 2.092 at the detector.
The naive prediction for the effects of $D^*$ decays on the numbers $N_{D^0}$ and $N_{D^+}$ at the detector
can be obtained by inserting the initial conditions at kinetic freezeout into Eqs.~\eqref{ND0,Dp}.
The naive prediction for the  ratio $N_{D^0}/N_{D^+}$ at the detector is  2.255.  
This is about 10\% larger than the ratio from solving the rate equations.
Thus  the rate equations in Eqs.~\eqref{dnD(*)/dt} must include reactions other than $D^*$ decays
whose effects are not negligible after kinetic freezeout.

\subsection{Asymptotic Evolution}
\label{sec:asymptotic}

As $\tau$ increases, the pion number density $\mathfrak{n}_\pi(\tau)$ decreases to 0 as $1/V(\tau)$.
As $\mathfrak{n}_\pi$ approaches 0, most of the reaction rates in Eqs.~\eqref{dnD(*)/dt}
approach the finite constant reaction rates in the vacuum.
The exceptions are
$\langle  v\sigma_{\pi D^{*0} \to \pi D^{\ast +}} \rangle$ and $\langle  v\sigma_{\pi D^{*+} \to \pi D^{\ast 0}} \rangle$,
which are given in Eqs.~\eqref{vsigmaD*0toD*+} and \eqref{vsigmaD*+toD*0}.
They have contributions with a factor of $1/\Gamma_0$ from a $D$-meson $t$-channel singularity.
Since $\Gamma_0$, which is given in Eq.~\eqref{GammaD0},
decreases to 0 in proportion to $\mathfrak{n}_\pi$ as $\mathfrak{n}_\pi \to 0$,
these reaction rates increase as $1/\mathfrak{n}_\pi$.
The limiting behaviors of the reaction rates  $\langle  v\sigma_{\pi D^{*b} \to \pi D^{\ast a}} \rangle$ as $\mathfrak{n}_\pi \to 0$
in  Eqs.~\eqref{vsigmaD*toD*} are
\begin{subequations}
    \bea
    \left\langle v\sigma_{\pi D^{*0} \to \pi D^{*0}} \right\rangle &\longrightarrow&
    \frac{1}{3\, \mathfrak{n}_\pi} \, \frac{(B_{00}\, \Gamma_{\ast 0})^2}{B_{00}\, \Gamma_{\ast 0} + B_{+0}\,\Gamma_{\ast +}} ,
    \label{vsigmapiD*0D*0:asymp}
    \\
    \left\langle v\sigma_{\pi D^{*0} \to \pi D^{*+}} \right\rangle &\longrightarrow&
    \frac{1}{3\, \mathfrak{n}_\pi} \,
    \frac{(B_{00}\, \Gamma_{\ast 0})\,(B_{+0}\,\Gamma_{\ast +})}
    {B_{00}\, \Gamma_{\ast 0} + B_{+0}\,\Gamma_{\ast +}} ,
    \label{vsigmapiD*0D*+:asymp}
    \\
    \left\langle v\sigma_{\pi D^{*+} \to  \pi D^{*0}} \right\rangle &\longrightarrow&
    \frac{1}{3\, \mathfrak{n}_\pi} \,
    \frac{(B_{00}\, \Gamma_{\ast 0})\, (B_{+0}\,\Gamma_{\ast +})}{B_{00}\, \Gamma_{\ast 0} + B_{+0}\,\Gamma_{\ast +}} ,
    \label{vsigmapiD*+D*0:asymp}
    \\
    \left\langle v\sigma_{\pi D^{*+} \to  \pi D^{*+}} \right\rangle &\longrightarrow&
    \frac{1}{3\, \mathfrak{n}_\pi} 
    \left(\frac{(B_{+0}\, \Gamma_{\ast +})^2}{B_{00}\, \Gamma_{\ast 0} + B_{+0}\,\Gamma_{\ast +}}
    + B_{++}\,\Gamma_{\ast +} \right) ,
    \label{vsigmapiD*+D*+:asymp}
    \eea
\end{subequations}
where $\Gamma_{\ast a}$ is the decay rate for $D^{*a}$ in the vacuum given in Eqs.~\eqref{Gamma*}
and $B_{ab}$ is the branching fraction for $D^{*a} \to D^b \pi$ given in Eqs.~\eqref{Br*ab}.
The factors of $1/\mathfrak{n}_\pi$ in these asymptotic reaction rates can
cancel explicit factors of $\mathfrak{n}_\pi$ in the rate equations.

At large times $\tau$, the only terms in the rate equation that survive are
1-body terms with a single factor of a number density $\mathfrak{n}_D$ or $\mathfrak{n}_{D^\ast}$.
There are 1-body terms from the decays $D^\ast \to D \pi$ and $D^\ast \to D \gamma$.
The $t$-channel singularities produce additional 1-body   terms.
If the 1-body terms from $D$-meson $t$-channel singularities are taken into account,
the asymptotic rate equations become
\begin{subequations}
    \begin{eqnarray}
        \mathfrak{n}_\pi \frac{d \ }{d \tau} \left( \frac{\mathfrak{n}_{D^+}}{\mathfrak{n}_\pi} \right)  &\longrightarrow&
        (1- B_{+0}) \, \Gamma_{*+}\, \mathfrak{n}_{D^{\ast  +}} ,
        \\
        \mathfrak{n}_\pi \frac{d \ }{d \tau} \left( \frac{\mathfrak{n}_{D^0}}{\mathfrak{n}_\pi} \right)   &\longrightarrow&
        \Gamma_{*0} \,\mathfrak{n}_{D^{\ast  0}}
        + B_{+0} \,\Gamma_{*+}\, \mathfrak{n}_{D^{\ast  +}} ,
        \\
        \mathfrak{n}_\pi \frac{d \ }{d\tau} \left( \frac{\mathfrak{n}_{D^{\ast +}}}{\mathfrak{n}_\pi} \right)  &\longrightarrow&
        -  \big( \Gamma_{*+} + \gamma \big)\,\mathfrak{n}_{D^{\ast  +}} +  \gamma \,\mathfrak{n}_{D^{*0}} ,
        \\
        \mathfrak{n}_\pi \frac{d \ }{d \tau} \left( \frac{\mathfrak{n}_{D^{\ast 0}}}{\mathfrak{n}_\pi} \right)  &\longrightarrow&
        -  \big( \Gamma_{*0} + \gamma \big) \,\mathfrak{n}_{D^{\ast  0}} + \gamma  \,\mathfrak{n}_{D^{*+}} ,
    \end{eqnarray}
    \label{dn/dt-asymptotic}%
\end{subequations}
where the rate $\gamma$ is
\beq
\gamma =
\frac{1}{1/( B_{00}\, \Gamma_{*0} ) +  1/(B_{+0}\, \Gamma_{*+})}
= 21.9~\mathrm{keV}.
\label{gamma}
\eeq
The terms in Eqs.~\eqref{dn/dt-asymptotic} with the factor $\gamma$ come from $D$-meson $t$-channel singularities.

The asymptotic rate equations in Eqs.~\eqref{dn/dt-asymptotic} can be solved analytically.
If the numbers of $D^0$, $D^+$, $D^{\ast 0}$, and $D^{\ast +}$
at kinetic freezeout are $(N_{D^0})_0$, $(N_{D^+})_0$, $(N_{D^{\ast 0}})_0$, and $(N_{D^{\ast +}})_0$,
the predicted asymptotic numbers of $D^0$ and $D^+$ are
\begin{subequations}
    \bqa
    N_0 &=&
    \big(N_0 \big)_0 +\big( N_{\ast 0} \big)_0 + B_{+0}  \big( N_{\ast +} \big)_0
    - \frac{(1-B_{+0})\,  \gamma }{ \Gamma_{*+}\Gamma_{*0}+  (\Gamma_{*+} \!+\!\Gamma_{*0})\, \gamma}
    \left[  \Gamma_{*+}\, \big( N_{\ast 0} \big)_0 \!-\! \Gamma_{*0} \,\big( N_{\ast +} \big)_0 \right],
    \nonumber\\
    \label{ND0-asymp}
    \\
    N_+ &=& \big( N_+ \big)_0 + (1 \!-\!B_{+0})\, \big( N_{\ast +} \big)_0
    + \frac{(1-B_{+0})\,  \gamma }{\Gamma_{*+}\Gamma_{*0}+  (\Gamma_{*+} \!+\! \Gamma_{*0})\, \gamma}
    \left[  \Gamma_{*+}\, \big( N_{\ast 0} \big)_0 \!-\! \Gamma_{*0} \,\big( N_{\ast +} \big)_0 \right],~~~~~~
    \label{ND+-asymp}
    \eqa
    \label{ND-asymp}%
\end{subequations}
where $\gamma$ is given in Eq.~\eqref{gamma}.
The coefficients of $(N_{\ast 0})_0$ and $(N_{\ast 0})_+$ in Eqs.~\eqref{ND-asymp}
depend only on $B_{+0}$, $B_{00}$, and the ratio $\Gamma_{\ast 0}/\Gamma_{\ast +}$.
The prediction for the difference between the numbers of $D^0$ and $D^+$ are
\bqa
N_0 - N_+ &=&
2\, B_{+0} \,  \big( N_{\ast +} \big)_0 + \big(N_0 - N_+\big)_0 +  \big( N_{\ast 0} - N_{\ast +} \big)_0
\nonumber\\
&&  - \frac{2 (1-B_{+0})\,  \gamma }{ \Gamma_{*+}\Gamma_{*0}+  (\Gamma_{*+} \!+\!\Gamma_{*0})\, \gamma}
\left[  \Gamma_{*+}\, \big( N_{\ast 0} \big)_0 - \Gamma_{*0} \,\big( N_{\ast +} \big)_0 \right].
\label{ND+-ND-:t}
\eqa
If  we impose the isospin-symmetry approximations $(N_0)_0 \approx ( N_+)_0$
and $( N_{\ast 0})_0 \approx( N_{\ast +})_0$, the difference reduces to
\beq
N_0 - N_+ \approx 2 \left( B_{+0}
- \frac{(1-B_{+0})\,  ( \Gamma_{*+}  - \Gamma_{*0})\, \gamma }{ \Gamma_{*+}\Gamma_{*0}+  (\Gamma_{*+} \!+\!\Gamma_{*0})\, \gamma}\right) \big( N_{\ast +} \big)_0 .
\label{ND+-ND-:isospin}
\eeq
The two terms in the parantheses come from $D^\ast$ decays and the $D$-meson $t$-channel singularity,
respectively.
The effect of $D$-meson $t$-channel singularities is to reduce the coefficient of $( N_{\ast +})_0$
from $1.35 \pm 0.01$, which includes the effects of $D^*$ decays only,  to $1.30 \pm 0.01$.  
The change in the coefficient is small but statistically significant.

We use our initial conditions on the ratios of the charm-meson/pion number densities
at kinetic freezeout to illustrate the effect of $t$-channel singularities on
the ratio $N_0/N_+$ of the observed numbers of $D^0$ and $D^+$.
The ratio  before $D^*$ decays is
 $(N_{D^0})_0/(N_{D^+})_0 = 1.053$. 
 The predicted numbers of $D^0$ and $D^+$ at the detector are given in Eqs.~\eqref{ND-asymp}.
Their ratio is predicted to increase to  $2.178 \pm 0.016$  at the detector,
where the error bar is from $B_{+0}$, $B_{00}$, $\Gamma_{\ast 0}$, and $\Gamma_{\ast +}$ only.
The naive ratio $N_{D^0}/N_{D^+}$ ignoring $t$-channel singularities, which is obtained using Eqs.~\eqref{ND0,Dp}, is $2.255 \pm 0.014$.   
The difference between the predicted ratio taking into account $t$-channel singularities 
and the naive prediction is $-0.077 \pm 0.006$,  
which differs from 0 by about 13  standard deviations.


\section{Conclusion}
\label{sec:conclusion}

The reactions $\pi D^* \to \pi D^*$ have $t$-channel singularities in 6 of the 10 scattering channels.
These reactions can proceed at tree level through the decay $D^* \to \pi D$ 
followed by the inverse decay $\pi D \to D^*$.
The $t$-channel singularity appears because the intermediate $D$ can be on shell.
The tree-level cross section diverges inside a narrow interval of the center-of-mass energy 
near the threshold, which is given by Eq.~\eqref{t-sing:piD} 
if the incoming and outgoing $\pi$ and  $D$ have the same flavor.
If the singularity is regularized by inserting the width $\Gamma$ of the $D$ into its propagator,
the cross section has a term with a factor of the $D$ lifetime $1/\Gamma$.
The resulting enormous cross section is unphysical,
because the lifetimes of the incoming and outgoing  $D^*$ 
are many orders of magnitude smaller than the $D$ lifetime.
A more physical regularization of the $t$-channel singularity in the reaction rate for $\pi D^* \to \pi D^*$
could perhaps be obtained by a resummation of loop diagrams.
There are $n$-loop diagrams with $n+1$ $D$ propagators, $n$ $D^*$ propagators, and $n$ pion propagators
in which all $2n+1$ charm-meson propagators 
are nearly on shell.
A resummation of these diagrams to all orders could produce a regularization of the
$t$-channel singularity that is determined by the $D^*$ width.

As pointed out  by Grzadkowski, Iglicki, and Mr\'owczy\'nski in Ref.~\cite{Grzadkowski:2021kgi},
the thermal widths of particles in a medium provide a physical regularization of $t$-channel singularities. 
In a hadronic medium, the $t$-channel singularity in the reaction rate for $\pi D^* \to \pi D^*$
is regularized by the thermal width of $D$.
A physical example of such a hadronic medium 
is the hadron gas produced by a central relativistic heavy-ion collision.
The effects of the hadron gas are particularly simple near and after kinetic freezeout,
because it can be accurately approximated by a pion gas.
The thermal widths of the charm mesons come primarily from coherent  pion forward scattering.
At leading order in the pion interactions, the thermal widths $\Gamma_+$ of $D^+$ 
and $\Gamma_0$ of $D^0$ are given by Eqs.~\eqref{GammaD}.
Before kinetic freezeout, the $D$ widths are determined by the decreasing temperature $T$.
After kinetic freezeout, the $D$ widths are determined by the kinetic freezeout temperature 
and the decreasing pion number density  $\mathfrak{n}_\pi$.

In a hadronic medium, the $t$-channel singularities in the reaction rates for $\pi D^* \to \pi D^*$
produce terms in the average reaction rates  $\langle v\sigma_{\pi D^* \to \pi D^*} \rangle$
inversely proportional to the thermal widths $\Gamma_+$ and $\Gamma_0$.
These terms come from $\pi D^*$ scattering in regions near the threshold, 
and they are sensitive to differences $\Delta -m_\pi$  
between $D^*-D$ mass differences in the medium and pion masses in the medium.
There are also nonsingular contributions from $\pi D^*$ scattering 
that are determined primarily by the temperature $T$ 
and are insensitive to the values of $\Delta -m_\pi$.
We found a simple prescription for the nonsingular reaction rates
that allowed the total reaction rate to be approximated by the sum of the nonsingular term 
and the $t$-channel singularity terms.
Our prescription for the nonsingular reaction rate is simply the rate in the limit $\Delta \to m_\pi$.
The resulting expressions for the reaction rates $\langle v\sigma_{\pi D^{*a} \to \pi D^{*b}} \rangle$
in the pion gas after kinetic freezeout are given in Eqs.~\eqref{vsigmaD*toD*}.
After kinetic freezeout, the most dramatic dependence of reaction rates for $\pi D^* \to \pi D^*$ 
on $\mathfrak{n}_\pi$ comes from a multiplicative factor of $1/\mathfrak{n}_\pi$.
This dependence for $\langle v\sigma_{\pi D^{*0} \to \pi D^{*+}} \rangle$
is made explicit in Eq.~\eqref{vsigmaD*0toD*+-npi}.

In rate equations for the evolution of number densities,
a reaction rate is multiplied by the number density of each of the incoming particles.
In the hadron gas after kinetic freezeout,  all the number densities are decreasing
in inverse proportion to the expanding volume $V(\tau)$.
Thus $n$-body reactions, which have $n \ge 2$ incoming particles, 
are suppressed compared to decays, which are 1-body reactions,
by the additional factors of the number densities.
A 2-body reaction whose rate is proportional to an inverse power of a number density provides an exception,
because its effects in the rate equation can be comparable to 1-body terms.
In particular, the effects of the reactions $\pi D^* \to \pi D^*$ can be comparable to 1-body terms
because of the multiplicative factor of $1/\mathfrak{n}_\pi$ in the  $t$-channel singularity term. 
Rate equations for the number density ratios 
$\mathfrak{n}_{D^a}/\mathfrak{n}_\pi$ and $\mathfrak{n}_{D^{\ast a}}/\mathfrak{n}_\pi$
are given in Eqs.~\eqref{dnD(*)/dt}.
The numerical solutions of these rate equations after kinetic freezeout 
for specific initials conditions motivated by the Statistical Hadronization Model
are illustrated in Fig.~\ref{fig:density-diff-evol-1}.
There is a small but significant difference between the asymptotic charm-meson ratios
and the naive predictions from Eqs.~\eqref{ND0,Dp},
which take into account only the decays of $D^{*0}$ and $D^{*+}$.

The asymptotic forms of the rate equations  in the limit $\mathfrak{n}_\pi \to 0$ 
are given in Eqs.~\eqref{dn/dt-asymptotic}.
The only rates that remain in that limit  are 1-body terms, but those terms come not only from $D^*$ decays 
but also from the $t$-channel singularities in $\pi D^* \to \pi D^*$ reactions. 
The rates are completely determined by $D^*$ decay rates and $D^*$ branching fractions in the vacuum.
From the analytic solutions of the asymptotic rate equations,
we deduced the simple approximations in Eqs.~\eqref{ND-asymp}
for the asymptotic numbers of $D^0$ and $D^+$
given the numbers of the $D^0$, $D^+$, $D^{*0}$, and $D^{*+}$ at kinetic freezeout.
The new terms from $t$-channel singularities are those with a factor of the rate $\gamma$.
The predicted deviations of charm-meson ratios from the naive predictions from Eqs.~\eqref{ND0,Dp}
are small but much larger than the statistical errors from the $D^*$ decay rates and branching fractions.
The analytic predictions  from Eqs.~\eqref{ND-asymp} give good approximations to numerical solutions 
of the more accurate rate equations in Eqs.~\eqref{dnD(*)/dt}.

There are other charm-meson reactions with $t$-channel singularities
including $\pi D^\ast \to \pi\pi D$ and $\pi\pi D \to \pi D^\ast$.
The reaction $\pi D^\ast \to \pi\pi D$ has a pion $t$-channel singularity from the decay
$D^\ast \to D \pi$ followed by the scattering $\pi \pi \to \pi \pi$.
In a hadronic medium, the $t$-channel singularity is regularized by the thermal width 
$\Gamma_\pi$ of the pion. 
In a pion gas, the  leading term in $\Gamma_\pi$ is proportional to $\mathfrak{n}_{\pi}$.
Our preliminary result for the $t$-channel singularity contribution 
to the reaction rate for $\pi D^\ast \to \pi\pi D$ 
can be reduced to  $\Gamma_{D^* \to D\pi}/\mathfrak{n}_\pi$, 
which is comparable to the $t$-channel singularity term in 
the reaction rate for $\pi D^{*0} \to \pi D^{*+}$ in Eq.~\eqref{vsigmaD*0toD*+-npi}. 
This suggests that the contributions from pion $t$-channel singularities 
may be comparable to those from $D$-meson $t$-channel singularities.

There have been previous studies of the effects of a thermal hadronic medium on charm mesons
\cite{Fuchs:2004fh,He:2011yi,Montana:2020lfi,Montana:2020vjg}.
In these studies, isospin splittings have been ignored 
and therefore the possibility of $t$-channel singularities has not been considered.
It might be worthwhile to look for other aspects of the thermal physics of charm mesons
in which the effects of $t$-channel singularities are significant.
One such aspect is the production of the exotic heavy hadrons $X(3872)$ and $T_{cc}^+(3875)$.
Their tiny binding energies relative to a charm-meson-pair threshold imply that they
are loosely bound charm-meson molecules.
In previous studies of the production of charm-meson molecules,   it has been assumed
that they are produced  before kinetic freezeout
\cite{Cho:2013rpa,MartinezTorres:2014son,ExHIC:2017smd,Chen:2021akx,Hu:2021gdg,Abreu:2022lfy,Yoon:2022voo}.
It is possible that $t$-channel singularities could have a significant effect on their production after kinetic freezeout. 

The problem of $t$-channel singularities is an unavoidable aspect of reactions involving unstable particles.
Unstable particles are ubiquitous in hadronic physics. In the Standard Model of particle physics,
the weak bosons and the Higgs are unstable particles.
Most models of physics beyond the Standard Model have unstable particles.
We have identified a simple aspect of charm-meson physics in which the effects of $t$-channel singularities
are significant.
This provides encouragement  to look for other effects of $t$-channel singularities
in hadronic, nuclear, and particle physics.

\begin{acknowledgments}
    KI would like to thank Ulrich Heinz for many helpful discussions during the early  stages of this project.
    This work was supported in part by the U.S.\  Department of Energy under grant DE-SC0011726,
    by the Ministry of Science, Innovation and Universities of Spain under grant BES-2017-079860,
    by the National Natural Science Foundation of China (NSFC) under grant 11905112,
    by the Natural Science Foundation of Shandong Province of China under grant ZR2019QA012,
    by the Alexander von Humboldt Foundation,
    and by NSFC and the Deutsche Forschungsgemeinschaft (DFG)
    through the Sino-German Collaborative Research Center TRR110
    (NSFC grant 12070131001, DFG Project-ID 196253076-TRR110).
\end{acknowledgments}

\appendix


\section{Feynman rules for HH$\bm{\chi}$EFT}
\label{sec:FeynmanRules}

In $\chi$EFT, the propagator for a pion with momentum $p$ and isospin indices $i,j$ is
\begin{equation}
    \boxed{\frac{i \, \delta^{ij}}{p^2-m_\pi^2  + i \epsilon}.}
\end{equation}
At LO  in $\chi$EFT, the only interaction parameter for pions is the pion decay constant $f_\pi$.
The four-pion vertex  is
\bea
&&\pi^i(p)  \pi^j(q) \to  \pi^m(p^\prime)  \pi^n(q^\prime):
\nonumber\\
&& \boxed{\frac{2 i}{f_\pi^2}
    \left[ s \, \delta^{ij} \delta^{mn} + t \, \delta^{im} \delta^{jn} + u \, \delta^{in} \delta^{jm}
        - \frac{3m_\pi^2 + Q^2}{3} (\delta^{ij} \delta^{mn} + \delta^{im} \delta^{jn} + \delta^{in} \delta^{jm}) \right],}
\label{4pivertex}
\eea
where the Mandelstam variables are $s = (p+q)^2$, $t=(p-p^\prime)^2$, and $u = (p-q^\prime)^2$
and $Q^2=p^2 + q^2 + p^{\prime \,2}  + q^{\prime\, 2}- 4 m_\pi^2$.

The interactions of charm mesons with pions can be described using heavy-hadron chiral effective field theory (HH$\chi$EFT).
In HH$\chi$EFT, the 4-momentum of a charm meson is expressed as the sum of $Mv$,
with $v$ a velocity 4-vector that satisfies $v^2 = 1$, and a residual 4-momentum $p$.
The propagator for a pseudoscalar charm meson $D$ with momentum $M v+p$ and  isospin indices $a,b$ is
\begin{equation}
    \boxed{\frac{i \, \delta_{ab}}{2(v \cdot p + i \epsilon)}.}
\end{equation}
In amplitudes that are sensitive to isospin splittings, $v \cdot p $ in the propagator for $D^a$
should be replaced by $v \cdot p - (M_a-M)$.
If $D^a$ can be on its mass shell,
the term $+ i \epsilon$ in the denominator should be replaced by
$ + i \Gamma_a/2$, where $\Gamma_a$ is the width of $D^a$.
The propagator for a vector charm meson $D^\ast$ with momentum $M v+p$ and  isospin indices $a,b$ is
\begin{equation}
    \boxed{ \frac{i \, \delta_{ab} \, (-g_{\mu\nu} + v_{\mu}v_{\nu})}{2(v \cdot p -\Delta + i \epsilon)},}
    \label{D*prop}
\end{equation}
where $\Delta = M_\ast - M$ is the mass difference between the vector and pseudoscalar meson.
The mass shell for $D^{(\ast)}$ is $v \cdot p = \Delta$.
In amplitudes that are sensitive to isospin splittings, $v \cdot p - \Delta$ in the propagator for $D^{*a}$
should be replaced by $v \cdot p - (M_{*a}-M)$.
If $D^{\ast a}$ can be on its mass shell,
the term $ + i \epsilon$ in the denominator should be replaced by
$ + i \Gamma_{\ast a}/2$, where $\Gamma_{\ast a}$ is the width of $D^{\ast a}$.

The interactions between charm meson and pions in HH$\chi$EFT at LO are determined by
the pion decay constant $f_\pi$ and a dimensionless coupling constant $g_\pi$.
The vertices  for $D^{(\ast)} \, \pi \to D^{(\ast)} \, \pi$ are
\begin{subequations}
    \bqa
    D^a \, \pi^i(q)  \longrightarrow  D^b\, \pi^j(q^\prime) :  && ~
    \boxed{+\frac{i}{2 f_\pi^2} \, v \!\cdot\! (q + q^\prime) \, [\sigma^i, \sigma^j]_{ab},}
    \\
    D^{\ast a}_\mu \, \pi^i(q) \longrightarrow  D^{\ast b}_\nu\, \pi^j(q^\prime): &&  ~
    \boxed{-\frac{i}{2 f_\pi^2} \,  g^{\mu \nu} \, v \!\cdot\! (q + q^\prime) \,  [\sigma^i, \sigma^j]_{ab}.}
    \eqa
\end{subequations}
The vertices for $D^{(\ast)} \to D^{(\ast)}\pi(q)$ are
\begin{subequations}
    \bqa
    D^{\ast a}_\mu \longrightarrow  D^b\,  \pi^i(q): &&  ~
    \boxed{+i \frac{\sqrt{2} \,g_\pi}{f_\pi} \,   \sigma^i_{ab} \, q^\mu, }
    \label{D*toDpi}
    \\
    D^a \,  \longrightarrow  D^{\ast b}_\mu\,  \pi^i (q): && ~
    \boxed{- i \frac{\sqrt{2} \,g_\pi}{f_\pi} \,  \sigma^i_{ab} \, q^\mu, }
    \label{DtoD*pi}
    \\
    D^{\ast a}_\mu \,  \longrightarrow  D^{\ast b}_\nu\,  \pi^i(q): && ~
    \boxed{+i\frac{\sqrt{2} g_\pi}{f_\pi} \, \sigma^i_{ab} \, \epsilon^{\mu \nu \alpha \beta}\, v_\alpha\, q_\beta. }
    \label{D*toD*pi}
    \eqa
\end{subequations}
The vertices for $D^{(\ast)}\pi(q) \to D^{(\ast)}$ are obtained by replacing $q$ by $-q$.
Our convention for the Levi-Civita tensor in Eq.~\eqref{D*toD*pi} is $\epsilon_{0123} =+1$.


\section{Integrals over the Momentum of a Thermal Pion }
\label{app:PionIntegral}

In this Appendix, we evaluate the integrals over the pion momentum
that appear in the on-shell charm-meson self energies in HH$\chi$EFT at LO.

\subsection{$\bm{i\, \epsilon}$ Prescriptions}
\label{sec:iepsilon}

The on-shell self energies for the charm mesons $D^a$ and $D^{\ast a}$  are given in
Eqs.~\eqref{PiDa-0} and \eqref{PiD*a-0}.
The thermal average that appears in these self energies is
\beq
\mathcal{F}_{cd} =
\left \langle \frac{q^2}{\omega_{cdq}\, (q^2 - q_{cd}^2 + i \epsilon)} \right\rangle,
\label{Fcd}
\eeq
where $\omega_{cdq} = \sqrt{m_{\pi cd}^2 + q^2}$,
$q_{cd}^2 = \Delta_{cd}^2- m_{\pi cd}^2$, $\Delta_{cd}$ is the $D^{\ast c}$-$D^d$ mass splitting,
and $m_{\pi cd}$ is the mass of the pion produced by the transition $D^{\ast c} \to D^d \pi^i$.
The angular brackets represent the average over the Bose-Einstein momentum distribution of the pion,
which is defined in Eq.~\eqref{<F>} as the ratio of momentum integrals.

The numerator of the thermal average $\mathcal{F}_{cd}$ in Eq.~\eqref{Fcd} can be expressed as an  integral
over a single momentum variable of the form
\beq
\mathcal{F}(\sigma) =
\lim_{\epsilon \to 0^+} \int_0^\infty \mathrm{d}q \,  \frac{F(q^2)}{q^2 - \sigma + i\epsilon} ,
\label{int-n}
\eeq
where $F(q^2)$ is a smooth, real-valued function that decreases as $q^4$ as $q \to 0$
and decreases exponentially to 0 as $q \to \infty$.
The real parameter $\sigma$, which can be positive or negative, is small compared to the scale of $q^2$ set by $F(q^2)$.
We would like to expand $\mathcal{F}(\sigma)$ in powers of $\sigma$.

The function $\mathcal{F}(\sigma)$ can be expressed as the sum of a principal-value integral  and the integral of a delta function:
\bqa
\mathcal{F}(\sigma)  &=& \int_0^\infty \mathrm{d}q \, F(q^2)
\left(\mathcal{P}  \frac{1}{q^2 - \sigma}  - i \pi \, \delta(q^2 - \sigma) \right)
\nonumber\\
&=& \int_0^\infty \mathrm{d}q \, \frac{F(q^2) - F(\sigma)}{q^2 - \sigma}
- i \, \frac{\pi}{2 \sqrt{\sigma}}\; F(\sigma)  \, \theta(\sigma).
\label{int-1}
\eqa
We have used an identity to express the principal-value integral in terms of an ordinary integral.
The Taylor expansion of the real part of $\mathcal{F}(\sigma)$ can be obtained
by expanding the integrand in the second line of Eq.~\eqref{int-1} as a Taylor expansion in $\sigma$:
\bqa
\mathrm{Re}\big[\mathcal{F}( \sigma)\big] &=&
\int_0^\infty \mathrm{d}q \, \frac{F(q^2) - F(0)}{q^2}
+ \sigma \int_0^\infty \mathrm{d}q \, \frac{F(q^2) - F(0) - F^\prime(0)\, q^2 }{q^4}
\nonumber\\
&& +\,  \sigma^2
\int_0^\infty \mathrm{d}q \, \frac{F(q^2) - F(0) - F^\prime(0) \, q^2 - \tfrac12\, F^{\prime \prime}(0)\,  q^4}{q^6}
+ \ldots.
\eqa
The first term $F(q^2)/(q^2)^n$ in each integrand can be obtained simply by expanding
the left side of Eq.~\eqref{int-1} in powers of $\sigma$.
The remaining terms in the integrand subtract
the divergent terms in the Laurent expansion of $F(q^2)/(q^2)^n$ in $q^2$.

\subsection{Integral over Momentum}
\label{app:IntMom}

The thermal average  $\mathcal{F}_{cd}$ is defined in Eq.~\eqref{Fcd}.
If $\Delta_{cd} > m_{\pi cd}$, its real part can be expressed in terms of a principal-value integral
that can be reduced to the form in the first term of the second line of Eq.~\eqref{int-1}:
\bqa
\mathrm{Re}\big[\mathcal{F}_{cd}\big] &=&
\frac{1}{2\pi^2\, \mathfrak{n}_\pi} \int_0^\infty dq
\left( \frac{q^4}{\omega_{cdq}} \mathfrak{f}_\pi(\omega_{cdq}) -  \frac{q_{cd}^4}{\Delta_{cd}} \mathfrak{f}_\pi(\Delta_{cd}) \right)
\frac{1}{q^2 - q_{cd}^2},
\label{ReFcd}
\eqa
where $\omega_{cdq}= \sqrt{m_{\pi  cd}^2 + q^2}$ and $q_{cd}^2 = \Delta_{cd}^2 - m_{\pi cd}^2$.
The denominator has a zero at $q_{cd} = \sqrt{\Delta_{cd}^2 - m_{\pi cd}^2}$.
The subtraction of the numerator makes the integral convergent.
If $\Delta_{cd} < m_{\pi cd}$, the subtraction at $\omega_{cdq} =  \Delta_{cd}$ is
not necessary because the integral is convergent.

The imaginary part of $\mathcal{F}_{cd}$ is nonzero only if $\Delta_{cd} > m_{\pi cd}$.
It can be evaluated analytically using a delta function as in Eq.~\eqref{int-1}:
\bqa
\mathrm{Im}\big[\mathcal{F}_{cd}\big] &=&
- \frac{1}{4\pi \, \mathfrak{n}_\pi}
\left[ \frac{ \mathfrak{f}_\pi(\Delta_{cd})}{\Delta_{cd}} \, q_{cd}^3 \right]
\theta \big(  \Delta_{cd} - m_{\pi cd} \big) .
\label{ImFcd}
\eqa

\subsection{Expansion in Isospin Splittings}
\label{sec:ExpandSplitting}

The thermal average over the Bose-Einstein distribution for a pion is defined in Eq.~\eqref{<F>}.
The thermal average  $\mathcal{F}_{cd}$ defined in Eq.~\eqref{Fcd}
depends on $\Delta_{cd}^2- m_{\pi cd}^2$, which is linear in isospin splittings.
The thermal average can be expanded in powers of $\Delta_{cd}^2 - m_{\pi cd}^2$
using the results presented in Section~\ref{sec:iepsilon}.
The real part can be expanded in integer powers of isospin splittings divided by $m_\pi$.
The leading term in the expansion of the real part of $\mathcal{F}_{cd}$ is
\bqa
\mathrm{Re}\big[\mathcal{F}_{cd}\big] &\approx&
\left \langle \frac{1}{\omega_q} \right\rangle .
\label{ReFcd:expand}
\eqa
The imaginary part of $\mathcal{F}_{cd}$ is nonzero only if $\Delta_{cd} > m_{\pi cd}$.
It can be expanded in half-integer powers of isospin splittings divided by $m_\pi$.
The leading term in the expansion of the imaginary part is
\bqa
\mathrm{Im}\big[\mathcal{F}_{cd}\big] &\approx&
\frac{\mathfrak{f}_\pi(m_\pi)}{4\pi\, \mathfrak{n}_\pi}
\left( -\frac{1}{m_\pi} \, q_{cd}^3 \right)
\theta \big(  \Delta_{cd} - m_{\pi cd} \big).
\label{ImFcdleading}
\eqa



\begin{thebibliography}{10}

    \bibitem{Grzadkowski:2021kgi}
    B.~Grzadkowski, M.~Iglicki and S.~Mr\'owczy\'nski,
    $t$-channel singularities in cosmology and particle physics,
Nucl.\ Phys.\ B \textbf{984}, 115967 (2022)
    [arXiv:2108.01757].

    \bibitem{Peierls:1961zz}
    R.F.~Peierls,
    Possible Mechanism for the Pion-Nucleon Second Resonance,
    Phys.\ Rev.\ Lett.\ \textbf{6}, 641-643 (1961).

    \bibitem{Melnikov:1996na}
    K.~Melnikov and V.G.~Serbo,
    New type of beam size effect and the $W$ boson production at $\mu^+ \mu^-$ colliders,
    Phys.\ Rev.\ Lett.\ \textbf{76}, 3263 (1996)
    [arXiv:hep-ph/9601221].

    \bibitem{Iglicki:2022jjf}
    M.~Iglicki,
    Thermal regularization of t-channel singularities in cosmology and particle physics: the general case,
  JHEP \textbf{06}, 006 (2023)
  [arXiv:2212.00561].

    \bibitem{Braaten:2022qag}
    E.~Braaten, R.~Bruschini, L.-P.~He, K.~Ingles and J.~Jiang,
    Evolution of charm-meson ratios in an expanding hadron gas,
    Phys.\ Rev.\ D \textbf{107},  076006 (2023)
    [arXiv:2209.04972].

    \bibitem{Weinberg:1978kz}
    S.~Weinberg,
    Phenomenological Lagrangians,
    Physica A \textbf{96}, 327-340 (1979).

    \bibitem{Burdman:1992gh}
    G.~Burdman and J.F.~Donoghue,
    Union of chiral and heavy quark symmetries,
    Phys.\ Lett.\ B \textbf{280}, 287 (1992).

    \bibitem{Wise:1992hn}
    M.B.~Wise,
    Chiral perturbation theory for hadrons containing a heavy quark,
    Phys.\ Rev.\ D \textbf{45}, R2188 (1992).

    \bibitem{Cheng:1992xi}
    H.Y.~Cheng, C.Y.~Cheung, G.L.~Lin, Y.C.~Lin, T.M.~Yan and H.L.~Yu,
    Chiral Lagrangians for radiative decays of heavy hadrons,
    Phys.\ Rev.\ D \textbf{47}, 1030 (1993)
    [arXiv:hep-ph/9209262].

    \bibitem{Bass:1998ca}
    S.A.~Bass,
    \textit{et al.},
    Microscopic models for ultrarelativistic heavy ion collisions,
    Prog.\ Part.\ Nucl.\ Phys.\ \textbf{41}, 255-369 (1998)
    [arXiv:nucl-th/9803035].

    \bibitem{Kolb:2003dz}
    P.F.~Kolb and U.W.~Heinz,
    Hydrodynamic description of ultrarelativistic heavy ion collisions,
    [arXiv:nucl-th/0305084].

    \bibitem{Gelis:2010nm}
    F.~Gelis, E.~Iancu, J.~Jalilian-Marian and R.~Venugopalan,
    The Color Glass Condensate,
    Ann.\ Rev.\ Nucl.\ Part.\ Sci.\ \textbf{60}, 463-489 (2010)
    [arXiv:1002.0333].

    \bibitem{Busza:2018rrf}
    W.~Busza, K.~Rajagopal and W.~van der Schee,
    Heavy Ion Collisions: The Big Picture, and the Big Questions,
    Ann.\ Rev.\ Nucl.\ Part.\ Sci.\ \textbf{68}, 339-376 (2018)
    [arXiv:1802.04801].

    \bibitem{Elfner:2022iae}
    H.~Elfner and B.~M\"uller,
    The exploration of hot and dense nuclear matter: Introduction to relativistic heavy-ion physics,
    [arXiv:2210.12056].

    \bibitem{ExHIC:2011say}
    S.~Cho \textit{et al.} [ExHIC],
    Studying Exotic Hadrons in Heavy Ion Collisions,
    Phys.\ Rev.\ C \textbf{84}, 064910 (2011)
    [arXiv:1107.1302].

    \bibitem{ExHIC:2017smd}
    S.~Cho \textit{et al.} [ExHIC],
    Exotic hadrons from heavy ion collisions,
    Prog.\ Part.\ Nucl.\ Phys.\ \textbf{95}, 279-322 (2017)
    [arXiv:1702.00486].

    \bibitem{Bjorken:1982qr}
    J.D.~Bjorken,
    Highly Relativistic Nucleus-Nucleus Collisions: The Central Rapidity Region,
    Phys.\ Rev.\ D \textbf{27}, 140-151 (1983)

    \bibitem{Ko:1998fs}
    C.M.~Ko, B.~Zhang, X.N.~Wang and X.F.~Zhang,
    Charmonium production from hot hadronic matter,
    Phys.\ Lett.\ B \textbf{444}, 237-244 (1998)
    [arXiv:nucl-th/9808032].

    \bibitem{Chen:2003tn}
    L.W.~Chen, V.~Greco, C.M.~Ko, S.H.~Lee and W.~Liu,
    Pentaquark baryon production at the Relativistic Heavy Ion Collider,
    Phys.\ Lett.\ B \textbf{601}, 34-40 (2004)
    [arXiv:nucl-th/0308006].

    \bibitem{Chen:2007zp}
    L.W.~Chen, C.M.~Ko, W.~Liu and M.~Nielsen,
    $D_{sJ}(2317)$ meson production at RHIC,
    Phys.\ Rev.\ C \textbf{76}, 014906 (2007)
    [arXiv:0705.1697].

    \bibitem{Abreu:2020ony}
    L.M.~Abreu,
    $X_J(2900)$ states in a hot hadronic medium,
    Phys.\ Rev.\ D \textbf{103}, 036013 (2021)
    [arXiv:2010.14955].

    \bibitem{Braun-Munzinger:2003pwq}
    P.~Braun-Munzinger, K.~Redlich and J.~Stachel,
    Particle production in heavy ion collisions,
    [arXiv:nucl-th/0304013].

    \bibitem{Andronic:2003zv}
    A.~Andronic, P.~Braun-Munzinger, K.~Redlich and J.~Stachel,
    Statistical hadronization of charm in heavy ion collisions at SPS, RHIC and LHC,
    Phys.\ Lett.\ B \textbf{571}, 36-44 (2003)
    [arXiv:nucl-th/0303036].

    \bibitem{Andronic:2021erx}
    A.~Andronic, P.~Braun-Munzinger, M.~K.~K\"ohler, A.~Mazeliauskas, K.~Redlich, J.~Stachel and V.~Vislavicius,
    The multiple-charm hierarchy in the statistical hadronization model,
    JHEP \textbf{07}, 035 (2021)
    [arXiv:2104.12754].

    \bibitem{ALICE:2019hno}
    S.~Acharya \textit{et al.} [ALICE],
    Production of charged pions, kaons, and (anti-)protons in Pb-Pb and inelastic $pp$ collisions at $\sqrt {s_{NN}}$ = 5.02 TeV,
    Phys.\ Rev.\ C \textbf{101}, 044907 (2020)
    [arXiv:1910.07678].

    \bibitem{Andronic:2017pug}
    A.~Andronic, P.~Braun-Munzinger, K.~Redlich and J.~Stachel,
    Decoding the phase structure of QCD via particle production at high energy,
    Nature \textbf{561}, no.7723, 321-330 (2018)
    [arXiv:1710.09425].

    \bibitem{Gasser:1986vb}
    J.~Gasser and H.~Leutwyler,
    Light Quarks at Low Temperatures,
    Phys.\ Lett.\ B \textbf{184}, 83-88 (1987).

    \bibitem{Goity:1989gs}
    J.L.~Goity and H.~Leutwyler,
    On the Mean Free Path of Pions in Hot Matter,
    Phys.\ Lett.\ B \textbf{228}, 517-522 (1989).

    \bibitem{Schenk:1993ru}
    A.~Schenk,
    Pion propagation at finite temperature,
    Phys.\ Rev.\ D \textbf{47}, 5138-5155 (1993).

    \bibitem{Toublan:1997rr}
    D.~Toublan,
    Pion dynamics at finite temperature,
    Phys.\ Rev.\ D \textbf{56}, 5629-5645 (1997)
    [arXiv:hep-ph/9706273].
 
    \bibitem{Fuchs:2004fh}
    C.~Fuchs, B.~V.~Martemyanov, A.~Faessler and M.I.~Krivoruchenko,
    D-mesons and charmonium states in hot pion matter,
    Phys.\ Rev.\ C \textbf{73}, 035204 (2006)
    [arXiv:nucl-th/0410065].

    \bibitem{He:2011yi}
    M.~He, R.~J.~Fries and R.~Rapp,
    Thermal Relaxation of Charm in Hadronic Matter,
    Phys.\ Lett.\ B \textbf{701}, 445-450 (2011)
    [arXiv:1103.6279].

    \bibitem{Montana:2020lfi}
    G.~Monta\~na, \`A.~Ramos, L.~Tolos and J.M.~Torres-Rincon,
    Impact of a thermal medium on $D$ mesons and their chiral partners,
    Phys.\ Lett.\ B \textbf{806}, 135464 (2020)
    [arXiv:2001.11877].

    \bibitem{Montana:2020vjg}
    G.~Monta\~na, \`A.~Ramos, L.~Tolos and J.M.~Torres-Rincon,
Pseudoscalar and vector open-charm mesons at finite temperature,
    Phys.\ Rev.\ D \textbf{102}, 096020 (2020)
    [arXiv:2007.12601].

\bibitem{Cho:2013rpa}
S.~Cho and S.H.~Lee,
Hadronic effects on the $X(3872)$ meson abundance in heavy ion collisions,
Phys.\ Rev.\ C \textbf{88}, 054901 (2013)
[arXiv:1302.6381].

\bibitem{MartinezTorres:2014son}
A.~Martinez Torres, K.P.~Khemchandani, F.S.~Navarra, M.~Nielsen and L.M.~Abreu,
On $X(3872)$ production in high energy heavy ion collisions,
Phys.\ Rev.\ D \textbf{90}, 114023 (2014)
[arXiv:1405.7583].

\bibitem{Chen:2021akx}
B.~Chen, L.~Jiang, X.H.~Liu, Y.~Liu and J.~Zhao,
$X(3872)$ Production in Relativistic Heavy-Ion Collisions,
Phys.\ Rev.\ C \textbf{105}, 054901 (2022)
[arXiv:2107.00969].

\bibitem{Hu:2021gdg}
Y.~Hu, J.~Liao, E.~Wang, Q.~Wang, H.~Xing and H.~Zhang,
Production of doubly charmed exotic hadrons in heavy ion collisions,
Phys.\ Rev.\ D \textbf{104},  L111502 (2021)
[arXiv:2109.07733].

\bibitem{Abreu:2022lfy}
L.~M.~Abreu, F.~S.~Navarra and H.~P.~L.~Vieira,
Multiplicity of the doubly charmed state $T_{cc}^+$ in heavy-ion collisions,
Phys.\ Rev.\ D \textbf{105}, 116029 (2022)
[arXiv:2202.10882].

\bibitem{Yoon:2022voo}
H.O.~Yoon, D.~Park, S.~Noh, A.~Park, W.~Park, S.~Cho, J.~Hong, Y.~Kim, S.~Lim and S.H.~Lee,
$X(3872)$ and $T_{cc}$: structures and productions in heavy ion collisions,
Phys.\ Rev.\ C \textbf{107},  014906 (2023)
[arXiv:2208.06960].

\end{thebibliography}
\end{document}